\documentclass [11pt] {report}

\usepackage{amssymb,latexsym,revsymb}
\usepackage{amsmath,amsbsy}
\usepackage{lscape}
\usepackage{indentfirst}
\usepackage{latexsym}
\usepackage{multirow}
\usepackage{tabls}
\usepackage{wrapfig}
\usepackage{slashbox}
\usepackage{longtable}
\usepackage{amsmath,amssymb,amsbsy}
\usepackage{graphicx,subfigure}
\usepackage{epsfig,bm}
\usepackage{graphicx,comment}
\usepackage{cite}
\usepackage{alltt}  %

\def\a{{\alpha}}

\def\d{{\delta}}
\def\D{{\Delta}}
\def\e{{\epsilon}}
\def\g{{\gamma}}

\def\l{{\lambda}}
\def\L{{\Lambda}}

\def\w{{\omega}}

\def\S{{\Sigma}}
\def\s{{\sigma}}
\def\t{{\tau}}
\def\th{{\theta}}

\def\X{{\Xi}}

\def\ol#1{{\overline{#1}}}

\def\Dslash{D\hskip-0.65em /}

\def\vslash{{\rlap \slash v}}

\def\CPT{{$\chi$PT$\;$}}

\def\diag{\text{diag}}

\def\bm{\beta_M}

\def\mp{{m_\pi}}
\def\fp{{f_\pi}}
\def\fps{{f_\pi^s}}
\def\fpt{{f_\pi^t}}
\def\delmd{{\partial_\mu}}
\def\delmu{{\partial^\mu}}
\def\delnd{{\partial_\nu}}
\def\delnu{{\partial^\nu}}

\def\mc#1{{\mathcal #1}}

\def\cL{{\mathcal L}}
\def\cO{{\mathcal O}}

\def\cM{{\mathcal M}}
\def\cD{{\mathcal D}}

\def\eqref#1{{(\ref{#1})}}

\def\ol#1{{\overline{#1}}}

\newcommand{\beq}{\begin{equation}}
\newcommand{\eeq}{\end{equation}}
\newcommand{\bea}{\begin{eqnarray}}
\newcommand{\eea}{\end{eqnarray}}

\newcommand{\nn}{\nonumber}
\newcommand{\benn}{\begin{displaymath}}
\newcommand{\eenn}{\end{displaymath}}

\newcommand{\tr}{{\rm tr}}

\def\mc#1{{\mathcal #1}}

\newcommand{\ee}{\mathbf e}

\def\slashchar#1{\ensuremath{                               %
   \setbox0=\hbox{${}#1{}$}       
   \dimen0=\wd0                                 
   \setbox1=\hbox{/} \dimen1=\wd1               
   \ifdim\dimen0>\dimen1                        
      \rlap{\hbox to \dimen0{\hfil/\hfil}}      
      {}#1{}                                    
   \else                                        
      \rlap{\hbox to \dimen1{\hfil${}#1{}$\hfil}}   
      /                                         
   \fi}}                                        %

\def\psit{{\tilde{\psi}}}

\begin{document}

\sloppy

\title{{\bf{\Huge TOPICS IN LATTICE QCD\\
AND \\
EFFECTIVE FIELD THEORY\\
}}}
\author{{\LARGE Michael I.~Buchoff}\\
\\
\\
\\
\\
A dissertation submitted in partial fulfillment of\\
the requirements for the degree of\\
\\
Doctor of Philosophy\\
\\
\\
Advisor: Paulo F.~Bedaque\\
Associate Professor  \\
Department of Physics\\
University of Maryland
}
\date{April 2010}

\maketitle

\thispagestyle{empty}
\cleardoublepage{\ }

\abstract{
Effective field theories provide a formalism for categorizing low-energy effects of a high-energy fundamental theory in terms of the low-energy degrees of freedom.  This process has been well established in mapping the fundamental theory of QCD in terms of the hadronic degrees of freedom, which allows for quantitative connections and predictions between hardronic observables.  

A more direct approach to performing the non-perturbative QCD calculations is through lattice QCD.  These computationally intensive calculations approximate continuum physics with a discretized lattice to extract hadronic phenomena from first principles.  However, as in any approximation, there are multiple systematic errors between lattice QCD calculation and actual hardronic  phenomena.  

To account for these systematic effects in terms of hadronic interactions, effective field theory proves to be useful.  However, the fundamental theory of interest here is lattice QCD, as opposed to the usual continuum QCD.  In this work, the basics of this process are outlined, and multiple original calculations are presented: effective field theory for anisotropic lattices, I=2 $\pi\pi$ scattering for isotropic, anisotropic, and twisted mass lattices.  Additionally, a usage of effective field theories and the employment of an isospin chemical potential on the lattice is proposed to extract several computationally difficult scattering parameters.  Lastly, recently proposed local, chiral lattice actions are analyzed in the framework of effective field theory, which illuminates various challenges in simulating such actions. 
}

\thispagestyle{empty}
\cleardoublepage{\ }

\thispagestyle{empty}

\pagenumbering{roman}
\setcounter{page}{0}
\setcounter{tocdepth}{3}

\tableofcontents


\chapter*{List of Abbreviations}      
\thispagestyle{plain}

\noindent {\bf QCD}:  Quantum chromodynamics. 

\bigskip
\noindent {\bf EFT}: Effective field theory.

\bigskip
\noindent {\bf LECs}: Low-energy constants.

\bigskip
\noindent {\bf \CPT}: Chiral perturbation theory.

\bigskip
\noindent {\bf L\CPT}: Lattice chiral perturbation theory.

\bigskip
\noindent {\bf HB\CPT }:  Heavy baryon chiral perturbation theory.

\bigskip
\noindent {\bf LO}: Leading order.

\bigskip
\noindent {\bf NLO}: Next-to-leading order.

\bigskip
\noindent {\bf N$\mathbf{{}^n}$LO}: Next-to-$($next-to-$)^{n-1}$-leading order, for $n \geq 1$.

\cleardoublepage{\ }

\thispagestyle{empty}

\setcounter{page}{1}
\pagenumbering{arabic}


\renewcommand{\thechapter}{1}

\chapter{Introduction} \label{ch:Intro}
Quantum chromodynamics (QCD) is the fundamental theory that governs hadrons (mesons and baryons) and hadronic interactions.  In QCD, fermions in the one-index fundamental representation of the $SU(3)$ gauge group\footnote{This is equivalent to two-index antisymmetric representation for an $SU(3)$ gauge group}, also known as quarks, are coupled with spin-1 bosons of the same gauge group, known as gluons.  This non-abelian $SU(3)$ gauge group, whose rank is referred to as color, contains 8 generators, each of which are associated with a gluon in the theory.  All hadrons are ``color neutral"; baryons contain three quarks in a state antisymmetric in color, and mesons contain a quark and anti-quark that form a color singlet.   The QCD Lagrangian including the three lightest quarks (up, down, strange) is given by
\beq\label{eq:QCD}
\mc L = -\frac{1}{4}F_{\mu\nu}^aF^{\mu\nu}_{a} + \sum_{f=1}^3 \ol \psi_{f i} \big(i\Dslash + m_{q_f}\big)_{i j} \psi_{f j} 
\eeq
where $\psi$ is the quark fields with flavor index $f$ and color index $i,j$, $D_{\mu,i j} = \partial_\mu\mathbb{I}_{i j}+igA_{\mu,i j}$ with $A_{\mu,i j} = A_\mu^c T^c_{i j}$ and $T^c_{i j}$ being the eight generators of $SU(3)$, and the kinetic and self interactions of the gauge fields are given by $F_{\mu\nu}^{a} = \partial_\mu A_\nu^{a} - \partial_\nu A_\mu^{a} +i  [A^\mu,A^\nu]^a$.
  
The most significant properties of QCD are that the theory is both asymptotically free and confining.  In other words, at short distance/high-energy, the coupling becomes perturbatively small and at long distances/low-energy, the coupling becomes strong.   A perturbative calculation of the QCD beta function at high energy yields a negative sign (for the number of flavors of quarks in the standard model), which leads to a decaying coupling constants as the energy increases (asymptotic freedom).   While not mathematically proven, QCD is believed to be confining, where it requires more and more energy to separate a quark or anti-quark in a hadron.  What makes this statement so difficult to prove is the fact that at long distances the theory is strongly coupled, and thus, the usual perturbative arguments break down.  As a result, another method is required to compute non-perturbative physics, and lattice QCD is the best option available.
 
\section{Lattice QCD}
In quantum field theory, there exists multiple methods for regulating divergent calculations.  One such method is to use a lattice; that is, to break up the continuum of space and time into many individual sites, which is often referred to as discretizing space and time.  For every finite sized lattice, there are two parameters introduced: the lattice spacing between sites and the total length of the lattice.  For a simple cubic lattice (hypercubic if time is included), these two parameters in a given direction are simply related by the number of sites in that direction.  By definition, this  particular regulator breaks the continuous translational symmetry to a discrete translational symmetry, but as we take the continuum limit (the spacing between sites goes to zero as the length of the lattice goes to infinity), we hope to recover the continuum result.

Non-perturbative lattice calculations take advantage of this regulator along with the path integral representation to perform calculations that do not require a perturbatively small parameter.  The path integral approach requires defining a partition function given by
\beq
Z = \int d [\ol\psi]d[\psi]d[A] e^{iS(\ol \psi, \psi, A)},
\eeq
where $d[\psi]= \prod_n d\psi_n$ represents all possible paths of $\psi$ and the action is related to the Lagrangian by $S(\ol \psi, \psi, A) = \int d^4x \mc L(\ol \psi, \psi, A)$.  Another useful quantity to define is the correlation function given by 
\beq
\langle f(\ol \psi, \psi, A) \rangle = Z^{-1}\int d [\ol\psi]d[\psi]d[A] f(\ol \psi, \psi, A) e^{iS(\ol \psi, \psi, A)},
\eeq
which is analogous to the definition of the expectation value in statistical mechanics.  Both of these equations require (infinitely) many multiple-dimensional integrals to compute, which is computationally difficult for non-perturbative systems.  Rather than the diagrammatic approach from perturbation theory, lattice calculations utilize a Monte Carlo integration method\footnote{For more detailed information, see Ref.~\cite{DeGrand:2006zz}.}, where the exponential of the Euclidean rotated\footnote{Transformation defined by going to imaginary time, $t \rightarrow i \tau$.  As a result the integrand $e^{iS} \rightarrow e^{-S}$.}  path integral acts as a probability distribution.  As a result of analyzing correlation functions in this framework, energy differences and matrix elements can be extracted.  The example of the scalar case, as in Ref.~\cite{Sharpe:1993wt}, is shown in Appendix \ref{ap:Scalar}.

There are several key issues that arise in non-perturbative lattice calculations.  One that is relevant to the results in this paper is the fermion doubling problem.  As mentioned previously, to perform numerical lattice calulations, the space and time directions are discretized, with the inverse of the lattice spacing between them acting as the ultraviolet cutoff of the theory.   However, depending on exactly how the action is discretized (in particular, the fermion terms), a single fermion attached to each node can describe multiple fermions in the continuum limit.  The most basic example of this effect is from the na\"ive fermion lattice action.  The fermionic part of this discretized action (without gauge interactions) is given by
\beq
S_{NF} = a^4 \sum_{x,\mu} \frac{1}{2a}\Big [ \ol \psi_x \gamma_\mu \psi_{x+\hat{\mu}}- \ol \psi_{x+\hat{\mu}} \gamma_\mu \psi_x \Big ],
\eeq
where, in conventional lattice notation, $a$ is the lattice spacing, $x$ is the location of a given site and $x+\hat{\mu}$ represents a site that is one lattice spacing away from $x$ in the $\mu$ direction\footnote{$\mu =1,\cdots,4$, where the fourth direction is the Euclidean rotation of the Minkowski time direction}, and $\psi_x$ and $\gamma_\mu$ are the fermionic fields and Dirac matricies, respectively.    This action is equivalent to taking the continuum kinetic action ($\ol \psi \gamma_\mu \partial_\mu \psi$)  and approximating the derivative in the na\"ive way ($\partial_x \psi(x) \approx \frac{1}{a}\big (\psi(x+a) - \psi(x)\big)$).  Using the discrete Fourier expansion
\beq
\psi_x = \frac{1}{L^4}\sum_p e^{i x \cdot p} \psi_p,
\eeq 
the action becomes
\beq
S_{NF} = \frac{1}{L^4}\sum_{p,\mu} \ol \psi_p \gamma_\mu \frac{1}{2a}(e^{i a p_\mu} -e^{-i a p_\mu}) \psi_p =  \frac{1}{L^4}\sum_{p,\mu} \ol \psi_p \gamma_\mu\Big[\frac{i}{a} \sin(a p_\mu) \Big]\psi_p,
\eeq
where $\frac{i}{a}\psi_p \gamma_\mu \sin(a p_\mu)$ is the reciprocal of the propagator.  Taking the continuum limit na\"ively ($a \rightarrow 0$) yields the correct continuum (Euclidean) behavior for the kinetic term ($\frac{i}{a} \sin(a p_\mu) \approx i p_\mu$).  However, it is important to note that the for every pole of the propagator, there is a particle.  In other words, a particle is present whenever $\sin(a p_\mu) = 0$.  Thus, values of $a p_\mu = (0, \pi)$ lead to two poles in each direction.  For a four-dimensional lattice, there are $2^4$ poles, which lead to 16 particles that are often referred to as ``doublers."  In this particular discretization, by representing a single fermion field at each node, the action is actually describing 16 fermions.  Since the goal of lattice QCD is to simulate two or three flavors, another discretization is required.

In order to achieve the correct continuum behavior of the theory being simulated on the lattice, an natural extension is to add another term to the na\"ive quark action that vanishes in the continuum limit.  As a result, the Wilson action was developed, which, in lattice terminology  is given by
\beq\label{eq:Wilson_Action}
S_{W} = a^4 \sum_{x,\mu} \frac{1}{2a}\Big [ \ol \psi_x (\gamma_\mu-r) \psi_{x+\hat{\mu}}- \ol \psi_{x+\hat{\mu}} (\gamma_\mu + r) \psi_x +r \ol \psi_x \psi_x \Big],
\eeq
where $r$ is the Wilson parameter.  While this action appears significantly different from the na\"ive quark action, it is equivalent to adding the discretized version of the term $a r \ol \psi \partial_\mu \partial_\mu \psi$.  Applying the same Fourier transform as before, 
\beq 
S_W = \frac{1}{L^4}\sum_{p,\mu} \ol \psi_p \Big [ \frac{i}{a}\gamma_\mu \sin(a p_\mu) + \frac{r}{a}\big(\cos(a p_\mu) - 1\big)\Big] \psi_p.
\eeq
This action only contains one pole (when $a p_\mu$ = 0) and reproduces the correct continuum behavior as the lattice spacing approaches zero.  Thus, this action describes only one particle.  If we included two or three flavors of this action, we can simulate two or three flavor QCD.

While this action appears to have removed the fermion doubling problem, it leads to a new problem.  The issue that arises from this discretization is that it unphysically breaks chiral symmetry\footnote{The Lagrangian $\mc L_F = \ol \psi D_F \psi$ is chirally symmetric if $\gamma_5 D_F \gamma_5 = -D_F$.  A quick way to see if an operator is chirally symmetric is to check whether it contains an odd number of $\gamma_\mu$'s.   If it has an even number of $\gamma_\mu$'s, it is not chirally symmetric (note, the number of $\gamma_5$'s does NOT alter chirality).}.  For a free action, this effect will vanish in the continuum limit.  However, QCD is a gauge theory with gauge interactions that can and will lead to radiative corrections.  Chiral symmetry restricts many operators that could be radiatively generated.  However, if chiral symmetry is broken in an unphysical way, as is the case with the Wilson action, this will lead to unphysical operators that will contaminate the final results.  What makes it even worse is that these operators can diverge in the continuum limit and require a difficult non-perturbative tuning to remove.  These issues will be explained in more detail in section \ref{ch:Aniso_EFT}.

\subsection{Chiral Symmetry in Lattice QCD}
One important question in Lattice QCD is whether or not it is possible to both resolve the fermion doubling problem and keep chiral symmetry intact.  Over the last two decades, this topic has been one of great interest and has led to several review articles \cite{Golterman:2000hr, Luscher:2000hn, Kaplan:2009yg,Poppitz:2010lq} as it is now considered a pivotal subject in lattice QCD.  

The Wilson lattice action is an example where chiral symmetry is sacrificed in order to remove the doublers.  However, this behavior is not unique to the Wilson lattice action and, in fact, Nielsen and Ninomiya proposed a no-go theorem for chiral symmetry on the lattice.  Specifically, the theorem states that a lattice action will contain doublers as long as three properties are held true: 1) Discrete translational symmetry, 2) Locality, 3) Chiral symmetry.  In other words, one or more of these three properties must be sacrificed in order to remove doublers.   Another way of understanding the Nilsen-Ninomiya theorem is in terms of the chiral anomaly.   For example, the na\"ive lattice action has an exact $U(1)_A$ symmetry which persist all the way to the continuum limit.  Continuum QCD, on the other hand, has an anomolus $U(1)_A$ symmetry that is broken by quantum corrections (in particular the one loop fermionic triangle diagrams).  Thus, in order to produce the correct continuum physics for QCD, a source for chiral symmetry breaking must be present to reproduce the correct flavor-singlet chiral anomaly defined in the standard way\footnote{As shown in Ref.~\cite{Sharatchandra:1981si}, if one defines a non-singlet axial current for the naive action, the correct chiral anomaly can be reproduced.  In other words, in the usual notation, the naive action has a flavor non-singlet chiral anomaly.}.  One way to do this is to break chiral symmetry explicitly, as in the Wilson action.  However, as mentioned before, this leads to unwanted radial corrections.  A more favorable discretization takes advantage of the Ginsparg-Wilson relation \cite{Ginsparg:1981bj}.

If a fermionic action has exact chiral symmetry, it obeys the relation
\beq
\gamma_5 D + D \gamma_5 = 0.
\eeq
Another relation akin to the one above is the Ginsberg-Wilson relation, which is given by
\beq
\gamma_5 D + D \gamma_5 = a D\gamma_5 D\quad,\quad (\gamma_5 D)^\dag = \gamma_5 D.
\eeq
This relation introduces a small source of chiral symmetry breaking at $\mc O(a)$, and in the continuum limit removes doublers, yields radiative corrections ala continuum QCD, and yields a source for the flavor-singlet chiral anomaly.  Two solutions to the Ginsparg-Wilson relation have been found, both of which are equivalent in the appropriate limit.  The first of such solutions is domain wall fermions \cite{Kaplan:1992bt,Shamir:1993zy,Furman:1994ky} where four-dimensional chiral fermions are given by a mass step function in an additional fifth direction.  In the limit of an infinite fifth direction, this yields a single four-dimensional Dirac fermion (a left-handed and right handed ``wall" in the fifth direction due to periodic boundary conditions).  In reality, simulations are performed with a finite fifth directions, but the residual chiral symmetry breaking is exponentially suppressed.  Another equally valid realization of Ginsparg-Wilson fermions are known as overlap fermions \cite{Narayanan:1992wx,Narayanan:1994gw,Neuberger:1997fp}, which satisfy the relations with a non-local four-dimensional operator.  In has been shown \cite{Neuberger:1997bg, Kikukawa:1999sy}, that including all the non-local contributions of the overlap operator is equivalent to the domain wall action with an infinite fifth direction (the two actions are related via a KaluzaÐKlein reduction).
 
While both domain wall and overlap lattice actions have favorable features, the major drawback to both actions are computational costs.  Either simulating an extra dimension or multiple non-local operators leads to computational costs about a factor  of ten greater than four-dimensional, local actions.  Thus, it would be beneficial if some cheap, local chiral action could be found, and this scenario of minimally doubled, non-orthogonal actions will be discussed in chapter \ref{ch:Non_orth}. 
 
\section{Effective Field Theory}
Throughout physics, there are multiple scales that arise when analyzing phenomena.  An effective field theory (EFT) is a theory that maps the symmetries of a more fundamental theory (usually at a higher energy scale) in terms of more relevant degrees of freedom (usually at a lower energy scale).  An example of this is the low-energy limit of the Standard Model, where the heavy, high-energy modes of the W and Z particles are ``integrated out" and act as small corrections to the low energy theory inversely proportional to powers of the W and Z mass. 

To construct an EFT from a more fundamental theory, the first step is to identify the symmetries and explicit symmetry breaking in the fundamental theory.  Next, for this approach to be controlled and fruitful, a seperation of scales is necessary.  For example, if a theory has a lower scale of $p$ and a higher scale of $\L$, then the expansion parameter in the EFT would be $p/\L$. In terms of the relevant degrees of freedom in the EFT, which depend on the particular energy ranges of interest,  one must write down all the possible operators that reflect the symmetries of the fundamental theory (the explicit symmetry breaking can be mapped through a spurion analysis) for a given order in the power counting parameter.    Each of these parameters have a coefficient known as low-energy constants (LECs) that are $\mc O(c/\L^n)$ in the power counting where $n$ is determined by  the order of the EFT and $c$ is $\mc O(1)$.  These LECs are not known \textit{a priori} and require external input to fix.  For non-perturbative theories, like QCD, the non-perturbative effects in the calculation are contained in these terms, and require a non-perturbative calculation, such as the lattice, to determine.

Throughout this work, the high-energy, fundamental theory is QCD (more specifically, lattice QCD) and the relevant degrees of freedom of interest in our EFT are hadrons, nucleons and mesons.  The particular EFT of interest will depend which limits we explore.  For example, if we are looking at characteristic momenta far below the mass of the particles of interest, non-relativistic field theory is appropriate.  Additionally, if  pion mass is small compared to the relevant cut-off scales, such as the chiral breaking scale, the pion mass can be viewed as a correction to the chiral limit ($m_\pi = 0$).  Most of this work will focus on these limits, and the resulting EFTs, namely chiral perturbation theory  and heavy baryon chiral perturbation theory  (HB\CPT). 

\subsection{Continuum Chiral Perturbation Theory}
Chiral perturbation theory (\CPT) is the mesonic EFT of QCD when the characteristic momenta are on the same order as the pion mass ($m_\pi \sim 135$ MeV), which are both well below the chiral breaking scale ($\Lambda_\chi \sim 4\pi^2 f_\pi \sim 1$ GeV).   Thus, the characteristic momenta, $p$ and the pion mass, $m_\pi$ act as the expansion parameters for continuum \CPT.   

As mentioned previously, the first step to constructing \CPT is to identify the symmetries of QCD.  The QCD lagrangian (Eq.~\eqref{eq:QCD}) has, in addition to the $SU(3)$ gauge symmetry, Lorentz symmetry, and is invariant under parity, charge conjugation, and time reversal.  However, the main symmetry of focus will be chiral symmetry.  In the massless limit, the QCD Lagrangian is chirally symmetric, while the mass term breaks the chiral symmetry explicitly.  This effect can be best seen by rewriting the fermionic part of the QCD lagrangian in terms the left and right-handed components of $\psi$:
\beq
\mc L_F = \ol \psi (\Dslash + m_q) \psi = \ol \psi_L \Dslash \psi_L + \ol \psi_R \Dslash \psi_R +\big(\ol \psi_L m_q \psi_R + \ol \psi_R m_q \psi_L\big),
\eeq
where
\bea
\psi_L &=&\Big( \frac{1 + \gamma_5}{2}\Big)\psi, \quad \psi_R = \Big(\frac{1 - \gamma_5}{2}\Big)\psi, \nn \\ \ol \psi_R &=&\ol \psi \Big(\frac{1 - \gamma_5}{2}\Big),\quad \ol \psi_L = \ol \psi \Big(\frac{1 + \gamma_5}{2}\Big),\nn
\eea
The chiral symmetric transformation in terms of these components is given by
\bea
\psi_L &=& L \psi_L, \quad \ol \psi_L = \ol \psi_L L^\dag \nn\\
\psi_R &=& R \psi_R, \quad \ol \psi_R = \ol \psi_R R^\dag, \nn
\eea
where the chiral flavor transformations $L$ and $R$ obey the relation $L^\dag L = 1$ and $R^\dag R = 1$.   The generators of the transformations are those of the $SU(N_F)$ flavor group.  In this work, we will only focus on either two or three lightest flavors, and these transformations obey $SU(2)$ and $SU(3)$ flavor, respectively.  

In the massless limit, the action is invariant under an $L$ and $R$ transformations that are independent of each other.  As a result, the action pocesses an $SU(N_F)_L \otimes SU(N_F)_R \otimes U(1)_B$, where the last $U(1)_B$ represents baryon number conservation.   Since QCD is a confining theory, the chiral condensate ($\langle \ol \psi \psi \rangle$) is non-zero, and thus this flavor symmetry is spontaneously broken.  Both phenomenogical and lattice evidence supports that this symmetry is spontaneously broken to the vector subgroup ($SU(N_F)_L \otimes SU(N_F)_R \rightarrow SU(N_F)_V$ leading to $N_F^2-1$ Nambu-Goldstone particles.  In three flavor QCD, these particles include the pions, kaons, and eta (the eta-prime is a flavor singlet whose mass differs from the eta due to the chiral anomaly).  With the knowledge in hand that non-zero chiral condensate is in the vector subgroup ($\langle \ol u u \rangle = \langle \ol d d \rangle = \langle \ol s s \rangle$), then it is adventagous to write
\beq
\langle \psi_{L i} \ol \psi_{R j} \rangle \equiv \Omega_{i j} = \omega \delta_{i j}, \quad \omega \neq 0.
\eeq
Performing a chiral transformation, this condensate becomes
\beq
\big(L \Omega R^\dag\big)_{i j} = \omega \big(L R^\dag\big)_{i j} \equiv \omega \Sigma_{i j}.
\eeq
If $L = R$, then $\Sigma_{i j} = \delta_{i j}$ leaving the vector subgroup unbroken as expected.  If $L \neq R$, then $\Sigma_{i j}$ represents a different vacuum from the vector subgroup.  As mentioned before, this vacuum, whose generators are that of $SU(N_F)$, is a matrix of the Nambu-Golstone particles (pion, kaon, eta, etc.) fields, which are the relevant degrees of freedom for \CPT.  In massless QCD, these different vacua are degenerate and each of these particles are massless.  However, since QCD does contain an explicit chiral breaking term proportional to the quark mass, this term ``tips" the potential, giving these psuedoscalar particles a mass.  To see how this occurs, it is important to follow the procedure for mapping the mass term into the chiral theory.

When constructing an effective field theory, one must write down all the possible operators at a given order that obey all the symmetries of the fundamental theory.  Thus, for mesonic observable in \CPT, the theory must be written in terms of the mesonic matrix $\Sigma_{i j}$.  Since there are no space-time indices in this matrix, the Lorentz symmetry is preserved as long as derivatives of these matrices are contracted in a Lorentz invariant way\footnote{In Chapter \ref{ch:Aniso_EFT} we will explore lattice theories that break the Euclidean equivance of Lorentz symmetry and the consequences.  As a notational point of emphasis, up and down contracted indices (such as $\partial_\mu\partial^\mu$) will represent Minkowski space and only down contracted indices (such as $\partial_\mu\partial_\mu$) will represent Euclidean space.}.  Under parity, $\Sigma \rightarrow \Sigma^\dag$, and an effective field theory of QCD should be invariant under this transformation\footnote{This is not the case for the twisted mass lattice action in Chapter \ref{ch:TM_pi_pi} or the convenient lattice action for finite isospin density in Chapter \ref{Iso_Chem}}.  Under chiral transformations, 
\beq
\Sigma \rightarrow L \Sigma R^\dag\quad,\quad \Sigma^\dag \rightarrow R \Sigma^\dag L^\dag.
\eeq
Thus, when constructing the chirally invariant \CPT Lagrangian without a mass with the monenta $p^2$ as the expansion parameter, the only combination at leading order (LO) is given by
\beq
\mc L^{m=0}_{LO} = \frac{f^2}{4}\tr\big(\partial_\mu \Sigma^\dag \partial^\mu \Sigma \big),
\eeq
where the coefficient $f$ can be related to the pion decay constant $f_\pi$.  Another possible operator is $\tr(\Sigma^\dag \Sigma)$, but since $\Sigma^\dag \Sigma = 1$, this term has no field dependence is just shifts the Lagrangian by an overall constant.  Thus, at LO in $p^2$, there is only one term when $m_q=0$.    To include the $m_q$ dependence, it is best to rewrite the mass term in QCD in a specific way:
\beq
\mc L_m = \ol \psi_L m_q \psi_R + \ol \psi_R m_q^\dag \psi_L,
\eeq
where $m_q$ is the diagonal mass matrix (in the $SU(2)$ isospin limit, $m_q = \ol m \mathbb{I}$).  This way of writing the mass term in QCD seems illogical, but it makes sense when $m_q$ is promoted to a spurion in order to account for its effects in \CPT.  Upgrading $m_q$ to the spurion\footnote{The $s$ stands for scalar part and $p$ stands for the parity violating pseudoscalar part and should not be confused with the momenta.} $s +ip$, this term  transforms under chirality the same way as $\Sigma$; namely $(s+ip) \rightarrow L (s+ip) R^\dag$.  Thus, we can now construct the next terms in the chiral Lagangian of the form
\beq
\tr\big((s+ip)\Sigma^\dag + (s-ip)\Sigma\big)
\eeq
In the real world and most lattice QCD simulations, $m_q$ is purely scalar, so we can set $s=B m_q$ and $p=0$, which correctly accounts for the chiral symmetry breaking effect of $m_q$. $B$ is proportional to the chiral condensate.  As a result, all the LO contributions to \CPT are given by
\beq\label{eq:chi_LO_2f}
\mc L_\chi^{LO} =  \frac{f^2}{8}\tr\big(\partial_\mu \Sigma^\dag \partial^\mu \Sigma \big)+\frac{B f^2}{4} \tr\big(m_q(\Sigma^\dag + \Sigma)\big).
\eeq
The matrix $\Sigma = e^{2i\phi}$ is
\beq
\phi = \begin{pmatrix} \frac{\pi^0}{\sqrt{2}} & \pi^+ \\ \pi^- & -\frac{\pi^0}{\sqrt{2}} \end{pmatrix}\quad, \quad m_q = \begin{pmatrix} m_u & 0 \\ 0 & m_d \end{pmatrix},
\eeq
for two flavors, and
\beq
\phi = \begin{pmatrix} \frac{\pi^0}{\sqrt{2}}+\frac{\eta}{\sqrt{6}} & \pi^+ & K^+ \\ \pi^-  & -\frac{\pi^0}{\sqrt{2}}+\frac{\eta}{\sqrt{6}} & K^0 \\ K^- & \ol K^0 & -\frac{2\eta}{\sqrt{6}}  \end{pmatrix}\quad, \quad m_q = \begin{pmatrix} m_u & 0 & 0\\ 0 & m_d & 0 \\ 0 & 0 & m_s  \end{pmatrix}.
\eeq
for three flavors.  The particular convention used in this Lagrangian is consistent with $f \sim 132$ MeV.

As an example of a LO two-flavor tree-level calculation for the pion mass, one must first expand  \eqref{eq:chi_LO_2f} in terms of pion fields, giving
\beq
\mc L_\chi^{LO} = \partial_\mu \pi^+ \partial^\mu \pi^- - B_0 (m_u + m_d) \pi^+\pi^-+\cdots,
\eeq
where the dots represent three or more pion interactions along with neutral pion interactions.  From this Lagrangian, the LO pion mass is given by
\beq
m_\pi^2 = B_0 (m_u + m_d),
\eeq
which is a famous result illustrating that $m_\pi^2 \propto m_q$.  Extending this analysis to three-flavor, the LO $K$ and $\eta$ mass are given by
\bea
m_{K^+}^2 &=& B_0 (m_u + m_s) \quad, \quad m_{K^0}^2 = B_0 (m_d + m_s) \nn\\
m_\eta^2 &=& B_0 \Bigg( \frac{4}{3}m_s + \frac{1}{3}m_u + \frac{1}{3}m_d \Bigg)= \frac{2}{3} m_{K^+}^2 +   \frac{2}{3} m_{K^0}^2-\frac{1}{3} m_\pi^2,
\eea
where the last equation yields the Gell-Mann-Okubo formula, which agrees with experiment up to 10\%.  The next-to-leading order (NLO) calculation of the mass requires one-loop corrections involving the LO terms in the Lagrangian and tree-level contributions from the NLO terms in the Lagrangian (terms proportional to $m_\pi^4$ or $p^4$).  Constructing the NLO requires the same procedure as the LO case, namely, writing down all possible independent operators that obey the overall symmetry.  However, the process is more complicated since the number of operators that can be written down has increased significantly.  In two-flavor \CPT, a simplification occurs ($(\Sigma^\dag + \Sigma)$ happens to be proportional to the identity matrix in flavor space) that leads to only four necessary operators at NLO:
\begin{align}
    \mc L_{cont} =& \frac{f^2}{8}\tr(\delmd \S \delmu \S^\dag) + \frac{Bf^2}{4}\tr(m_q \S^\dag + \S m_q)  \nonumber\\
    &+\frac{\ell_1}{4}\big[\tr(\delmd \S \delmu \S^\dag)\big]^2+ \frac{\ell_2}{4} \tr(\delmd \S \delnd \S^\dag) \tr(\delmu \S^\dag \delnu \S) \nonumber\\
    &+\frac{(\ell_3+\ell_4)B^2}{4}\big[\tr(m_q \S^\dag + \S m_q)\big]^2 + \frac{\ell_4B}{4}\tr(\delmd \S \delmu \S^\dag)\tr(m_q \S^\dag + \S m_q) ,
\end{align} 
where the LECs $\ell_1 - \ell_4$ are the original Gasser-Leutwyler coefficients defined in Ref.~\cite{Gasser:1983yg}. This Lagrangian will be one of the main focuses in Chapter \ref{ch:Aniso_pi_pi}.  For three-or more flavors, this simplification does not occur and 8 operators are necessary\footnote{If the action were to contain external, dyamical gauge fields, there would be more operators needed.}.  The three-flavor NLO Lagrangian will not be the focused on in this thesis, but will be included in Appendix \ref{ap:Three_flav} for completeness.

\subsection{Symanzik Action from Lattice QCD}
As mentioned previously, Lattice QCD is a discretized approximation to QCD.  For certain lattice actions in the continuum limit (the lattice spacing approaches zero),  one expects to achieve continuum QCD.  However, it is impossible to simulate with zero lattice spacings, so it would be convenient to develop a continuum EFT for the lattice action at small lattice spacings that would account for both the continuum theory and the finite $a$ corrections that are proportional to powers of the lattice spacings.    This EFT is referred to as the Symanzik action \cite{Symanzik:1983dc,Symanzik:1983gh, Sheikholeslami:1985ij, Luscher:1996sc}.

To illustrate how the Symanzik action is developed, the Wilson lattice action, Eq.~\eqref{eq:Wilson_Action} will be used.  To begin the process, the symmetries of the Wilson lattice action (now acting as our fundamental theory) must be mapped onto the continuum EFT.  Many of the symmetries are similar to continuum QCD in Euclidean space, but there are two key differences.  The first one, that was mentioned previously, is that the action breaks chiral symmetry.  Second, the lattice does not contain a continuum translation symmetry, but rather contains a discretized translational symmetry.  To be more specific, the $O(4)$ rotational symmetry of the Euclidean continuum (the Euclidean version of Lorentz symmetry) is broken to the hypercubic group.  Thus, whatever Symanzik action we write down must account for these broken symmetries.

When constructing an EFT, any possible operator can and will be generated unless a symmetry prevents them.  As a result, a convenient way of organizing the Symanzik action is as follows:
\beq\label{eq:Sym}
\mc L_{Sym} = \frac{1}{a}\mc L^{(3)} +   \mc L^{(4)} + a  \mc L^{(5)} + a^2 \mc L^{(6)} + \cdots,  
\eeq
where $a$ is the lattice spacing, and $L^{(n)}$ represents the linear combination of all the linearly independent operators with mass dimension $n$.  The dots represents all other powers of positive $n$ (there are no operators allowed operators the Wilson action for mass dimensions of 1 or 2).  Each $L^{(n)}$ can be written as
\beq
\mc L^{(n)} = \sum_i c_i^{(n)} \mc O_i^{(n)},
\eeq
with $\mc O_i^{(n)}$ represents the set of allowed independent operators at mass dimension $n$ and $c_i^{(n)}$ are the undetermined, unitless coefficients of these operators.  As in the continuum \CPT case, some effort has to be made in order to show that one has the minimum set of operators (often, operators can be related through change of variables).  Focusing on just the fermionic part of the action, the operators needed through $\mc O(a)$ are given by
\bea
&&3: \quad \mc O_1^{(3)} = \ol \psi \psi \nn \\
&&4: \quad \mc O_1^{(4)} = \ol \psi \Dslash \psi \nn\\
&&5: \quad  \mc O_1^{(5)} = i \ol \psi \sigma_{\mu\nu} F_{\mu\nu} \psi.
\eea 
There is only one dimension-5 operator needed, which is referred to as the Sheikholeslami-Wohlert \cite{Sheikholeslami:1985ij} term for short.  Since there is only one term at $\mc O(a)$, one approach is to add this operator to the actual lattice action to remove all the $\mc O(a)$ effects in the calculation\footnote{This approach is often referred to as ``$\mc O(a)$ improved" or ``Clover improved" Wilson action.  Since the coefficient of the Sheikholeslami-Wohlert term depends on non-perturbative physics, this process requires a non-trivial, non-perturabtive fine tuning of the coefficients in the lattice action.}.  At $\mc O(a^2)$, there are 14 additional operators (which are catalogued in Ref.~\cite{Bar:2003mh}), which consist of both bilinears and four-fermion operators.  The most notable of these operators is $\ol \psi \gamma_\mu D_\mu D_\mu D_\mu \psi$, which is the first term that breaks the $O(4)$ rotation symmetry to the hypercubic group.  

One important issue that becomes clear from the Symanzik action is the complication that arise when taking the continuum limit (when $a \rightarrow 0$).  Ultimately, the desired continuum limit is QCD.  However, as can be seen in Eq.~\eqref{eq:Sym}, there are non-physical terms that can either remain or diverge in the continuum limit.  There are three categories these operators fall in: ``Relevant" operators that have a mass dimension less than four and diverge in the continuum limit, ``Marginal" operators that have a mass dimension of four and remain finite in the continuum limit, and ``Irrelevant" operators that have a mass dimension greater than four and vanish in the continuum limit.  For unwanted, unphysical operators, irrelevant operators can be suppressed by reducing the lattice spacings.  However, relevant and marginal operators require an additional non-perturbative fine tuning.  For example, the Wilson action has one relevant and no marginal operators.  The one relevant operator $\sim \frac{1}{a}\ol \psi \psi$ has the same form as the QCD mass term.  Effectively, this relevant operator acts as a large (on the order of the ultraviolet cutoff of the inverse lattice spacings) additive renormalization of the quark mass.  However, in reality, we want to simulate QCD at finite and small quark masses.  Thus, when simulating the action, one needs to add an additional mass term to the action to cancel this divergence, which requires a non-perturbative tuning (such as verifying the measurement of the pion mass in the chiral limit is zero).  A similar process would be required for each additional relevant or marginal operator an action might generate.  It should be noted, however, that each additional fine tuning becomes significantly more difficult to satisfy.  Thus, too many fine tunings of a lattice action can prevent a lattice action from being practically useful.

\subsubsection{Lattice Chiral Perturbation Theory}
Just as \CPT was the EFT of continuum QCD, lattice chiral perturbation theory (L\CPT) is the EFT to the continuum Symanzik action in the previous section.  The power-counting for L\CPT that will be analyzed throughout this thesis is $p^2\sim m_q \sim a$, known as the generic small mass (GSM) power-counting.   Another power-counting of interest when performing calculations when quark masses are much smaller than lattice spacings, namely $m_q\sim a^2$, is known as the Aoki regime.  From L\CPT, one can calculate the analytic lattice spacing dependence of mesonic observables, which aid in continuum extrapolations.  This approach has been a large focus for categorizing these systematic effects in lattice QCD \cite{Lee:1999zxa,Bernard:2001yj,Aubin:2003mg,Sharpe:1998xm,Rupak:2002sm,Bar:2003mh}.

For a Wilson lattice action whose relevant operator has been fine tuned, the new terms as compared to continuum QCD are (primarily) the Sheikholeslami-Wohlert term and the $O(4)$ breaking term.  The $O(4)$ breaking term,  $\ol \psi \gamma_\mu D_\mu D_\mu D_\mu \psi$,  first appears at $\mc O(a^2p^4)$, which in the GMS power-counting, is at a high order ($N^3LO$) and is a small effect in most standard lattice calculations.  The greater focus will be on the Sheikholeslami-Wohlert term, which obeys the same symmetries as the QCD mass term.  Thus, many of the operators in L\CPT are analogous to the mass terms in \CPT with different LECs (indicated by $w$ or $W$).  The resulting theory can be given in terms the continuum \CPT Lagrangian and the contribution from an isotropic lattice spacing,
\begin{align} \label{eq:L_iso}
    \mc L_{iso} =& \mc L_{cont}+ \D\mc L_{iso}.
\end{align}
\begin{align} \label{eq:dl_iso}
    \D\mc L_{iso} =& \frac{a_sWf^2}{4}\tr(\S^\dag + \S) +\frac{(w_3+w_4)a_sWB_0}{4}\tr(m_q \S^\dag + \S m_q)\tr(\S^\dag + \S) \nonumber\\  
    &+\frac{w_3^\prime (a_sW)^2}{4}\big[\tr(\S^\dag + \S)\big]^2+ \frac{w_4a_sW}{4}\tr(\delmd \S \delmu \S^\dag)\tr(\S^\dag + \S).
\end{align}  
This Lagrangian (along with its anisotropic counterpart) and observables calculated is the focus of Chapter \ref{ch:Aniso_pi_pi}.
 
\section{Organization}
The structure of this thesis is organized in the following way.  Chapter \ref{ch:Aniso_EFT}, which is primarily based on Ref.~\cite{Bedaque:2007xg} with P.~Bedaque and A.~Walker-Loud, develops the EFT for anisotropic lattice actions; namely, lattice actions with temporal lattice spacings different from spacial lattice spacings.  Chapter \ref{ch:Aniso_pi_pi}, based on Ref.~\cite{Buchoff:2008ve}, utilizes this anisotrpic EFT to calculate analytic extrapolation formulae for several observables, including pion masses, decay constants, and, most notably, phase shifts for I=2 $\pi\pi$ scattering.  Chapter \ref{ch:TM_pi_pi}, based on Ref.~\cite{Buchoff:2008hh} with A.~Walker-Loud and J-W.~Chen, also calculates extrapolation formulae for phase shifts for $\pi\pi$ scattering, but for the twisted mass lattice action, whose results were used  in practice to extrapolate to the continuum limit in Ref.~\cite{Feng:2009ij, Feng:2009ck}.  Chapter \ref{Iso_Chem}, based on Ref.~\cite{Bedaque:2009yh} with P.~Bedaque and B.~Tiburzi, proposes a novel method of extracting several scattering parameters between baryons and mesons, which is currently prohibitively expensive in computing time due to the presence of disconnected contributions.  Chapter \ref{ch:Non_orth}, based on Ref.~\cite{Buchoff:2008ei} and previous work in Ref.~\cite{Bedaque:2008xs, Bedaque:2008jm} with P.~Bedaque,  B.~Tiburzi, and A.~Walker-Loud , EFT methods are used to pin down unphysical effects from lattice spacings for several non-orthogonal lattice actions, including the recent graphene-inspired lattice action proposed by Creutz \cite{Creutz:2007af}.  As a final note, the reader should be aware that some notation has been altered from chapter to chapter to remove ambiguity of symbols (for example, in Chapter \ref{ch:TM_pi_pi}, the lattice spacing is referred to as $b$ as opposed to the usual $a$ since $a$ in that chapter is reserved for the scattering length).  As a result the conventions are defined in each chapter.


\renewcommand{\thechapter}{2}

\chapter{Effective Field Theory for the Anisotropic Wilson Lattice Action}\label{ch:Aniso_EFT}

\section{Overview}

All lattice QCD calculations are performed with finite lattice spacings and finite spacial extent (both of which are examples of lattice artifacts).  In addition to the extrapolations from the unphysically large quark masses that present lattice QCD calculations require, extrapolations from a discretized lattice theory to the desired continuum result is also necessary.  In this work, we will focus mainly on the effects from the finite lattice spacings.  The formalism that is employed here is a generalization of chiral perturbation theory that includes both quark mass and lattice spacing dependence.  This patricular formalism, which has been utilized to calculate a plethora of  lattice observables \cite{Creutz:1996bg,Sharpe:1998xm,Rupak:2002sm,Bar:2003mh,Beane:2003xv,Tiburzi:2005vy}, is valid for a specific power counting of the lattice spacing effects, specifically these effects are on the same order as the contribution from the quark mass ($p^2 \sim B m_q \sim W a$).   These resulting formula often depend on LECs which are not known \textit{a priori}.  These LECs, which depend on the details of the short distance physics of the non-perturbative calculation, can be separated into two categories: ones that contribute to the actual low energy physics of interest (physical contributions), and ones that are a result of the finite lattice spacing of the particular lattice calculation (unphysical contributions).  The goal of this formalism of \CPT is to isolate these unphysical contributions, so they can be removed in the proper continuum extrapolation.  

As mentioned in Appendix \ref{ap:Scalar}, the lowest energy of the system can be extrapolated from the long time behavior of the Euclidean correlation functions.  Such correlation functions can suffer from signal-to-noise degeneration for certain baryonic observables at long times (see Ref.~\cite{Bedaque:2007pe} for more details) and can be contaminated by excited states at short times.  Thus, one method to tame such issues is to utilize an anisotropic lattice where the lattice spacing in time differs from the lattice spacing in space by a proportionality factor $\xi$, whose relation is given by $a_t =  a_s/\xi$.  Such lattices can give more temportal resolution. This is highly beneficial for multiple calculations such as heavy systems consisting of light quarks (for example, nucleon-nucleon calculations) where few time slices remain after the excited state contamination dies off.  In other words, anisotropic lattices can give additional information, which might lead to an earlier identification of the ground state plateau. Anisotropic lattices have been used extensively in the study of heavy quarks and quarkonia~\cite{Alford:1996nx,Klassen:1998ua}, glueballs~\cite{Morningstar:1997ff,Morningstar:1999rf}, and excited state baryon spectroscopy~\cite{Basak:2006ww}.  Also, anisotropic lattices may aid in the study of nucleon-nucleon~\cite{Fukugita:1994ve,Beane:2006mx} and hyperon-nucleon~\cite{Beane:2006gf} interactions on the lattice.  

Since  anisotropic lattices are in production and currently being used \cite{Lin:2009zr, Beane:2009ys, Beane:2009rt, Beane:2009fr}, the goal of this work is to create an effective field theory that encompasses the details and lattice artifacts present in these lattices.  This work allows for one to perform a systematic analysis to identify and remove the new unphysical lattice artifacts that these anisotropic lattices present. We begin by constructing the anisotropic version of the Symanzik action~\cite{Symanzik:1983dc,Symanzik:1983gh} for both the Wilson~\cite{Wilson:1974sk} and $\mathcal{O}(a)$ improved Wilson~\cite{Sheikholeslami:1985ij} actions in Sec.~\ref{sec:Ani_Wilson_action}.  Then in Sec.~\ref{sec:aniso_EFT}, we construct the chiral Lagrangians~\cite{Weinberg:1978kz,Gasser:1983yg,Gasser:1984gg} relevant for these anisotropic lattices for both mesons and baryons, focussing on the new effects arising from the anisotropy.  We also provide extrapolation formulae for these hadrons with their modified dispersion relations.  In Sec.~\ref{sec:Aoki_regime}, we highlight an important feature of anisotropic actions; for a fixed spatial lattice spacing and bare fermion mass, if the isotropic action is in the QCD-phase, this does not guarantee the anisotropic action is outside the Aoki-phase~\cite{Aoki:1983qi}.

\section{Anisotropic Wilson lattice Action}\label{sec:Ani_Wilson_action}

The starting point for our discussion is the anisotropic lattice action and its symmetries, from which we will construct the continuum effective Symanzik action~\cite{Symanzik:1983dc,Symanzik:1983gh}, which will then allow us to construct the low-energy EFT describing the hadronic interactions including the dominant lattice spacing artifacts~\cite{Sharpe:1998xm}.  For the isotropic Wilson (and $\mc{O}(a)$ improved~\cite{Sheikholeslami:1985ij} Wilson) action, this program has been carried out to $\mc{O}(a^2)$ for the mesons~\cite{Sharpe:1998xm,Rupak:2002sm,Bar:2003mh} and baryons~\cite{Beane:2003xv,Tiburzi:2005vy}.  This work is a generalization of the previous work, extending the low-energy Wilson EFT to include the dominant lattice artifacts associated with the anisotropy.

The $\mc{O}(a)$-improved anisotropic lattice action, in terms of dimensionless fields, is given by~\cite{Klassen:1998fh,Chen:2000ej}
\begin{align}
S^\xi =&S_1^\xi+S_2^\xi,\\
S_1^\xi=& \beta \sum_{t,i<j} \frac{1}{\xi_0} P_{ij}(U) 
		+\beta \sum_{t,i} \xi_0 P_{ti}(U) 
	+\sum_{n} \bar{\psi}_n \bigg[ a_t m_0 + W_t (U) +\frac{\nu}{\xi_0} W_i(U) \bigg] \psi_n,\\
S_2^\xi=&	-\bar{\psi}_n \bigg[ c_t\, \sigma_{ti} \hat{F}_{ti}(U) 
		+\sum_{i<j} \frac{c_r}{\xi_0}\, \sigma_{ij} \hat{F}_{ij}(U) \bigg] \psi_n\, .
\end{align}
Here, $\xi_0$ is the bare anisotropy, $P_{ij}$ and $P_{ti}$ are space-space and space-time plaquettes of the gauge links $U$.  The bare (dimensionless) quark mass is $a_t m_0$, $W_t(U)$ and $W_i(U)$ are Wilson lattice derivatives and $\nu$ is a parameter which must be tuned to obtain the correct the ``speed of light."  The fields $\hat{F}_{ti}(U)$ and $\hat{F}_{ij}(U)$ are lattice equivalents of the gauge field-strength tensor in the space-time and space-space directions.  $S_1^\xi$ is the unimproved
Wilson action and $S_2^\xi$ is the anisotropic generalization of the Sheikholeslami-Wohlert term. The coefficients, $c_t$ and $c_r$, appearing in $S_2^\xi$ are needed for $\mc{O}(a)$ improvement of the anisotropic Wilson lattice action~\cite{Sheikholeslami:1985ij}.  At the classical level, they have been determined to be~\cite{Chen:2000ej}
\begin{align}\label{eq:ct_cr}
	c_t &= \frac{1}{2}\left( \frac{1}{\xi} + \nu \right)\, ,
	\nonumber\\
	c_r &= \nu\, ,
\end{align}
where $\xi = a_s / a_t$ is the renormalized anisotropy.%
\footnote{These parameters can also be tadpole improved with no more effort than in the isotropic action~\cite{Liao:2001yh}.} 
The choice of using $\xi$ as opposed to $\xi_0$ is conventional and the difference amounts to a slightly different value of $\nu$.

This anisotropic lattice theory retains all the symmetries of the Wilson action, except for the hypercubic invariance, and thus respects parity, time-reversal, translational invariance, charge-conjugation and cubic invariance.  In addition, for suitably tuned bare fermion masses, $a_t m_0,$ the theory has an approximate chiral symmetry, $SU(N_f)_L \otimes SU(N_f)_R$, which spontaneously breaks to the vector subgroup.%
%
%

%
%
\subsection{Anisotropic Symanzik Action \label{sec:quarks}}

We begin by constructing the Symanzik Lagrangian for the {\it unimproved} anisotropic Wilson $S_1^\xi$ lattice action.  This will allow us to set our conventions and introduce a new basis of improvement terms which is advantageous to studying the new lattice artifacts which are remnants of the anisotropy.  In terms of dimensionful fields, the anisotropic Symanzik action is given by
\begin{align}
	S_{Sym}^\xi &=\int d^4x  \mc{L}_{Sym}^\xi\, ,
	\nonumber\\
	\mc{L}_{Sym}^\xi &= \mc{L}_{Sym}^{\xi(4)} + a_s \mc{L}_{Sym}^{\xi(5)}
		+ a_s^2 \mc{L}_{Sym}^{\xi(6)}\, ,
\end{align}
where we have conventionally chosen to use the spatial lattice spacing as our Symanzik expansion parameter.  In terms of the dimensionful fermion fields
\begin{equation}
	q \sim \frac{1}{a_s^3} \psi \quad,\quad 
	\bar{q} \sim \frac{1}{a_s^3} \bar{\psi} \, ,
\end{equation}
the anisotropic Lagrangian is given through $\mc{O}(a)$ by
\begin{align}\label{eq:sym_a0}
\mc{L}_{Sym}^{\xi} &= \bar{q} \left[ \Dslash +m_q \right] q
	+a_s\, \bar{q} \Big[ \bar c_t\, \sigma_{ti} F_{ti} +\bar c_r\, \sum_{i<j} \sigma_{ij} F_{ij} \Big] q\, .
\end{align}
 We have assumed that the parameter $\nu$ has been tuned in such a way as to make the breaking of $O(4)$ symmetry to vanish in the continuum limit. Otherwise, the quark kinetic term would separate into two terms, with an additional free parameter appearing in eq.~(\ref{eq:sym_a0}).  To clearly identify the new lattice spacing effects associated with the anisotropy, it is useful to work with the basis,
\begin{align}
\mc{L}_{Sym}^{\xi} &= \bar{q}\, \left[ \Dslash +m_q \right]\, q
	+a_s \frac{\bar c_r}{2}\, \bar{q}\, \sigma_{\mu\nu} F_{\mu\nu}\, q
	+a_s(\bar c_t-\bar c_r)\, \bar{q}\, \sigma_{ti} F_{ti}\, q\, ,
\end{align}
from which we recognize the first $\mc{O}(a)$ term as the Sheikholeslami-Wohlert~\cite{Sheikholeslami:1985ij} term which survives the isotropic limit, $c_{SW} = \bar c_r / 2$.  The second $\mc{O}(a)$ term $c_{SW}^\xi = (\bar c_t-\bar c_r)$, is an artifact of the anisotropy and the focus of this work. 
We can classify the effects from this operator and subsequent anisotropy operators at higher orders into two categories: those which contribute to physical quantities in a fashion similar to the lattice spacing artifacts already present, and those which introduce new hypercubic breaking effects.  The first type of effect will be difficult to distinguish from the already present lattice spacing artifacts which survive the isotropic limit.%
\footnote{Assuming a given set of lattice calculations are close to the continuum limit, and that a range of anisotropies is employed, one can disentangle the lattice artifacts associated with the new anisotropic operator from those which survive the isotropic limit.} 
The second category of effects are unique to anisotropic actions, and therefore more readily identifiable from correlation functions.

A useful manner to quantify these new anisotropic effects is to recognize that the anisotropy introduces a direction into the theory, which we can denote with the four-vector
\begin{equation}\label{eq:u_aniso}
	u^\xi_\mu = (1, \mathbf{0})\, .
\end{equation}
 This allows us to re-write the anisotropic Lagrangian in a beneficial form for the spurion analysis,
\begin{align}\label{eq:L_W^xi}
\mc{L}_W^{\xi} &	= \bar{q} \left[ \Dslash +m_q \right] q
	+a_s\, \bar{q}\, \Big[ c_{SW}\, \sigma_{\mu\nu} F_{\mu\nu}
		+c_{SW}^{\xi} u^\xi_\mu u^\xi_\nu\, \sigma_{\mu\lambda} F_{\nu\lambda} 
	\Big] q\, .
\end{align}
We then promote both $a_s c_{SW}$ and $a_s c_{SW}^\xi u^\xi_\mu u^\xi_\nu$ to spurion fields, transforming under chiral transformations in such a way as to conserve chiral symmetry,
\begin{align}
	&a_s c_{SW} \longrightarrow L (a_s c_{SW}) R^\dagger \ ,&
	& (a_s c_{SW})^\dagger \longrightarrow R (a_s c_{SW})^\dagger L^\dagger &
	\\
	&a_s c_{SW}^\xi u^\xi_\mu u^\xi_\nu \longrightarrow 
		L (a_s c_{SW}^\xi u^\xi_\mu u^\xi_\nu) R^\dagger \ ,&
	&(a_s c_{SW}^\xi u^\xi_\mu u^\xi_\nu)^\dagger \longrightarrow 
		R (a_s c_{SW}^\xi u^\xi_\mu u^\xi_\nu)^\dagger L^\dagger &
\end{align}
By constraining $a_s c_{SW}$ and $a_s c_{SW}^\xi u^\xi_\mu u^\xi_\nu$ and their hermitian conjugates to be proportional the flavor identity, they both explicitly break the $SU(N_F)_L \otimes SU(N_F)_R$ chiral symmetry down to the vector subgroup, just as the quark mass term.  In addition, we promote $a_s c_{SW}^\xi u^\xi_\mu u^\xi_\nu$ to transform under hypercubic transformations, so as to conserve hypercubic symmetry,
\begin{equation}
	a_s c_{SW}^\xi u^\xi_\mu u^\xi_\nu \longrightarrow 
		a_s c_{SW}^\xi u^\xi_\rho u^\xi_\sigma\, \Lambda_{\mu \rho} \Lambda_{\nu \sigma}\, .
\end{equation}
By constraining $u^\xi_\mu = (1, \mathbf{0})$, this spurion explicitly breaks the hypercubic symmetry of the action down to the cubic sub-group.  

It is worth noting that, in fact, at $\mc{O}(a)$, the anisotropic Symanzik action retains an accidental $O(3)$ symmetry in the spatial directions.  Close to the continuum, isotropic lattice actions retain an accidental Euclidean $O(4)$ (Lorentz) symmetry as the operators required to break this symmetry are of higher dimension and thus become irrelevant in the continuum limit~\cite{Symanzik:1983dc,Symanzik:1983gh}.  This phenomena is observed in the isotropic limit where the $O(4)$ symmetry is broken by the operator $a^2 \bar{q}\, \g_\mu D_\mu D_\mu D_\mu\, q$.   For unimproved anisotropic Wilson fermions, the breaking of the hypercubic to cubic symmetry (which can be viewed as the breaking of the accidental $O(4)$ to the accidental $O(3)$ symmetry) occurs one order lower in the lattice spacing, at $\mc{O}(a)$, and therefore this will likely be a larger lattice artifact than the $O(4)$ breaking of the isotropic action.  We now perform a similar analysis for the $\mc{O}(a^2)$ Symanzik action.

%
%
\subsubsection{$\mc{O}(a^2)$ Symanzik Lagrangian}
In Ref.~\cite{Bar:2003mh}, the complete set of $\mc{O}(a^2)$ operators in the isotropic Symanzik action for Wilson fermions was enumerated, including the quark bi-linears and four-quark operators.  From an EFT point of view, it is useful to classify these operators in three categories, those operators which do not break any of the continuum (approximate) symmetries, those which explicitly break chiral symmetry and those which break Lorentz symmetry.  Most of the $\mc{O}(a^2)$ operators belong to the first category.  Because of their nature, they are the most difficult to determine and ultimately lead to a polynomial dependence in the lattice spacing of all correlation functions computed on the lattice (which can be parameterized as a polynomial dependence in $a$ of all the coefficients of the chiral Lagrangian).  The second set of operators, those which explicitly break chiral symmetry, can be usefully parameterized within an EFT framework as is commonly done with chiral Lagrangians extended to include lattice spacing artifacts~\cite{Sharpe:1998xm,Rupak:2002sm,Bar:2003mh,Beane:2003xv,Tiburzi:2005vy}.  The last set of operators which break Lorentz symmetry, can also be usefully studied in an EFT framework.  In the meson Lagrangian, these effects are expected to be small as they do not appear until $\mc{O}(p^4 a^2)$~\cite{Bar:2003mh}, while in the heavy baryon Lagrangian, these effects appear  at $\mc{O}(a^2)$~\cite{Tiburzi:2005vy}.  To distinguish these Lorentz breaking terms from the general lattice spacing artifacts appearing at $\mc{O}(a^2)$, one must study the dispersion relations of the hadrons, and not only their ground states.  This is also generally true of all the anisotropic lattice artifacts.

In the construction of the anisotropic action, it is also beneficial to sperate the operators into several categories along the lines of those in the isotropic action mentioned above.  We do not show all of the new operators, as their explicit form will not be needed, but instead provide a representative set of the new anisotropic operators which illustrate the new lattice-spacing artifacts.  In the first category, we begin with operators which in the isotropic limit do not break any of the continuum symmetries.  Using the notation of Ref.~\cite{Bar:2003mh}, and using a superscript-${}^\xi$ to denote the new operators due to the anisotropy, we find for example 
\begin{align}
&O_3^{(6)} \longrightarrow \{ O_3^{(6)}\ ,\ {}^\xi O_3^{(6)} \}
	= \{ \bar{q}\, D_\mu \Dslash D_\mu \,q\ ,\ \bar{q}\, D_t \Dslash D_t\, q \}&
	\nonumber\\
&O_{11}^{(6)} \longrightarrow \{ O_{11}^{(6)}\ ,\  {}^\xi O_{11}^{(6)} \}
	= \{ ( \bar{q}\, \g_\mu q ) (\bar{q}\, \g_\mu q )\ ,\ ( \bar{q}\, \g_t q ) (\bar{q}\, \g_t q ) \} &\, .
\end{align}
In the second category, operators which explicitly break chiral symmetry, we find
\begin{equation}
O_{13}^{(6)} \longrightarrow \{ O_{13}^{(6)}\ ,\ {}^\xi O_{13}^{(6)} \}
	= \{ ( \bar{q}\, \s_{\mu\nu} q ) (\bar{q}\, \s_{\mu\nu} q )\ ,\  ( \bar{q}\, \s_{ti} q ) (\bar{q}\, \s_{ti} q )\} \, ,
\end{equation}
from which we observe that there is an operator which both breaks chiral and hypercubic symmetry.  The last category of operators stems from the Lorentz breaking operator in the isotropic limit,
\begin{equation}
O_4^{(6)} \longrightarrow \{ O_4^{(6)}\ ,\ {}^\xi O_4^{(6)} \}
	=\{ \bar{q}\, \g_{\mu} D_\mu D_\mu D_\mu\, q\ ,\  \bar{q}\, \g_{i} D_i D_i D_i\, q\} \, ,
\end{equation}
from which we note that there is only one operator which breaks the accidental $O(3)$ symmetry down to the cubic group, ${}^\xi O_4^{(6)}$, and therefore the dominant $O(3)$ breaking artifacts, in principle, can be completely removed from the theory by studying the dispersion relation of only one hadron, for example the pion.  Each of these new operators can be written in their spuriously hypercubic-invariant form by making use of the anisotropic  vector we introduced in Eq.~\eqref{eq:u_aniso},
\begin{align}\label{eq:asq_aniso}
	&{}^\xi O_3^{(6)} = u^\xi_\mu u^\xi_\nu\, \bar{q}\, D_\mu \Dslash D_\nu\, q,&
	\nonumber\\
	&{}^\xi O_{11}^{(6)} = u^\xi_\mu u^\xi_\nu\, ( \bar{q}\, \g_\mu q ) (\bar{q}\, \g_\nu q ),&
	\nonumber\\
	&{}^\xi O_{13}^{(6)} = u^\xi_\mu u^\xi_\nu\, (\bar{q}\, \s_{\mu\lambda} q) (\bar{q}\, \s_{\nu\lambda} q),&
	\nonumber\\
	&{}^\xi O_4^{(6)} = 
		\bar{\d}^\xi_{\mu\mu^\prime}
		\bar{\d}^\xi_{\mu\nu^\prime}
		\bar{\d}^\xi_{\mu\rho^\prime}
		\bar{\d}^\xi_{\mu\sigma^\prime}
		\bar{q}\, \g_{\mu'} D_{\nu'} D_{\s'} D_{\rho'}\, q,& 
\end{align}
and similarly for the rest of the dimension-6 anisotropic operators, ${}^\xi O_{1-8}^{(6)}$, and ${}^\xi O_{11-18}^{(6)}$.  In this equation, for we have defined
\begin{equation}
	\bar{\d}^\xi_{\mu\nu} \equiv \delta_{\mu\nu} - u^\xi_\mu u^\xi_\nu\, .
\end{equation}
Most of these operators do not break chiral symmetry, and therefore are present for chirally symmetric fermions such as domain-wall~\cite{Kaplan:1992bt,Shamir:1993zy,Furman:1994ky} and overlap~\cite{Narayanan:1992wx,Narayanan:1994gw,Neuberger:1997fp} fermions.  We now proceed to construct the anisotropic chiral Lagrangian.

%
%
\section{Anisotropic Chiral Lagrangian \label{sec:aniso_EFT}}

Now that we have the complete set of Symanzik operators through $\mc{O}(a^2)$ relevant to the anisotropic Wilson action and the $\mc{O}(a)$ improved version thereof, we can construct the equivalent operators in the chiral Lagrangian which encode these new anisotropic artifacts.  We begin with the meson Lagrangian and then move to the heavy baryon Lagrangian.  

%
%
\subsection{Meson Chiral Lagrangian \label{sec:mesons}}

We construct the chiral Lagrangian using a spurion analysis of the quark level Lagrangian given in Eqs.~\eqref{eq:L_W^xi} and \eqref{eq:asq_aniso}.  We generally assume a power counting
\begin{equation}
	m_q \sim a \Lambda^2\, ,
\end{equation}
but work to the leading order necessary to parameterize the dominant artifacts from the anisotropy.  At LO, the meson Lagrangian is given by%
\footnote{We remind the reader we are working in Euclidean spacetime.  We are using the convention $f\sim 132$~MeV.} 
\begin{align}\label{eq:MesonsLO}
\mc{L}_\phi^\xi =&\ 
	\frac{f^2}{8} \tr \left( \partial_\mu \Sigma \partial_\mu \Sigma^\dagger \right)
	-\frac{f^2}{4} \tr \left( m_q\mathbb{B} \Sigma^\dagger + \Sigma (m_q\mathbb{B})^\dagger \right) 
	\nonumber\\&
	-\frac{f^2}{4} \tr \left( a_s \mathbb{W} \Sigma^\dagger + \Sigma (a_s \mathbb{W})^\dagger \right)
	-\frac{f^2}{4} \tr \left( a_s \mathbb{W}^\xi \Sigma^\dagger + \Sigma (a_s \mathbb{W}^\xi)^\dagger \right)\, .
\end{align}
By taking functional derivatives with respect to the spurion fields in both the quark level and chiral level actions, one can show~\cite{GellMann:1968rz}
\begin{equation}
	\mathbb{B} = \lim_{m_q\rightarrow 0} \frac{ | \langle \bar{q} q \rangle |}{f^2}\, ,
\end{equation}
and similarly the new dimension-full chiral symmetry breaking parameters are defined as,
\begin{align}
	\mathbb{W} &= \lim_{m_q \rightarrow 0}\ 
		c_{SW} \frac{ \langle \bar{q} \sigma_{\mu\nu} F_{\mu\nu} q \rangle}{f^2}\, ,
	\\
	\mathbb{W}^\xi &= \lim_{m_q \rightarrow 0}\ 
		c_{SW}^\xi \frac{ \langle \bar{q} \sigma_{ti} F_{ti} q \rangle}{f^2}
	\nonumber\\
		&= \lim_{m_q \rightarrow 0}\ 
		c_{SW}^\xi u^\xi_\mu u^\xi_\nu 
		\frac{ \langle \bar{q} \sigma_{\mu\lambda} F_{\nu\lambda} q \rangle}{f^2}\, .
\end{align}
This anisotropic Lagrangian, Eq.~\eqref{eq:MesonsLO}, has one more operator than in the isotropic limit~\cite{Sharpe:1998xm,Rupak:2002sm}, which is simply a reflection that there are now two distinct $\mc{O}(a)$ operators in the Symanzik action.  However, for a fixed anisotropy, $\xi = a_s /a_t$, these two $\mc{O}(a)$ operators are indistinguishable and can be combined into one.  Inserting the tree level values of the coefficients in the Symanzik action~\cite{Chen:2000ej}, one finds
\begin{align}
	\mathbb{W} &\propto c_{SW} = \nu\, ,
	\\
	\mathbb{W}^\xi &\propto c_{SW}^\xi = \frac{1}{2} \left( \frac{a_s}{a_t} -\nu \right)\, ,
\end{align}
where the speed-of-light parameter, $\nu$, is determined in the tuning of the anisotropic action.  The first clear signal of the anisotropy begins at the next order, $\mc{O}(a p^2)$ for the unimproved action and $\mc{O}(a^2 p^2)$ for the improved action.  We first discuss the unimproved case.

These next set of operators are only present for the unimproved action.  In the isotropic limit, it was shown there are five additional operators at this order~\cite{Rupak:2002sm}
\begin{align}\label{eq:Wilson_Oam}
\mc{L}_{\phi,am} =&\ 2 W_4 \tr \left( \partial_\mu \S \partial_\mu \S^\dagger \right)
		\tr \left( a_s \mathbb{W} \Sigma^\dagger + \Sigma (a_s \mathbb{W})^\dagger \right)
	\nonumber\\& 
	+2W_5 \tr \left( \partial_\mu \S \partial_\mu \S^\dagger 
		\left[ a_s \mathbb{W} \Sigma^\dagger + \Sigma (a_s \mathbb{W})^\dagger \right] \right)
	\nonumber\\& 
	+4 W_6 \tr \left( m_q\mathbb{B} \Sigma^\dagger + \Sigma (m_q\mathbb{B})^\dagger \right)
		\tr \left( a_s \mathbb{W} \Sigma^\dagger + \Sigma (a_s \mathbb{W})^\dagger \right)
	\nonumber\\&Ä
	+4 W_7 \tr \left( m_q\mathbb{B} \Sigma^\dagger - \Sigma (m_q\mathbb{B})^\dagger \right)
		\tr \left( a_s \mathbb{W} \Sigma^\dagger - \Sigma (a_s \mathbb{W})^\dagger \right)
	\nonumber\\&
	+4 W_8 \tr \left( m_q\mathbb{B} \S^\dagger a_s \mathbb{W} \S^\dagger
		+\S (m_q \mathbb{B})^\dagger \S (a_s \mathbb{W})^\dagger \right)\, .
\end{align}
For the anisotropic action, there are an additional five operators similar to the above five with the replacement of LECs $W_{4-8} \rightarrow W_{4-8}^\xi$ and a simultaneous replacement of the condensate $\mathbb{W} \rightarrow \mathbb{W}^\xi$.  These five new operators, as with the $\mc{O}(a)$ operators, are indistinguishable from those in Eq.~\eqref{eq:Wilson_Oam} at a fixed anisotropy.  There are two additional operators at this order however, which introduce new effects associated with the anisotropy,
\begin{align}
\mc{L}_{\phi,am}^{\xi} =&\ W_1^\xi \tr \left( \partial_\mu \S \partial_\nu \S^\dagger \right)
		\tr \left( u^\xi_\mu u^\xi_\nu \left[ 
			a_s \mathbb{W}^\xi \Sigma^\dagger + \Sigma (a_s \mathbb{W}^\xi)^\dagger \right] \right)
	\nonumber\\& 
	+W_2^\xi \tr \left( \partial_\mu \S \partial_\nu \S^\dagger\, 
		u^\xi_\mu u^\xi_\nu \left[ 
			a_s \mathbb{W}^\xi \Sigma^\dagger + \Sigma (a_s \mathbb{W}^\xi)^\dagger \right] \right)\, .
\end{align}
When the anisotropic  vectors are set to their constant value, $(u^\xi_\mu)^T = (1,\mathbf{0})$, one sees that these operators lead to a modification of the pion (meson) dispersion relation,
\begin{multline}\label{eq:pion_dispersion_Oa}
	(E_\pi^2+p_\pi^2)\left( 1 +W  \frac{a_s \mathbb{W}}{f_\pi^2} \right)  \longrightarrow 
	\\
	(E_\pi^2+p_\pi^2) \left( 1 
		+W \frac{a_s\mathbb{W}}{f_\pi^2} +W^\xi  \frac{a_s\mathbb{W}^\xi}{f_\pi^2} \right)
	+E_\pi^2\, \tilde{W}^\xi  \frac{a_s \mathbb{W}^\xi}{f_\pi^2}\, ,
\end{multline}
where 
\begin{align}
	&W = 32(N_f W_4 +W_5)\, ,&
	&W^\xi = 32(N_f W_4^\xi +W_5^\xi)\, ,&
	\nonumber\\
	&\tilde{W}^\xi = 16(N_f W_1^\xi + W_2^\xi)\, ,
\end{align}
and $N_f$ is the number of fermion flavors.  With the $\mc{O}(a)$ improved anisotropic action, these effects all vanish and the leading lattice artifacts begin at $\mc{O}(a^2)$.  The chiral Lagrangian at this next order in the isotropic limit was determined in Ref.~\cite{Bar:2003mh} for which there were three new operators.  In the anisotropic theory, there are an additional six operators but just as with the $\mc{O}(a)$ Lagrangian, the three new anisotropic operators can not be distinguished from those which survive the isotropic limit unless multiple values of the anisotropy are used.  The Lagrangian at this order is
\begin{align}
\mc{L}_{\phi,a^2}^{\xi} =& -4W_6^\prime\, 
		\Big[ a_s \mathbb{W}\, \tr \left( \S +\S^\dagger \right) \Big]^2
	-4\hat{W}_6^\xi\,  
		\Big[ a_s \mathbb{W}^\xi\, \tr \left( \S +\S^\dagger \right) \Big]^2
	\nonumber\\&
	-4W_7^\prime\, 
		\Big[ a_s \mathbb{W}\, \tr \left( \S -\S^\dagger \right) \Big]^2
	-4\hat{W}_7^\xi\,  
		\Big[ a_s \mathbb{W}^\xi\, \tr \left( \S -\S^\dagger \right) \Big]^2
	\nonumber\\&
	-4W_8^\prime\, (a_s \mathbb{W})^2\, 
		\tr \left( \S\S +\S^\dagger \S^\dagger \right)
	-4\hat{W}_8^\xi\, (a_s \mathbb{W}^\xi)^2\, 
		\tr \left( \S\S +\S^\dagger \S^\dagger \right)
	\nonumber\\&
	-4\bar{W}_6^\xi\, (a_s \mathbb{W})(a_s \mathbb{W}^\xi)
		\Big[ \tr \left( \S +\S^\dagger \right) \Big]^2
	-4\bar{W}_7^\xi\, (a_s \mathbb{W})(a_s \mathbb{W}^\xi)
		\Big[ \tr \left( \S -\S^\dagger \right) \Big]^2
	\nonumber\\&
	-4\bar{W}_8^\xi\, (a_s \mathbb{W})(a_s \mathbb{W}^\xi)\, 
		\tr \left( \S\S +\S^\dagger \S^\dagger \right)\, .
\end{align}
The last three operators in this Lagrangian vanish for an $\mc{O}(a)$-improved action as they are directly proportional to the product of the two $\mc{O}(a)$ terms in the Symanzik Lagrangian, Eq.~\eqref{eq:L_W^xi}.  The other six operators in this Lagrangian receive contributions both from products of the $\mc{O}(a)$ Symanzik operators as well as from terms in the $\mc{O}(a^2)$ Symanzik Lagrangian.%
\footnote{For conventional reasons~\cite{Bar:2003mh} we have normalized these operators to the square of the condensates appearing at $\mc{O}(a)$, but one should not confuse this to mean that these operators vanish for the $\mc{O}(a)$-improved action.} 
Therefore they are still present for an $\mc{O}(a)$-improved action but the numerical values of their LECs, $W_{6-8}^\prime$ and $\hat{W}_{6-8}^\xi,$ will be different in the improved case.   

At $\mc{O}(a^2 p^2)$, there are nine new operators, three of which survive the isotropic limit, and three of which explicitly break the accidental $O(4)$ symmetry down to $O(3)$,
\begin{align}\label{eq:asq_psq}
\mc{L}_{\phi,a^2p^2}^{\xi} =&\
	Q_1 (a_s \mathbb{W})^2 \tr \left( \partial_\mu \S \partial_\mu \S^\dagger \right)
	+Q_2 (a_s \mathbb{W})^2 
		\tr \left( \partial_\mu \S \partial_\mu \S^\dagger \right) \tr \left( \S +\S^\dagger \right)
	\nonumber\\& 
	+Q_3 (a_s \mathbb{W})^2 
		\tr \left( \partial_\mu \S \partial_\mu \S^\dagger \left[ \S +\S^\dagger \right] \right)
	+Q_1^\xi (a_s \mathbb{W}^\xi)^2 \tr \left( \partial_\mu \S \partial_\mu \S^\dagger \right)
	\nonumber\\& 
	+Q_2^\xi (a_s \mathbb{W}^\xi)^2 
		\tr \left( \partial_\mu \S \partial_\mu \S^\dagger \right) \tr \left( \S +\S^\dagger \right)
	+Q_3^\xi (a_s \mathbb{W}^\xi)^2 
		 \tr \left( \partial_\mu \S \partial_\mu \S^\dagger \left[ \S +\S^\dagger \right] \right)
	 \nonumber\\& 
	+\hat{Q}_1^\xi (a_s \mathbb{W}^\xi)^2 u^\xi_\mu u^\xi_\nu\ 
		\tr \left( \partial_\mu \S \partial_\nu \S^\dagger \right)
	+\hat{Q}_2^\xi (a_s \mathbb{W}^\xi)^2 u^\xi_\mu u^\xi_\nu\ 
		\tr \left( \partial_\mu \S \partial_\nu \S^\dagger \right)
		\tr \left( \S +\S^\dagger \right)
	\nonumber\\&
	+\hat{Q}_3^\xi (a_s \mathbb{W}^\xi)^2 u^\xi_\mu u^\xi_\nu\ 
	 \tr \left( \partial_\mu \S \partial_\nu \S^\dagger \left[ \S +\S^\dagger \right] \right)\, .
\end{align}
The first operator in each set of three with coefficients $Q_1$, $Q_1^\xi$ and $\hat{Q}_1^\xi$ are modifications of the LO kinetic operator, while the remaining operators additionally break chiral symmetry.  The first operator in Eq.~\eqref{eq:asq_psq} is an example of the operators mentioned before which do not break any of the (approximate) lattice symmetries, and are therefore amount to polynomial renormalizations of the continuum LECs.  In this example, we see with the replacement
\begin{equation}
	f^2 \rightarrow f^2 + 8 Q_1 (a_s \mathbb{W})^2\, ,
\end{equation}
the entire effects from this $\mc{O}(a^2 p^2)$ operator are renormalized away to all orders in the EFT.  In the anisotropic theory, this also works for the operator with LEC $Q_1^\xi$, as this operator does not explicitly break $O(4)$ symmetry.  This provides a further example of effects which arise because of the anisotropy but contribute to the low-energy dynamics in an isotropic fashion.  The operators with explicit anisotropic vectors will modify the dispersion relation as in Eq.~\eqref{eq:pion_dispersion_Oa} but at $\mc{O}(a^2)$.  For the $\mc{O}(a)$-improved action, these operators will provide the dominant modification to the pseudo-Goldstone dispersion relations.

The last set of operators we wish to address for the meson Lagrangian are those which explicitly break the accidental $O(3)$ symmetry down to the cubic group.  There are two operators in the meson chiral Lagrangian but they stem from only one quark level operator at $\mc{O}(a^2)$ and therefore all the LO $O(3)$ breaking effects for all hadrons can be removed with the inclusion and tuning of one new operator in the action, ${}^\xi O_4^{(6)}$ from Eq.~\eqref{eq:asq_aniso}.  The $O(3)$ breaking operators in the meson chiral Lagrangian are
\begin{align}
\mc{L}_{\phi,O(3)}^\xi =&\ C_1^\xi (a_s \mathbb{W}^\xi)^2 
		\bar{\d}^\xi_{\mu\mu^\prime}
		\bar{\d}^\xi_{\mu\nu}
		\bar{\d}^\xi_{\mu\rho}
		\bar{\d}^\xi_{\mu\sigma}\ 
		\tr \left( \partial_{\mu^\prime} \S \partial_\nu \S^\dagger \right)
		\tr \left( \partial_\rho \S \partial_\s \S^\dagger \right)
	\nonumber\\&
	+C_2^\xi (a_s \mathbb{W}^\xi)^2 
			\bar{\d}^\xi_{\mu\mu^\prime}
	\bar{\d}^\xi_{\mu\nu}
	\bar{\d}^\xi_{\mu\rho}
	\bar{\d}^\xi_{\mu\sigma}\ 
	\tr \left( \partial_{\mu^\prime} \S \partial_\nu \S^\dagger \partial_\rho \S \partial_\s \S^\dagger \right)\, ,
\end{align}

%
%
\subsubsection{Aoki Regime \label{sec:Aoki_regime}}

\begin{figure}[t]
\center
\includegraphics[width=0.6\textwidth]{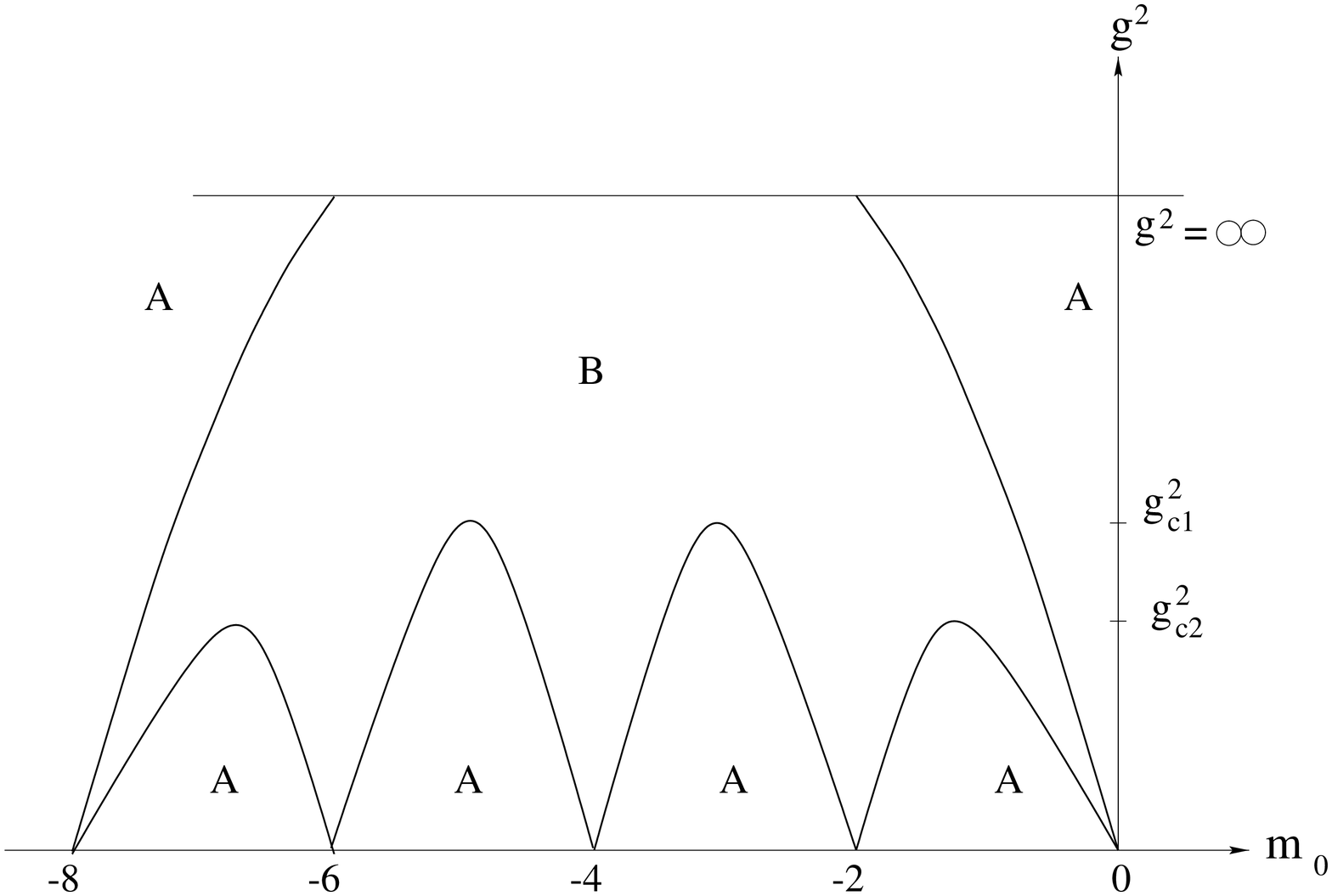} 
\caption{\label{fig:Aoki} Proposed Aoki phase diagram in Ref.~\cite{Sharpe:1998xm} for isotropic lattice action.  Region A represents continuum-like QCD and region B represents the parity and flavor broken phase.  Isotropic actions are generally tuned to lie between the first two-fingers of the Aoki-regime in the $m_0 - g^2$ plane~\cite{Antonio:2008zz}.  However, the anisotropic diagram could look quite different than this picture, and caution should be taken to ensure the correct phase. }
\end{figure}

  Aoki first pointed out the possibility that at finite lattice spacing, lattice actions can undergo spontaneous symmetry breaking of flavor and parity in certain regions of parameter space~\cite{Aoki:1983qi}, as in Fig.~\ref{fig:Aoki}. 
 
%
%
  
 In Ref.~\cite{Sharpe:1998xm}, Sharpe and Singleton addressed this possibility within an EFT framework by extending the meson chiral Lagrangian to include lattice spacing contributions.  We begin by summarizing the discussion of Sharpe and Singleton which will allow us to highlight the new effects that arise from the anisotropy.  For clarity of discussion, we consider the unimproved two-flavor theory in the isospin limit.  Following the notation of Ref.~\cite{Sharpe:1998xm}, the non-kinetic part of the chiral potential can be written
\begin{equation}\label{eq:chiral_potential}
	\mc{V}_\chi = -\frac{c_1}{4} \tr \left( \S +\S^\dagger \right)
		+\frac{c_2}{16} \left[ \tr \left( \S + \S^\dagger \right) \right]^2\, ,
\end{equation}
where $c_1$ and $c_2$ are functions of the quark mass $m_q$ and the lattice spacing $a$,
\begin{align}
	c_1 &\sim \L^4 \left( \frac{m_q}{\L} + a \L \right) +\dots\, ,
	\nonumber\\
	c_2 &\sim \L^4 \left( \frac{m_q^2}{\L^2} +m_q a +a^2 \L^2 \right) +\dots\, .
\end{align}
The ``$\dots$" denote higher order terms in the quark mass and lattice spacing and we are ignoring dimensionless numbers of $\mc{O}(1)$.  Assuming a power counting $m_q \sim a\L^2$, the contributions to the vacuum from the $c_2$ term are suppressed compared to the $c_1$ term, and the vacuum is in the continuum phase with the chiral condensate aligned with unity.  The Aoki-phase can occur when there is a fine-tuning of the quark mass and lattice spacing contributions to $c_1$ such that overall size of $c_1$ becomes comparable to $c_2$.  Parameterizing the $\S$-field as
\begin{equation}
	\S = A +i \mathbf{\tau} \cdot \mathbf{B}\, ,
\end{equation}
with the constraint $A^2 +\mathbf{B}^2 =1$, the potential, Eq.~\eqref{eq:chiral_potential} becomes
\begin{equation}
	\mc{V}_\chi = -c_1 A +c_2 A^2\, .
\end{equation}
If the minimum of the potential occurs for $-1 < A_{0} < 1$, then the vacuum
\begin{equation}
\S_0 = \langle\, \S\, \rangle 
	= A_0 +i \mathbf{\tau} \cdot \mathbf{B_0}\, ,
\end{equation} 
develops a non-zero value of $\mathbf{B}_0$, spontaneously breaking both parity and the remnant vector-chiral symmetry, $SU(2)_V \rightarrow U(1)$~\cite{Sharpe:1998xm}, giving rise to one massive pseudo-Goldstone pion and two massless Goldstone pions.  If the minimum of the potential occurs for $A_{min} \leq -1$ or $A_{min} \geq 1$ then the vacuum lies along (or opposite) the identitiy, $|A_0| =1, |\mathbf{B}_0|=0$ with the same symmetry breaking pattern as QCD.

For unimproved Wilson fermions in the isotropic limit, the leading contributions to $c_1$ are
\begin{equation}
	c_1 = f^2 \Big( m_q \mathbb{B} +a_s \mathbb{W} \Big)\, .
\end{equation}
In the anisotropic theory, there is an additional contribution to the LO potential, such that
\begin{align}
c_1 \rightarrow c_1^\xi &= 
	c_1 + f^2 a_s \mathbb{W}^\xi
	\nonumber\\&
	= f^2 \Big( m_q \mathbb{B} 
	+a_s \mathbb{W} 
	+a_s \mathbb{W}^\xi \Big)\, .
\end{align}
If the two terms which contribute to 
$\mathbb{W} \propto 2 \langle \bar{q} \s_{ti} F_{ti} q + \bar{q} \s_{ij} F_{ij} q \rangle$ are of opposite sign, then $\mathbb{W}^\xi \propto \langle \bar{q} \s_{ti} F_{ti} q \rangle$ may be the dominant lattice spacing contribution to $c_1$.  Therefore, the anisotropic theory may be in the Aoki-regime even when the isotropic limit of the theory (with the same $a_s$) is not, and \textit{vice versa}.  This same discussion holds for $\mc{O}(a)$-improved actions as well but with a different power counting. 

\bigskip
This analysis carries important consequences for anisotropic actions with domain-wall and overlap fermions as well.  These actions are generally tuned to lie between the first two-fingers of the Aoki-regime in the $m_0 - g^2$ plane~\cite{Antonio:2008zz} as in Fig.~\ref{fig:Aoki}.  The optimal value of the bare fermion mass, which provides the least amount of residual chiral symmetry breaking, may be shifted in the anisotropic theory relative to the isotropic value, and furthermore the allowed values of the coupling for which there is a QCD-phase may shift as well.

%
%
\subsection{Heavy Baryon Lagrangian \label{sec:baryons}}

\begin{figure}[t]
\center
\begin{tabular}{cccc}
\includegraphics[width=0.18\textwidth]{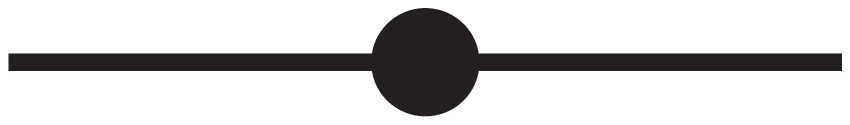} &
\includegraphics[width=0.18\textwidth]{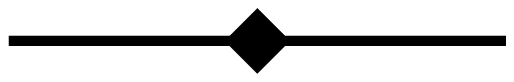} &
\includegraphics[width=0.26\textwidth]{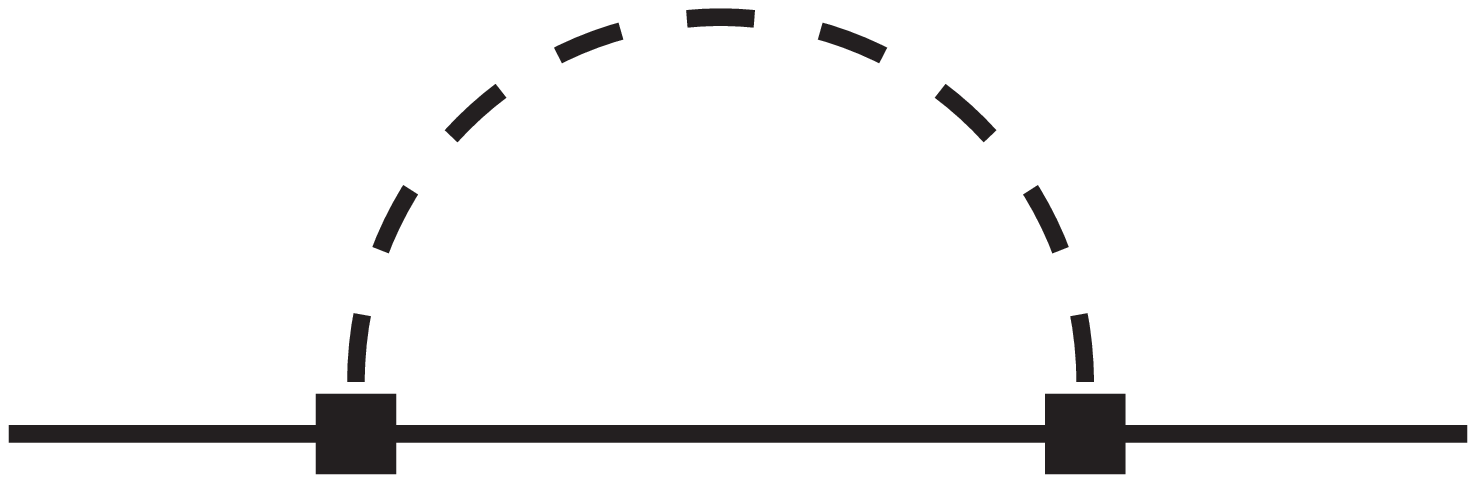} &
\includegraphics[width=0.26\textwidth]{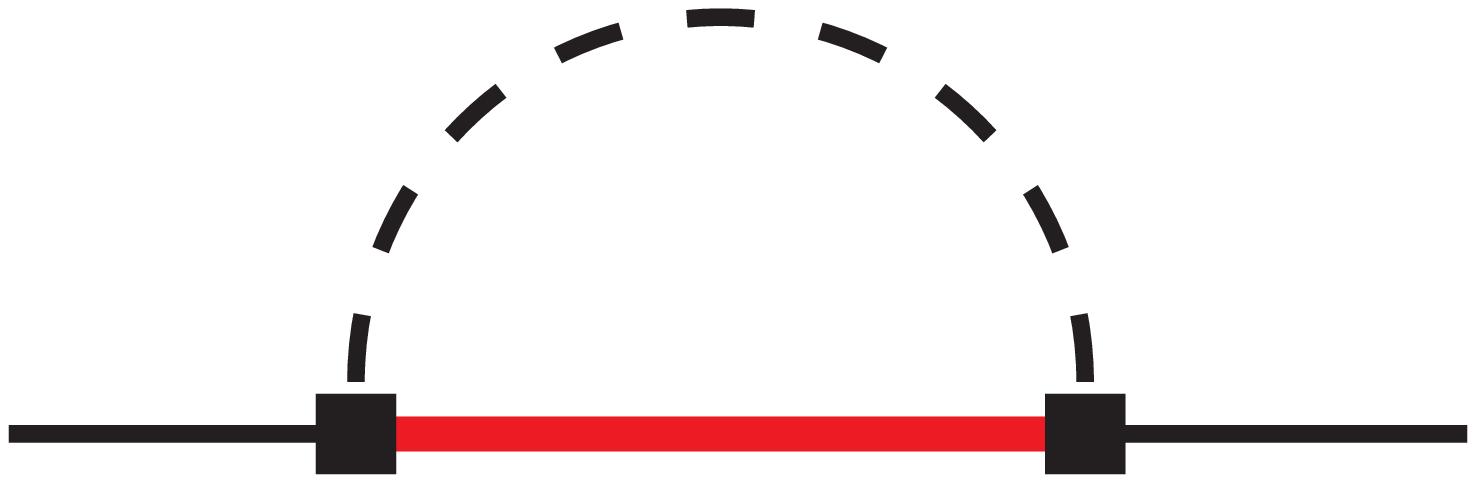} \\
(a) & (b) & (c) & (d)
\end{tabular}
\caption{\label{fig:NmassNLO} Diagrams contributing to the nucleon mass at LO ((a) and (b)) and NLO ((c) and (d)).  Figure (a) is an insertion of the leading quark mass term proportional to $\s_M$.  Figure (b) is an insertion of the lattice spacing terms proportional to $\s_W$ and $\s^\xi_W$.  The loop graphs arise from the pion-nucleon and pion-nucleon-delta couplings.  These loop graphs generically scale as $m_\pi^3$ but also depend upon $\D$, and away from the continuum limit upon the lattice spacing as well.  All vertices in these graphs are from the Lagrangian in Eq.~\eqref{eq:NTphiLag}.}
\end{figure}

We now construct the operators in the baryon Lagrangian which encode the leading lattice artifacts from the anisotropy.  We use the heavy baryon formalism~\cite{Jenkins:1990jv,Jenkins:1991es} and explicitly construct the two-flavor theory in the isospin limit including nucleons, delta-resonances and pions.  The extension of this to include the octet and decuplet baryons is also possible.  This construction builds upon previous work in which the heavy baryon Lagrangian has been extended in the isotropic limit to include the $\mc{O}(a)$~\cite{Beane:2003xv} and $\mc{O}(a^2)$~\cite{Tiburzi:2005vy} lattice artifacts for various baryon observables.%
\footnote{In those works~\cite{Beane:2003xv,Tiburzi:2005vy}, the heavy baryon Lagrangian was extended explicitly for Wilson fermions and in the latter also for a mixed action scheme with chirally symmetric valence fermions.  In this work, we explicitly address the anisotropic effects for these two scenarios.  One can also include the anisotropic effects for twisted mass~\cite{Frezzotti:2000nk} and staggered fermions~\cite{Kogut:1974ag,Susskind:1976jm} following the heavy baryon construction in Refs.~\cite{WalkerLoud:2005bt} and \cite{Bailey:2006zn,Bailey:2007iq} respectively.  However, we are unaware of specific plans to generate anisotropic twisted mass or staggered fermions so we do not pursue this here.} 
The Lagrangian is constructed as a perturbative expansion about the static limit of the baryon, treating it as a heavy, static source as with heavy quark effective theory~\cite{Georgi:1990um,Manohar:2000dt}.  In building the heavy baryon chiral Lagrangian it is useful to introduce a new chiral field,%
\footnote{This field is generally denoted as $\xi$ but to avoid confusion with the anisotropy parameter, we use $\s$.} 
\begin{equation}
	\s = \sqrt{\S} = \exp \left( \frac{i \phi}{f} \right)\, ,
\end{equation}
which transforms under chiral rotations as
\begin{equation}\label{eq:V}
	\s \rightarrow L \s V^\dagger = V \s R^\dagger\, ,
\end{equation}
with $V$ defined by Eq.~\eqref{eq:V}.  One then dresses the baryon fields with $\s$ such that under chiral transformations, the nucleon and delta fields transform as
\begin{align}
	N_i &\rightarrow V_{ij} N_j\, ,
	\nonumber\\
	T_{ijk} &\rightarrow V_{ii^\prime}V_{jj^\prime}V_{kk^\prime} T_{i^\prime j^\prime k^\prime}\, ,
\end{align}
where $N$ is a two-component flavor field and $T$ is a flavor-symmetric rank-three tensor, normalized such that
\begin{align}
	&&
	& N_1 = p,&
	& N_2 = n,& &&
	\nonumber\\
	&T_{111} = \D^{++},&
	&T_{112} = \frac{1}{\sqrt{3}} \D^{+},&
	&T_{122} = \frac{1}{\sqrt{3}} \D^{0},&
	&T_{222} = \D^{-}.&
\end{align}
Both the nucleon and delta are treated as heavy matter fields and are constrained by the relations
\begin{align}
	N &= \frac{1+\vslash}{2} N\, ,
	\nonumber\\
	T_\mu &= \frac{1 +\vslash}{2} T_\mu
\end{align}
where $v_\mu$ is the four velocity of the baryon which can be chosen in its rest frame to be $v_\mu = (1, \mathbf{0})$.  This has the effect of projecting onto the particle component of the field in the rest frame of the baryon.  The spin-3/2 fields can be described with a Rarita-Schwinger field, which in the heavy baryon formalism gives rise to the constraints
\begin{equation}
	v \cdot T = 0\quad, \quad S \cdot T = 0\, ,
\end{equation}
where $S_\mu$ is a covariant spin vector~\cite{Jenkins:1990jv,Jenkins:1991es}.  The simultaneous inclusion of the nucleon and delta fields introduces a new parameter into the Lagrangian, the delta-nucleon mass splitting in the chiral limit,
\begin{equation}\label{eq:Delta}
	\Delta = m_T - m_N \Big|_{m_q = 0}\, ,
\end{equation}
which is generally counted as $\D \sim m_\pi$ in the chiral power counting~\cite{Jenkins:1990jv,Jenkins:1991es,Hemmert:1997ye}.  Because this mass parameter is a chiral singlet, it leads to a modification of all the LECs in the heavy baryon Lagrangian.  These particular effects can be systematically accounted for be treating all LECs as polynomials in $\D$~\cite{WalkerLoud:2004hf,Tiburzi:2004rh,Tiburzi:2005na,WalkerLoud:2006sa}.  In the isospin limit, the heavy baryon Lagrangian, including the leading quark mass terms, the leading lattice spacing terms and the leading baryon-pion couplings is given by
\begin{align}\label{eq:NTphiLag}
	\mc{L}_{NT\phi} =&\ \bar{N} v \cdot D N 
		+a_t \sigma_{E} \bar{N} (v\cdot u^\xi u^\xi\cdot D) N
		-2 \sigma_M \bar{N} N \tr ( \mc{M}_+ ) 
	\nonumber\\&
		+ \bar{T}_\mu \big[ v \cdot D +\Delta \big] T_\mu
		+ a_t \bar{\sigma}_E \bar{T}_\mu (v\cdot u^\xi u^\xi\cdot D) T_\mu
		-2 \bar{\sigma}_M \bar{T}_\mu T_\mu \tr ( \mc{M}_+ ) 
	\nonumber\\&
		+2 g_A \bar{N} S \cdot \mc{A} N
		-2g_{\D\D} \bar{T}_\mu S \cdot \mc{A} T_\mu
		+g_{\D N} \big[ \bar{T}^{kji}_\mu \mc{A}_{i,\mu}^{\ i^\prime} \epsilon_{j i^\prime} N_k +h.c. \big]\, .
	\nonumber\\&
		- 2\sigma_W \bar{N} N \tr ( \mc{W}_+ )
		- 2\sigma_W^\xi \bar{N} N \tr ( \mc{W}^\xi_+ )
		-2 \bar{\sigma}_W \bar{T}_\mu T_\mu \tr ( \mc{W}_+ ) 
		-2 \bar{\sigma}_W^\xi \bar{T}_\mu T_\mu \tr ( \mc{W}^\xi_+ ) 
\end{align}
In this Lagrangian, the chiral symmetry breaking spurions are given by
\begin{align}
	\mc{M}_+ &= \frac{1}{2} \left( \s m_q^\dagger \s + \s^\dagger m_q \s^\dagger \right)\, ,
	\nonumber\\
	\mc{W}_+ &= \frac{a_s}{2} \left( \s \mathbb{W}^\dagger \s + \s^\dagger \mathbb{W} \s^\dagger \right)\, ,
	\nonumber\\
	\mc{W}^\xi_+ &= \frac{a_s}{2} \left( \s (\mathbb{W}^\xi)^\dagger \s 
		+ \s^\dagger \mathbb{W}^\xi \s^\dagger \right)\, ,
\end{align}
the chiral covariant derivative is
\begin{align}
	(D^\mu N)_i &= \partial^\mu N_i + \mc{V}^\mu_{ij}N_j\, ,
	\nonumber\\
	(D^\mu T)_{ijk} &= \partial^\mu T_{ijk} + \mc{V}^\mu_{i i^\prime} T_{i^\prime j k}
		+ \mc{V}^\mu_{j j^\prime} T_{ij^\prime k}
		+ \mc{V}^\mu_{k k^\prime} T_{ijk^\prime}
\end{align} 
and the vector and axial fields are respectively given by
\begin{align}
	\mc{V}_\mu &= \frac{1}{2} \left( \s \partial_\mu \s^\dagger + \s^\dagger \partial_\mu \s \right)\, ,
	\nonumber\\
	\mc{A}_\mu &= \frac{i}{2} \left( \s \partial_\mu \s^\dagger - \s^\dagger \partial_\mu \s \right)\, .
\end{align}
Relative to the Wilson extension of the heavy baryon Lagrangian in the isotropic limit~\cite{Beane:2003xv,Tiburzi:2005vy} there is one additional mass operator for the nucleon and delta fields.  There are additionally extra derivative operators which give rise to the leading modification of the dispersion relation for the baryon fields.  As in the meson chiral Lagrangian, Eq.~\eqref{eq:MesonsLO}, at fixed anisotropy, these two additional mass operators (with coefficients $\s_W^\xi$ and $\bar{\s}_W^\xi$) can not be distinguished from their counterparts which survive the isotropic limit (with coefficients $\s_W$ and $\bar{\s}_W$).  At this order in the lattice spacing, $\mc{O}(a)$, the baryon-pion couplings, $g_A$, $g_{\D N}$ and $g_{\D\D}$, are not modified~\cite{Beane:2003xv}.  This Lagrangian, Eq.~\eqref{eq:NTphiLag} gives rise to the LO and NLO mass corrections to the baryon masses.  For example, in Fig.~\ref{fig:NmassNLO} we display the graphs contributing to the nucleon mass at $\mc{O}(m_q)$, Fig.~\ref{fig:NmassNLO}(a), $\mc{O}(a)$, Fig.~\ref{fig:NmassNLO}(b) and $\mc{O}(m_q^{3/2})$, Fig.~\ref{fig:NmassNLO}(c) and Fig.~\ref{fig:NmassNLO}(d).
%
%
The expressions for the quark mass dependence of the nucleon and delta masses can be found for example in Ref.~\cite{Tiburzi:2005na}, and the lattice spacing dependent corrections to these masses for Wilson fermions in the isotropic limit can be found in Refs.~\cite{Beane:2003xv,Tiburzi:2005vy}.

At $\mc{O}(am_q)$, there are two types of mass corrections, those from loop graphs from the lattice spacing dependent operators in Eq.~\eqref{eq:NTphiLag} and tree level terms from operators in the $\mc{O}(am_q)$ Lagrangian.  These corrections will be identical in form to those present in the isotropic limit, but with different numerical values due to the new anisotropic operators which do not explicitly break the hypercubic symmetry.  The loop graphs for the nucleon and delta mass corrections are depicted in Ref.~\cite{Tiburzi:2005vy} in Figs.~1 and 2 respectively.  In the isospin limit, there are two tree level operators for the nucleon and deltas each,
\begin{align}
	\mc{L}_{NT\phi}^{(am)} =&
		-n_{MW} \bar{N} N \tr(\mc{M}_+) \tr(\mc{W}_+)
		-n_{MW}^\xi \bar{N} N \tr(\mc{M}_+) \tr(\mc{W}^\xi_+)
	\nonumber\\&
		+t_{MW} \bar{T}_\mu T_\mu \tr(\mc{M}_+) \tr(\mc{W}_+)
		+t_{MW}^\xi \bar{T}_\mu T_\mu \tr(\mc{M}_+) \tr(\mc{W}^\xi_+)\, ,
\end{align}
which also function as counterterms for divergences from the above mentioned loop graphs.  As with the mesons, we must consider the $\mc{O}(a^2)$ effects to see the first explicit breaking of the anisotropy.  Even in the isotropic limit, the lattice spacing corrections to the baryons at this order proved to be quite interesting, with the presence of the first operators which arise from the $O(4)$ breaking down to the hypercubic group at this order~\cite{Tiburzi:2005vy}.  There are several generic $\mc{O}(a^2)$ operators which do not explicitly break the hypercubic group,
\begin{align}
	\mc{L}_{NT\phi}^{a^2} =& 
		-n_{WW} \bar{N} N \tr(\mc{W}_+) \tr(\mc{W}_+)
		-n_{WW}^\xi \bar{N} N \tr(\mc{W}^\xi_+) \tr(\mc{W}^\xi_+)
	\nonumber\\&
		+t_{WW} \bar{T}_\mu T_\mu \tr(\mc{W}_+) \tr(\mc{W}_+)
		+t_{WW}^\xi \bar{T}_\mu T_\mu \tr(\mc{W}^\xi_+) \tr(\mc{W}^\xi_+)
	\nonumber\\&
		-\bar{n}_{WW}^\xi \bar{N} N \tr(\mc{W}_+) \tr(\mc{W}^\xi_+)
		+\bar{t}_{WW}^\xi \bar{T}_\mu T_\mu \tr(\mc{W}_+) \tr(\mc{W}^\xi_+)\, ,
\end{align}
where the last two operators vanish for the $\mc{O}(a)$ improved action and the first two survive the isotropic limit.  The $O(4)$ breaking operators appear at this order in the heavy baryon Lagrangian because in the heavy baryon theory, there is an additional vector, the four-velocity of the baryon one can use to construct operators.  For the mesons, the only vector is $\partial_\mu$ and so the $O(4)$ breaking operators in the meson Lagrangian don't appear until $\mc{O}(a^2 p^4)$~\cite{Bar:2003mh}.  These heavy baryon $O(4)$ breaking operators are~\cite{Tiburzi:2005vy}
\begin{equation}\label{eq:baryon_O4}
	\mc{L}_{NT\phi}^{O(4)} = a_s^2 n_{4}\, \bar{N} v_\mu v_\mu v_\mu v_\mu N
		+a_s^2 t_{4}\, \bar{T}_\rho v_\mu v_\mu v_\mu v_\mu T_\rho \, .
\end{equation}
Similar to these, the $O(3)$ breaking operators appear at this order,
\begin{align}\label{eq:baryon_O3}
	\mc{L}_{NT\phi}^{O(3)} =\ 
		a_s^2 n_3\, \bar{N} 
		\bar{\d}^\xi_{\mu\nu}
		\bar{\d}^\xi_{\mu\nu}
		\bar{\d}^\xi_{\mu\nu}
		\bar{\d}^\xi_{\mu\nu}\ 
	 N 
		+a_s^2 t_3\, \bar{T}_\rho 
		\bar{\d}^\xi_{\mu\nu}
		\bar{\d}^\xi_{\mu\nu}
		\bar{\d}^\xi_{\mu\nu}
		\bar{\d}^\xi_{\mu\nu}\ 
		 T_\rho\, .
\end{align}
However, even though these two sets of operators explicitly break the $O(4)$ and $O(3)$ symmetries respectively, they do not lead to modifications of the dispersion relations until higher orders~\cite{Tiburzi:2005vy}, and thus function as mass corrections.  There is a subtlety of the heavy baryon Lagrangian related to reparameterization invariance (RPI)~\cite{Luke:1992cs}, which  constrains coefficients of certain operators in the heavy baryon Lagrangian.  For example, in the continuum limit, the LO kinetic operator of Eq.~\eqref{eq:NTphiLag} is related to a higher dimensional operator in such a way as to provide the correct dispersion relation,%
\footnote{In this equation, $D_\perp^2 = D^2 - (v\cdot D)^2$.}
\begin{equation}
 \mc{L} = \bar{N} v\cdot D N \longrightarrow
	\bar{N} v\cdot D N + \bar{N} \frac{D_\perp^2}{2M_N} N\, ,
\end{equation}
such that the energy of the non-relativistic nucleon is given by
\begin{equation}
	E = M_N + \frac{\mathbf{p}^2}{2M_N} + \dots
\end{equation}
For the anisotropic theory, this dispersion relation is then modified by an additional operator.  While in the continuum limit, RPI constrains the coefficient in front of the $1/M_N$ operator, the anisotropic action gives rise to modifications of this relation.
\begin{multline}
 \mc{L} = \bar{N} v\cdot D N + \bar{N} \frac{D_\perp^2}{2M_N} N \longrightarrow\
 \\
 \mc{L}^\xi = \bar{N} v\cdot D N 
	+\bar{N} \frac{D_\perp^2}{2M_N} N
	+a_t \sigma_E \bar{N} v\cdot u^\xi u^\xi \cdot D N
	+a_s \sigma_{KE} \bar{N} (u^\xi \cdot D)^2 N\, .
\end{multline}
The resulting nucleon dispersion relation from this action yields
\begin{multline}
E_N = M_N  + \frac{\mathbf{p}^2}{2M_N}  \longrightarrow\
	\\
	E_N (1+a_t \s_E)  = M_N(1+a_t \s_E) 
		-2N_f a_s ( \sigma_W +\s_W^\xi)
		+ \frac{\mathbf{p}^2}{2M_N} 
		+a_s \sigma_{KE}\, (E_N-M_N)^2 \, .
\end{multline}
For the $\mc{O}(a)$-improved action, the form of this dispersion relation stays the same with $\{a_t,a_s\} \rightarrow \{a_t^2, a_s^2\}$.  Comparing this dispersion relation to that of the pion, Eq.~\eqref{eq:pion_dispersion_Oa}, it is clear that even if the pion dispersion relation were tuned to be continuum like, the nucleon dispersion relation would still contain lattice spacing artifacts.  This is a simple consequence of the fact that the LECs in the heavy baryon Lagrangian (both the physical and unphysical) have no relation to those in the meson chiral Lagrangian.

%
%
\section{Discussion \label{sec:concl}}

In this chapter, the low-energy hadronic effective field theory including the finite lattice spacing effects from anisotropic lattice action was developed.  This theory allows for one to calculate extrapolation formulas for hadronic processes including pions, nucleons, and deltas.  The main feature that this extension of effective field theory includes is the effects of the hypercubic symmetry breaking present in these anisotropic lattices.  Additional extensions of this effective field theory could be made for mixed action lattices (lattices in with different actions for the valence fermions) as in Refs.~\cite{Chen:2005ab,Chen:2006wf,Orginos:2007tw,Chen:2007ug}.  Additionally, we point out that the that these anisotropic lattices can have an altered Aoki-phase as compared to the isotropic case.  Thus, if the isotropic action is in a QCD phase, the anisotropic theory might still be in an Aoki-phase.  Overall, this work will aid in the continuum extrapolation of correlation functions computed with anisotropic lattices, which will be computed in the following chapter for the pion mass, the pion decay constant, and scattering lengths and effective ranges for I=2 $\pi\pi$ scattering.



\renewcommand{\thechapter}{3}

\chapter{Isotropic and Anisotropic Lattice Spacing Corrections for I=2 $\pi\pi$ Scattering from Effective Field Theory}\label{ch:Aniso_pi_pi}

\section{Overview}

Numerical scattering calculations in lattice QCD are being performed by several collaborations.  These calculations are performed through the analysis of two hadrons in finite volume  \cite{Huang:1957im,Hamber:1983vu,Luscher:1990ux,Luscher:1986pf}.  One such scattering that has gained much attention in the field is I=2 $\pi\pi$ scattering.  The two-pion system is both numerically the simplest and the best understood theoretically.  In fact, in 1966, the scattering of two pions at low energies was calculated at leading order in \CPT by Weinberg~\cite{Weinberg:1966kf}.  The subsequent orders in the chiral expansion have been worked out next-to-leading order by Gasser and Leutwyler~\cite{Gasser:1983yg} and next-to-next-to-leading order ~\cite{Knecht:1995tr,Bijnens:1995yn,Bijnens:1997vq}.  Each new order introduces more undetermined LECs.  To have predictive power, these LECs must be determined either by comparison with experiment or lattice QCD calculational results.  

Such numerical lattice QCD calculations (usually involving phase shifts and scattering lengths) have been calculated using Wilson lattice actions \cite{Gupta:1993rn,Fukugita:1994na,Fukugita:1994ve,Fiebig:1999hs,Aoki:1999pt,Liu:2001ss,Aoki:2001hc,Aoki:2002in,Aoki:2002sg,Aoki:2002ny,Ishizuka:2003nb,Yamazaki:2004qb,Du:2004ib,Aoki:2004wq,Aoki:2005uf,Li:2007ey} along with a several other lattice actions  \cite{Sharpe:1992pp,Kuramashi:1993ka, Juge:2003mr,Feng:2009rc}.  Additionally, there are currently only two fully dynamical, 2+1 flavor calculations of I=2 $\pi\pi$ scattering, which use mixed lattice actions \cite{Beane:2005rj,Beane:2007xs} . Regardless of the action, unphysical lattice artifacts due to the finite lattice spacings exist in the numerical results of these calculations.   Therefore, measures should be taken in order to remove these effects so that the results from the lattice can best represent the continuum limit. The analysis in this chapter is applicable for both isotropic and anisotropic Wilson actions.

Effective field theory  provides a framework by which one can remove these unphysical effects.  Lattice spacing effects were first made explicit in chiral perturbation theory by Sharpe and Singleton \cite{Sharpe:1998xm}.  For the Wilson action, the chiral breaking terms that depend on the lattice spacing can be accounted for in a similar way to the chiral breaking quark mass.  Such methods have been extended to mixed-action, partially quenched theories for mesons through $\mc O(a)$ and $\mc O(a^2)$ \cite{Rupak:2002sm,Bar:2003mh,Chen:2006wf}, and baryons through $\mc O(a)$ \cite{Beane:2003xv} and $\mc O(a^2)$ \cite{Tiburzi:2005vy,Tiburzi:2005is,Chen:2007ug}. Additionally, Ref.~\cite{Chen:2006wf} carries out multiple meson scattering calculations (including I=2 $\pi\pi$ scattering) for mixed lattice actions and shows that for actions with chiral valence fermions, mesonic scattering parameters in terms of the lattice-physical parameters will have no counter-terms dependent on lattice spacing through next-to-leading order.   Alternatively, this work calculates these lattice spacing effects\footnote{These effects are in addition to the finite volume corrections to I=2 $\pi\pi$ scattering from Ref.~\cite{Bedaque:2006yi}} for I=2 $\pi\pi$ scattering for the chiral breaking Wilson fermions in both valance and sea sectors, and shows that these finite lattice spacing effects first appear in the next-to-leading order counter-terms for this action.

Many collaborations are now using anisotropic lattices (lattices with different temporal and spacial lattice spacings) as opposed the the usual isotropic lattices.  Such lattices can probe higher energy states (inverse time spacings $a_t^{-1} \sim 6$ GeV) and allow for a greater resolution (more data points).  However, anisotropic lattices lead to new lattice artifacts, including terms that explicitly break hypercubic symmetry.  Recent work has derived these anisotropic lattice artifacts in \CPT for $\mc O(a)$ and $\mc O(a^2)$ for mesons and baryons \cite{Bedaque:2007xg}.  There are several I=2 $\pi\pi$ scattering results published for anisotropic Wilson lattices \cite{Liu:2001ss,Du:2004ib,Li:2007ey}, which can benefit from removing these additional lattice artifacts.  

This chapter presents the I=2 $\pi\pi$ scattering results from the isotropic \CPT and the anisotropic \CPT.  The pion mass and decay constant are also determined in this context. Sec.~\ref{sec:Lat_Scat} presents scattering on the lattice and defines the relevant quantity, $k \cot \d_0$, used to make comparisons between \CPT and the actual lattice calculation.  Next, in Sec.~\ref{sec:continuum}, the continuum scattering theory from \CPT is formulated in the context of this work (originally worked out before \cite{Weinberg:1966kf,Gasser:1983yg,Bijnens:1997vq,Colangelo:2001df}).   Then, in Sec.~\ref{sec:iso}, the continuum result is extended for the isotropic Wilson lattice and finally, in Sec.~\ref{sec:aniso}, the result is extended for the anisotropic Wilson lattice.

%
%
\section{Scattering on the Lattice}\label{sec:Lat_Scat}

The Euclidean two-hadron correlation function in infinite volume gives no information about the Minkowski scattering amplitude (except at kinematic thresholds) \cite{Maiani:1990ca}.  However, when the correlation functions of two hadrons in a finite box are analyzed, the resulting energy levels are given by the sum of the energies of these two hadrons plus an additional energy of interaction, $\D E$, which is related to the scattering phase shift, $\d_l$ \cite{Huang:1957im,Hamber:1983vu,Luscher:1990ux,Luscher:1986pf}.   The $l$ subscript here represents the partial wave contribution of the phase shift.  In the infinite volume, the relation between the total scattering amplitude, $T(s,\th)$, and the partial waves amplitude, $t_l(s)$, is given by 

\begin{equation}\label{eq:Partial_Wave }
T(s,\th) = \sum_{l=0}^\infty (2l+1)P_l(\cos \th)t_l(s),
\end{equation}
where $s=4(\mp^2+k^2)$, and $k$ is the magnitude of the 3-momentum of the incoming particle in the center-of-mass frame.  The partial scattering amplitude $t_l(s)$ is related to the phase shift, $\d_l$ by

\begin{equation}\label{eq:t_to_delta }
t_l(s) = 32\pi \sqrt{\frac{s}{s-4\mp^2}}\frac{1}{2i}[e^{2i\d_l(s)}-1]=32\pi \sqrt{\frac{s}{s-4\mp^2}}\frac{1}{\cot \d_l - i}.
\end{equation}

These relations allow one to compare the calculated scattering amplitude (in \CPT) to the lattice calculation of $\d_l$.  The s-wave ($l = 0$) scattering amplitude is the dominant contribution to the total scattering amplitude in most low energy scattering processes and gives the cleanest signal in the lattice calculation.   The s-wave projection of the continuum scattering amplitude, $t_0(s)$, is

\begin{equation}\label{s_proj}
t_0(s) = \frac{1}{2}\int_{-1}^1 T(s,\th)  \: d(\cos \th).
\end{equation}

This s-wave scattering amplitude will be the scattering amplitude analyzed throughout the rest of this paper.  Following the discussion in Ref.~\cite{Bedaque:2006yi}, through one loop order in perturbation theory in Minkowski space, $t_0(s)$ can be written as

\begin{equation}\label{t0_bubbles}
t_0(s) \simeq t_0^{(LO)}(s) + t_0^{(NLO,R)}(s) + i t_0^{(NLO,I)}(s) \simeq \frac{(t_0^{(LO)}(s))^2}{t_0^{(LO)}(s) - t_0^{(NLO,R)}(s) - i t_0^{(NLO,I)}(s)},
\end{equation}
where $t_0^{(LO)}(s)$ is the leading order s-wave scattering amplutide, and $t_0^{(NLO,R)}(s)$ ($t_0^{(NLO,I)}(s)$) is the real (imaginary) part of the next-to-leading order s-wave scattering amplitude.  At this point, it is advantageous to introduce a $K$-matrix, which is defined through one loop as

\begin{equation}\label{k_mat}
K(s)  \equiv \frac{(t_0^{(LO)}(s))^2}{t_0^{(LO)}(s) - t_0^{(NLO,R)}(s)} .
\end{equation}

Taking the real part of the reciprocal of Eq.~\eqref{eq:t_to_delta } and Eq.~\eqref{t0_bubbles} and comparing to Eq.~\eqref{k_mat}, one gets the relation

\begin{equation}\label{k_rel}
\frac{1}{K(s)} = Re\bigg(\frac{1}{t_0(s)}\bigg) = \frac{1}{32\pi}\sqrt{\frac{s-4\mp^2}{s}}\cot\d_0(s), 
\end{equation}
where

\begin{equation}\label{re_rec_t}
Re\bigg(\frac{1}{t_0(s)}\bigg) = \frac{Re \;  t_0(s)}{\big(Re \;  t_0(s)\big)^2 + \big(Im \;  t_0(s)\big)^2} \approx \frac{1}{t_0^{(LO)}(s)}\bigg(1-\frac{t_0^{(NLO,R)}(s)}{t_0^{(LO)}(s)} \bigg).
\end{equation}
It is worth noting that when keeping terms though one loop, $Im \; t_0(s)$ does not contribute (it contributes at the next order).  Combining Eq.~\eqref{k_mat}, Eq.~\eqref{k_rel}, and Eq.~\eqref{re_rec_t}, one arrives at the continuum result

\begin{equation}\label{kcot_cont_amp}
k\cot\d_0(s) = 16\pi \sqrt{s} \: Re\bigg(\frac{1}{t_0(s)}\bigg) \approx 16\pi \sqrt{s}\frac{1}{t_0^{(LO)}(s)}\bigg(1-\frac{t_0^{(NLO,R)}(s)}{t_0^{(LO)}(s)} \bigg).
\end{equation}

As previously mentioned, lattice scattering calculations are performed in Euclidean space at finite volume.  The Euclidean amputated four-point correlator from the lattice, $\mc \tau_0(s)$,  is given by

\begin{equation}\label{lat_amp}
\mc \tau_0(s) \simeq \frac{(t_0^{(LO)}(s))^2}{t_0^{(LO)}(s) - t_0^{(NLO,R)}(s)-\D t_0(s) - \frac{(t_0^{(LO)})^2}{16\pi^2 L \sqrt{s}}\mc S \Big(\frac{(s-4\mp^2) L^2}{4 \pi^2} \Big)},
\end{equation}
where $\D t_0(s)$ represents all of the non-physical lattice artifacts (lattice spacing errors, finite volume errors ,etc.), $s$ is related to the energy of interaction, $\D E$, and $\mc S$ is a universal function of $s$ \cite{Luscher:1990ux,Beane:2003yx,Beane:2003da}.  If both pions in the box start with no external momentum, then $s=(\D E +2m_\pi)^2$.  In this paper, the only effect from lattice artifacts that will be included in $\D t_0$ is the lattice spacing effect.  Manipulating Eq.~\eqref{lat_amp}:

\begin{align}\label{lat_amp_pole}
\mc \tau_0(s) &\simeq \frac{1}{\frac{1}{K(s)}-\frac{\D t_0(s)}{(t_0^{(LO)})^2} - \frac{1}{16\pi^2 L \sqrt{s}}\mc S \Big(\frac{(s-4\mp^2) L^2}{16 \pi^2} \Big)} \nonumber\\
&=\frac{16\pi\sqrt{s}}{k\cot \d_0(s) - 16\pi \sqrt{s} \frac{\D t_0(s)}{(t_0^{(LO)})^2} - \frac{1}{\pi L}\mc S \Big(\frac{(s-4\mp^2) L^2}{16 \pi^2} \Big)}. 
\end{align}

The energy states are given by the poles of Eq.~\eqref{lat_amp_pole}, which are given by \cite{Bedaque:2006yi}

\begin{equation}\label{eq:LR }
k\cot \d_0 + \D(k\cot \d_0) = \frac{1}{\pi L} \mc S \bigg(\frac{(s-4\mp^2)L}{16\pi^2}\bigg),
\end{equation}
where
\begin{equation}\label{eq:D_t }
\D(k\cot \d_0) = -16\pi \sqrt{s}  \frac{\D t_0(s)}{(t_0^{(LO)})^2}.
\end{equation}

In general, most lattice calculations give their results in terms of the scattering length, $a_{\pi\pi}^{I=2}$.  One can extract the scattering length and the effective range, $r_{\pi\pi}^{I=2}$,  via the expansion of $k\cot \d_0$:
\begin{equation}\label{eq:cot_expand }
k\cot \d_0 = \frac{1}{a_{\pi\pi}^{I=2}} + \frac{1}{2}r_{\pi\pi}^{I=2}k^2 + \cdots.
\end{equation}

It is important to note that the prescription given above for finding the scattering length and effective range implies that the lattice artifacts, $\D(k\cot \d_0)$, have already been subtracted \textit{before} the expansion.  In this paper, continuum results are given in terms of $k\cot \d_0$ and lattice artifacts are given in terms of $\D(k\cot \d_0)$.

For results given in terms of the scattering length and effective range, one can relate Eq.~\eqref{eq:cot_expand } to the left hand side of Eq.~\eqref{eq:LR } to arrive at

\begin{equation}\label{eq:cot_expand_art }
k\cot \d_0+\D(k\cot \d_0) = \bigg(\frac{1}{a_{\pi\pi}^{I=2}}+\D\Big(\frac{1}{a_{\pi\pi}^{I=2}}\Big)\bigg) + \frac{1}{2}\bigg(r_{\pi\pi}^{I=2}+\D r_{\pi\pi}^{I=2} \bigg)k^2 + \cdots,
\end{equation}
where 

\begin{equation}\label{eq:recip_a_art }
\D\Big(\frac{1}{a_{\pi\pi}^{I=2}}\Big)= \D(k\cot \d_0)|_{k^2=0},
\end{equation}
and

\begin{equation}\label{eq:r_art }
\D r_{\pi\pi}^{I=2} = 2 \frac{d\big(\D(k\cot \d_0)\big)}{dk^2}\bigg|_{k^2=0}.
\end{equation}

While these relations are not too complicated, they do add additional steps to the calculation when compared to working with only $k\cot \d_0$ and $\D(k\cot \d_0)$.  Therefore, if one wants to extract the scattering length and effective range from the lattice calculation, one should first subtract  $\D(k\cot \d_0)$ from the right hand side of Eq.~\eqref{eq:LR } and then expand to determine the individual parameters\footnote{Current numerical calculations can only determine $k\cot \d_0$ for a limited number of $k$ values.  This leads to inaccuracies in expansions of $k^2$ and adds difficulty to finding the effective range.}.  This paper relates $k\cot \d_0$ and $\D(k\cot \d_0)$ to the effective field theory of the lattice.

%
%
\section{Chiral Perturbation Theory Results for $k\cot \d_0$ and $\D(k\cot \d_0)$}\label{sec:t_dt_CPT}

In leading order and next-to-leading-order chiral perturbation theory, it is necessary to introduce several undetermined LECs in order properly account for corrections and counter-terms.   The number of independent LECs in the continuum depends on whether there are two flavors or more.  For I=2 $\pi\pi$ scattering being calculated here, only two flavor \CPT ($SU(2)_L\otimes SU(2)_R$ chiral symmetry) is needed for extrapolation.

%
%
\subsection{Continuum}\label{sec:continuum}

From the continuum \CPT Lagrangian, one can predict numerous results for different low energy processes involving hadrons.  However, the extent of the accuracy of these predictions are ultimately tied to how well the LECs are known.  For this reason, there has been much effort in the lattice community to try to determine these values \cite{Bernard:2006zp,Boucaud:2007uk,DelDebbio:2006cn,DelDebbio:2007pz,Leutwyler:2007ae}.

The continuum Lagrangian in \CPT is determined order by order in $Bm_q$ and $p^2$.  The Lagrangian through $\mc O(p^4)$ for two flavors is given by \cite{Gasser:1983yg}

\begin{align} \label{eq:L_cont}
    \mc L_{cont} =& \frac{f^2}{8}\tr(\delmd \S \delmu \S^\dag) + \frac{Bf^2}{4}\tr(m_q \S^\dag + \S m_q)  \nonumber\\
    &+\frac{\ell_1}{4}\big[\tr(\delmd \S \delmu \S^\dag)\big]^2+ \frac{\ell_2}{4} \tr(\delmd \S \delnd \S^\dag) \tr(\delmu \S^\dag \delnu \S) \nonumber\\
    &+\frac{(\ell_3+\ell_4)B^2}{4}\big[\tr(m_q \S^\dag + \S m_q)\big]^2 + \frac{\ell_4B}{4}\tr(\delmd \S \delmu \S^\dag)\tr(m_q \S^\dag + \S m_q) ,
\end{align}
where $f \sim 132$ MeV, $\ell_{1-4}$ are the original Gasser-Leutwyler coefficients defined in Ref.~\cite{Gasser:1983yg} and

\begin{align}\label{eq:def}
    &\S = \exp \Big(\frac{2i\phi}{f}\Big)\, ,&
    &\phi = \left(\begin{array}{cc}\frac{\pi_0}{\sqrt{2}} & \pi^+ \\ \pi^- & -\frac{\pi_0}{\sqrt{2}}\end{array}\right)\, ,&
    &m_q = \left(\begin{array}{cc}\bar{m} & 0 \\0 & \bar{m}\end{array}\right).
\end{align}
At LO, the resulting condensate is

\begin{equation}\label{qq_cond }
B = \lim_{m_q\rightarrow 0} \frac{ | \langle \bar{q} q \rangle |}{f^2}. 
\end{equation}

From  Eq.~\eqref{eq:L_cont}, one can calculate the physical values for the mass of the pion  ($\mp$) and the pion decay constant ($\fp$) to LO and NLO.  These expressions are given by

\begin{eqnarray}\label{eq:cont_f_m}
    \mp^2 &=& m^2 + \frac{1}{3f^2}[4\mp^2i\mc I(\mp) - m^2i\mc I(\mp)] + 4\ell_3 \frac{m^4}{\fp^2} \\
    &&\nonumber \\
    \fp &=& f \Big [1-\frac{2}{\fp^2}i\mc I(\mp) + 2\ell_4\frac{m^2}{\fp^2} \Big ] 
\end{eqnarray}
where $m^2$ and $\mc I(\mp)$ are defined below in Eq.~\eqref{eq:int_def}.  When evaluating the scattering amplitude from \CPT, one has the option of either expressing the answer in terms of the bare parameters ($f$ and $m$) or in terms of lattice-physical parameters ($\fp$ and $\mp$).  Throughout this paper, the bare parameters will always be eliminated from the scatting amplitude.  

\begin{figure}[t]\label{fig:feynman}
\begin{tabular}{ccccc}
	\includegraphics[width=0.23\textwidth]{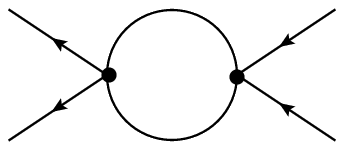} & $\;\;\;\;$ & \includegraphics[width=0.23\textwidth]{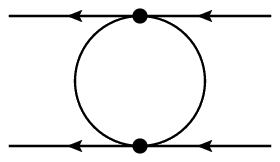} & $\;\;\;\;$ & \includegraphics[width=0.23\textwidth]{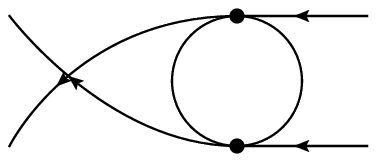} \\
	(a) & $\;\;$ & (b) & $\;\;$ & (c) \\
\end{tabular}
\vspace{2mm}
\begin{tabular}{cccc}
	\includegraphics[width=0.16\textwidth]{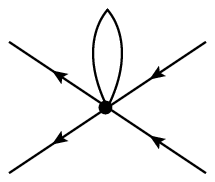} & $\;\;\;\;\;\;$ & \includegraphics[width=0.16\textwidth]{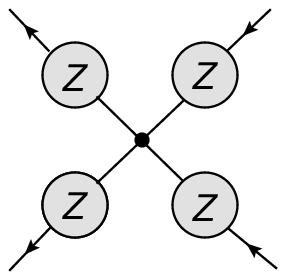}  \\
	(d) & $\;\;\;\;\;\;\;\;\;\;\;\;\;\;$ & (e)  \\
	\end{tabular}
\caption{The one-loop diagrams which contributing to the $\pi\pi$ scattering amplitude (same diagrams that contribute to the finite volume effects in Ref.~\cite{Bedaque:2006yi}).  The top three represent the s-channel (a), t-channel (b), and u-channel diagrams (c).  The diagram resulting from the six-pion vertex is given in (d) and the wavefunction corrections are in diagram (e).  These are the only diagrams for continuum, isotropic, and anisotrpoic Wilson actions.}
\label{fig:one-loop}
\end{figure}

The continuum I=2 $\pi\pi$ scatting length at infinite volume is given by

\begin{align}\label{eq:cont_scatt_amp}
    T_{cont}= -\frac{2}{\fp^2} \bigg\{ s &- 2\mp^2 - \frac{2(3s-4\mp^2)}{3\fp^2}i\mc I(\mp) + \frac{(s-2\mp^2)^2}{\fp^2}i\mc J(\mp,p_s) \nonumber \\
    &+\frac{1}{3\fp^2}\big [ 3(t^2-\mp^4)+t(t-s)-2t\mp^2+4s\mp^2-2\mp^4 \big ] i\mc J(\mp,p_t) \nonumber \\
    &+ \frac{1}{3\fp^2}\big [ 3(u^2-\mp^4)+u(u-s)-2u\mp^2+4s\mp^2-2\mp^4 \big ] i\mc J(\mp,p_u) \nonumber \\
    &-\frac{1}{9(4\pi\fp)^2}\big [ 2s^2+6s\mp^2 -8\mp^4 -t^2 - u^2 \big] \nonumber \\
    &-\frac{4\ell_1}{\fp^2} \big[(t -2\mp^2)^2 + (u -2\mp^2)^2 \big ] \nonumber\\
    &- \frac{2\ell_2}{\fp^2}\big [2(s-2\mp^2)^2 + (t-2\mp^2)^2 + (u-2\mp^2)^2 \big] \nonumber \\
    &-8\ell_3 \frac{\mp^4}{\fp^2} + 4\ell_4 \frac{\mp^2(s-2\mp^2)}{\fp^2} \bigg\}
\end{align}
where
\begin{eqnarray}\label{eq:int_def}
m^2 & = & 2B\bar{m} \nonumber\\
\mc I(\mp) & = & \int_\mathcal{R} \frac{d^4k}{(2\pi)^4}\frac{1}{k^2-m^2} \nonumber\\
\mc J(\mp,P) & = & \int_\mathcal{R} \frac{d^4k}{(2\pi)^4}\frac{1}{[(k+P)^2-m^2]}\frac{1}{[k^2-m^2]}. 
\end{eqnarray}

This scatting amplitude includes all the partial wave contributions (this $T$ is the same as the $T(s,\th)$ that appears in Eq.~\eqref{eq:Partial_Wave }).  When projecting on the s-wave, expanding through $\mc O(k^2/\mp^2)$ and using Eq.~\eqref{kcot_cont_amp}, the result for $k\cot \d_0$(for the regularization and renormalization scheme defined in Ref.~\cite{Bijnens:1997vq})  is

\begin{align}\label{cont_k}
k\cot \d_0 \approx -\frac{8\pi\fp^2}{\mp} \Bigg\{& \bigg(1-\frac{\mp^2}{(4\pi\fp)^2} \Big[3\ln\Big(\frac{\mp^2}{\mu^2} \Big) -1 + \ell_{\pi\pi}^a(\mu)\Big]\bigg) \nonumber\\
&-\frac{1}{2} \bigg(3+\frac{\mp^2}{(4\pi\fp)^2} \Big[\frac{17}{3}\ln\Big(\frac{\mp^2}{\mu^2} \Big) +\frac{31}{3} + \ell_{\pi\pi}^r(\mu)\Big]\bigg)\frac{k^2}{\mp^2}+ \cdots \Bigg\},
\end{align}

where $\ell_{\pi\pi}^a(\mu)$ and  $\ell_{\pi\pi}^r(\mu)$ are linear combinations of the Gasser-Leutwyler coefficients given by \cite{Bijnens:1997vq}
\begin{align}\label{eq:def_l}
    \ell_{\pi\pi}^a(\mu) =& -4(4\pi)^2\Big(4\big(\ell_1^R(\mu) + \ell_2^R(\mu) \big) +\big(\ell_3^R(\mu)-\ell_4^R(\mu)\big)\Big),\nonumber\\
    \ell_{\pi\pi}^r(\mu) =& 4(4\pi)^2\big(12\ell_1^R(\mu)+4\ell_2^R(\mu)+7\ell_3^R(\mu)-3\ell_4^R(\mu)\big).
\end{align}
The superscript $R$ represents the renormalized Gasser-Leutwyler coefficients with scale-dependence.

In order for these continuum predictions from \CPT to be useful in a physical context, one needs to determine the values for $\ell_{\pi\pi}^a(\mu)$ and  $\ell_{\pi\pi}^r(\mu)$, which are undetermined from \CPT alone. While numerous values have been quoted for the Gasser-Leutwyler coefficients \cite{Leutwyler:2007ae}, it is still beneficial to determine these values with more precesion.  Therefore, it is prudent to use various lattice calculations at different pion masses determine these values.  However, since the lattice calculations are performed with discretized space and time, one needs to remove these lattice artifacts to extract the continuum result.

%
%
\subsection{Isotropic Discretization}\label{sec:iso}

To calculate finite lattice spacing corrections to I=2 $\pi\pi$ scattering for the isotropic Wilson action, one can follow the same steps done in the continuum case, but starting from a Lagrangian which includes these lattice spacing artifacts. The analysis on lattice spacing effects was done for the Symanzik action by Sheikholeslami and Wohlert \cite{Sheikholeslami:1985ij}.  From this analysis, the Lagrangian was made explicit in \CPT by Sharpe and Singleton \cite{Sharpe:1998xm}, followed by B\"ar, Rupak, and Shoresh \cite{Rupak:2002sm,Bar:2003mh}.  The power counting they used for this \CPT Lagrangian is
\begin{equation}\label{eq:pow_iso }
a_sW \sim Bm_q \sim p^2 \sim \e,
\end{equation}
where $a_s$ is the lattice spacing (same spacing in space and time direction) and $W$ is a condensate defined below.  The simplified two flavor Lagrangian to $\mc O(\e^2)$ is
     
\begin{align} \label{eq:L_iso}
    \mc L_{iso} =& \mc L_{cont}+ \D\mc L_{iso}.
\end{align}
\begin{align} \label{eq:dl_iso}
    \D\mc L_{iso} =& \frac{a_sWf^2}{4}\tr(\S^\dag + \S) +\frac{(w_3+w_4)a_sWB_0}{4}\tr(m_q \S^\dag + \S m_q)\tr(\S^\dag + \S) \nonumber\\  
    &+\frac{w_3^\prime (a_sW)^2}{4}\big[\tr(\S^\dag + \S)\big]^2+ \frac{w_4a_sW}{4}\tr(\delmd \S \delmu \S^\dag)\tr(\S^\dag + \S)
\end{align}

This Lagrangian is similar to Eq.~\eqref{eq:L_cont} with one new term at LO and three new terms at NLO.  All new terms are proportional to $a_s$ or $a_s^2$, which will vanish in the continuum limit as $a_s \rightarrow 0$.  At LO, there is a new condensate given by
\begin{equation}\label{SW_cond }
W = \lim_{m_q \rightarrow 0}c_{SW} \frac{ \langle \bar{q} \sigma_{\mu\nu} F^{\mu\nu} q \rangle}{f^2}.
\end{equation}
The new LECs at NLO are given by $w_3$, $w_3^\prime$, and $w_4$.  All these new terms obey the same symmetries as the Lagrangian in the continuum case and break chiral symmetry in a similar way to the quark mass.  It should also be noted that these new LECs depend on $a_s\ln a_s$ as well (as opposed to the Gasser-Leutwyler coefficients that have no dependence on the mass term). 

Furthermore, Ref.~\cite{Aoki:2007es} showed that the axial current (needed to calculate $\fp$) has an additional term at this order in \CPT given by 
\begin{eqnarray}
    \D A^a_\mu = 2aw_A \partial_\mu \tr\big(\s^a(\S-\S^\dag)\big).
\end{eqnarray}
This term (which can also be inferred from Ref.~\cite{Sharpe:2004ny}) leads to modifications of the LECs as well as the coefficient in front of the chiral logarithm in $\fp$.  Thus, this coefficient has dependence on the lattice artifacts at NNLO.  Ref.~\cite{Aoki:2007es} also points out that the condition for fixing the renormalization factor, $Z_A$, of the lattice currents needs to be mapped onto \CPT\footnote{The condition for fixing $Z_A$ is chosen by individual lattice calculations.  Ref.~\cite{Aoki:2007es} shows an example of this in \CPT by using the chiral Ward identities in infinite volume, which leads to  $\fp$ being free  lattice artifacts until NNLO.}.
   From this Lagrangian, $\fp$ and $\mp$ through NLO are \cite{Rupak:2002sm,Bar:2003mh,Aoki:2007es}

\begin{eqnarray}
    \mp^2 &=& (m^2 + 2a_sW) + \frac{1}{3f^2}[4\mp^2i\mc I(\mp) - (m^2+2a_sW)i\mc I(\mp)]\nonumber \\
    &&+ 4\ell_3 \frac{m^4}{\fp^2}+ 8w_3 \frac{a_sWm^2}{\fp^2}+16w_3^\prime \frac{(a_sW)^2}{\fp^2} \label{eq:iso_m}\\
    &&\nonumber \\
    \fp &=& f \Big [1-\frac{2}{\fp^2}i\mc I(\mp) + 2\ell_4\frac{m^2}{\fp^2}+4w_{eff}\frac{a_sW}{\fp^2} \Big ], \label{eq:iso_f}
\end{eqnarray}
where $w_{eff}$ in $\fp$ includes $w_4$ and $w_A$ and can vary based on the given renormalization condition for the axial current.  

To acquire the continuum result of these quantities, one needs to remove all the terms with dependence on $a_s$ or $a_s^2$.   The resulting I=2 $\pi\pi$ scatting amplitude with the physical parameters restored is given by

\begin{equation}\label{eq:T_iso}
T_{iso}=T_{cont}+\D T_{iso},
\end{equation}
where the $\mp$ and $\fp$ in $T_{cont}$ are given by Eq.~\eqref{eq:iso_m} and Eq.~\eqref{eq:iso_f} and $\D T_{iso}$ is given by

\begin{align}\label{eq:iso_scatt_amp}
    \D T_{iso}= -\frac{2}{\fp^2} \bigg\{&-16(w_3-2\ell_3)\frac{a_sW\mp^2}{\fp^2} -32(w_3^\prime-w_3+\ell_3)\frac{(a_sW)^2}{\fp^2}\nonumber\\
    &+8(w_{eff}-\ell_4)\frac{a_sW(s-2\mp^2)}{\fp^2}\bigg\}.
\end{align}

It is worth noting that by restoring the physical parameters, $m^2$ does not appear, and $a_sW$ only appears in the terms containing the LECs (the $\ell$ and $w$ terms).  This is a bit different than the continuum case where one could eliminate $m^2$ with only $\mp^2$.  Now, one eliminates $m^2$ with $(\mp^2-2a_sW)$, and thus, several LECs are multiplied by factors of $a_sW$.  In addition, all of the continuum results without LECs remain unchanged since each vertex will only contribute $\mp^2$ when the physical parameters are restored.  Using the relation in Eq.~\eqref{eq:D_t }, the resulting artifact for the isotropic Wilson lattice, $\D(k\cot \d_0)_{iso}$ is given by

\begin{align}\label{iso_k}
\D(k\cot \d_0)_{iso} \approx \frac{\mp}{2\pi} \Bigg\{& \bigg(w_{\pi\pi}^a(\mu)\frac{a_sW}{\mp^2}+w_{\pi\pi}^{\prime a}(\mu)\frac{(a_sW)^2}{\mp^4}\bigg) \nonumber\\
&-\frac{1}{2} \bigg(w_{\pi\pi}^r(\mu)\frac{a_sW}{\mp^2}+7w_{\pi\pi}^{\prime a}(\mu)\frac{(a_sW)^2}{\mp^4}\bigg)\frac{k^2}{\mp^2}+ \cdots \Bigg\},
\end{align}
where

\begin{align}\label{eq:def_w}
    w_{\pi\pi}^a(\mu) =& -8(4\pi)^2\big(w_3^R(\mu)-w_{eff}^R(\mu) -2\ell_3^R(\mu)+\ell_4^R(\mu)\big),\nonumber\\
    w_{\pi\pi}^{\prime a}(\mu) =&-16(4\pi)^2\big(w_3^{\prime R}(\mu)-w_3^R(\mu)+\ell_3^R(\mu)\big),\nonumber\\
    w_{\pi\pi}^r(\mu) =&-8(4\pi)^2\big(7w_3^R(\mu)-3w_{eff}^R(\mu)-14\ell_3^R(\mu)+3\ell_4^R(\mu)\big).
\end{align}

As seen in the results, the artifacts from the isotropic lattice that are present in the final form are either linear or quadratic in $a_s$.  By using results that differ in lattice spacing, one can pick off the coefficients of these artifacts and remove them from the final result.  If one is working with a perfectly clover-improved Wilson lattice, this would remove all $\mc O(a_s)$ effects leaving only the $\mc O(a_s^2)$ effects.  It is also important to note that there is no physical information gained by determining specific LECs that are a result of the isotropic lattice spacing (the individual $w$ terms) unlike determining specific Gasser-Leutwyler coefficents.  Therefore, the useful coefficient to extract is the linear combination of these terms so they can be removed from the final result. 

At this point, one can compare these lattice spacing effects for the Wilson action to those found for the mixed action (with chiral valence fermions) in Ref.~\cite{Chen:2005ab,Chen:2006wf}.  For this mixed action case, when in terms of the lattice-physical parameters, there is no lattice spacing dependence through the NLO counter-terms.  In contrast, when the lattice-physical parameters are restored in the Wilson action, these effects first appear at the NLO counter-terms.  Thus, these additional effects that are not present at NLO in the mixed action calculation will need to be removed for the Wilson action calculation.

%
%
\subsection{Anisotropic Discretization}\label{sec:aniso}

With the recent formulation of \CPT for the anisotropic lattice \cite{Bedaque:2007xg}, one can begin calculating corrections to various quantities of interest on the lattice.  This process is, in general, very similar to calculations in the continuum and isotropic lattices, but one now picks up additional terms that are a result of having different spacial and temporal spacings.  To help extract these effects in a more simplistic notation, the superscript $\xi$ has been added to all the new terms resulting from this anisotropy.  In practice, the anisotropic lattice picks up two new non-perturbative parameters: the parameter $\xi = a_s/a_t$ which is the measure of anisotropy and the parameter $\nu$, which is used to correct the ``speed of light" \cite{Klassen:1998fh,Chen:2000ej,Umeda:2003pj}.  By setting both parameters to 1, the isotropic limit is recovered.  In addition to the $W$ condensate defined in Eq.~\eqref{SW_cond }, we pick up a $W^\xi$ condensate that is given by
\begin{equation}\label{SW_xi_cond }
W^\xi = \lim_{m_q \rightarrow 0}c_{SW}^\xi (u^\xi)^\mu (u^\xi)_\nu \frac{ \langle \bar{q} \sigma_{\mu\l} F^{\nu\l} q \rangle}{f^2}.
\end{equation}
where $u_\mu^\xi$ is a vector that breaks hypercubic invariance.  It is important to note that the convention chosen for $u_\mu^\xi$ appears in the anisotropic \CPT and its observables.  For convenience, we choose this vector to be  $u_\mu^\xi = (1,\textbf{0})$. The condensates and the anisotropic paramaters are related at the classic level by (with Wilson coefficients $r_s=r_t=1$) 
\begin{eqnarray}\label{W_prop}
    W&\propto& c_{SW} \propto \nu,   \\
    W^\xi &\propto& c_{SW}^\xi \propto \frac{1}{2}\bigg(\frac{a_t}{a_s}-\nu \bigg).  
\end{eqnarray}
In the isotropic limit when $\xi$ and $\nu$ are set to 1, the isotropic condensate will remain and the anisotropic condensate will vanish.  Using a similar notation throughout, all terms that appear with a $\xi$ superscript will vanish when $a_s = a_t$ and $\nu = 1$.

The power counting convention used in Eq.~\eqref{eq:pow_iso } for the anisotropic Lagrangian is

\begin{equation}\label{eq:pow_aniso }
a_sW \sim a_sW^\xi \sim Bm_q \sim p^2 \sim \e.
\end{equation}
Writing this new Lagrangian in the form of Eq.~\eqref{eq:L_iso}, the two-flavor anisotropic \CPT Lagrangian through $\mc O(\e^2)$ is
\begin{align} \label{eq:L_aniso}
    \mc L_{aniso} =& \mc L_{cont}+ \D\mc L_{iso}+\D\mc L_{aniso}.
\end{align}
where

\begin{align} \label{eq:L_aniso}
    \D\mc L_{aniso} =& \frac{a_sW^\xi f^2}{4}\tr(\S^\dag + \S)+\frac{(w_3^\xi+w_4^\xi+w_1^\xi) a_sW^\xi B_0}{4}\tr(m_q \S^\dag + \S m_q)\tr(\S^\dag + \S)\nonumber\\ 
    &+\frac{\hat{w}_3^\xi (a_sW^\xi)^2}{4}\big[\tr(\S^\dag + \S)\big]^2+\frac{\bar{w}_3^\xi (a_sW)(a_sW^\xi)}{4}\big[\tr(\S^\dag + \S)\big]^2\nonumber\\
    &+ \frac{w_4^\xi a_sW^\xi}{4}\tr(\delmd \S \delmu \S^\dag)\tr(\S^\dag + \S)+\frac{w_1^\xi a_sW^\xi}{4}u^\mu u_\nu (\delmd \S \delnu \S^\dag)\tr(\S^\dag + \S).
\end{align}

In addition to the anisotropic condensate $W^\xi$ mentioned above at LO, there are five new LECs at NLO as a result of this anisotropy.  Four of these new LECs obey the same symmetry structure as the isotropic terms, however the $\w_1^\xi$ term additionally breaks hypercubic invariance.  Therefore, this term only corrects the time derivative, but not the spacial one (for the convention of $u_\mu^\xi$ chosen).  As a result, $\fp$ is parameterized by two constants; $\fpt$, which is $\fp$ measured in time, and $\fps$, which is $\fp$ measured in space.  This leads to one correction for the space-measured $\fps$ and a separate    correction for the $\fpt$.  The pion mass ($\mp$), the time-measured pion decay constant ($\fpt$), and the space-measured pion decay constant ($\fps$) through NLO are

\begin{eqnarray}
    \mp^2 &=& (m^2 + 2a_sW+2a_sW^\xi) + \frac{1}{3f^2}[4\mp^2i\mc I(\mp) - (m^2+2a_sW+2a_sW^\xi)i\mc I(\mp)]\nonumber \\
    &&+ 4\ell_3 \frac{m^4}{\fpt^2}+ 8w_3 \frac{a_sWm^2}{\fpt^2}+16w_3^\prime \frac{(a_sW)^2}{\fpt^2}\nonumber \\
    &&+8w_3^\xi \frac{a_sW^\xi m^2}{\fpt^2}+16\hat{w}_3^\xi \frac{(a_sW^\xi)^2}{\fpt^2} + 16\bar{w}_3^\xi \frac{(a_sW)(a_sW^\xi)}{\fpt^2}  \label{eq:aniso_m}\\
    &&\nonumber \\
    \fpt &=& f \Big [1-\frac{2}{\fpt^2}i\mc I(\mp) + 2\ell_4\frac{m^2}{\fpt^2}+4w_4\frac{a_sW}{\fpt^2} +4(w_{eff}^\xi+w_1^\xi)\frac{a_sW^\xi}{\fpt^2} \Big ]  \label{eq:aniso_f}\\
    &&\nonumber \\
    \fps &=& f \Big [1-\frac{2}{\fpt^2}i\mc I(\mp) + 2\ell_4\frac{m^2}{\fpt^2}+4w_4\frac{a_sW}{\fpt^2} +4w_{eff}^\xi\frac{a_sW^\xi}{\fpt^2} \Big ],
\end{eqnarray}
where, as in the isotropic case, the $w_{eff}^\xi$ depends on the  renormalization condition for the axial current (this LEC is the same for both $\fpt$ and $\fps$).

For the rest of this section, all calculations are given in terms of $\mp$ and $\fpt$.  As mentioned before, how one accounts for the effect of this hypercubic breaking term depends on the convention.  In the convention used here, only $\fpt$ sees the effect of this term and $\fps$ does not.  The scatting amplitude is given by

\begin{equation}\label{ }
T_{iso}=T_{cont}+\D T_{iso}+\D T_{aniso}
\end{equation}
where the $\mp$ and $\fpt$ in $T_{cont}$ are given by Eq.~\eqref{eq:aniso_m} and Eq.~\eqref{eq:aniso_f} and $\D T_{aniso}$ is given by

\begin{align}\label{eq:aniso_scatt_amp}
    \D T_{aniso}= -\frac{2}{(\fpt)^2} \bigg\{&-16(w_3^\xi-2\ell_3)\frac{a_sW^\xi\mp^2}{(\fpt)^2} -32(\hat{w}_3^\xi-w_3^\xi+\ell_3)\frac{(a_sW^\xi)^2}{(\fpt)^2}\nonumber\\
    &-32(\bar{w}_3^\xi-w_3-w_3^\xi+2\ell_3)\frac{(a_sW)(a_sW^\xi)}{(\fpt)^2}\nonumber\\
    &+8(w_{eff}^\xi+w_1^\xi-\ell_4)\frac{a_sW^\xi(s-2\mp^2)}{(\fpt)^2}-16w_1^\xi \frac{a_sW^\xi k^2}{(\fpt)^2}\bigg\}.
\end{align}

Most of the effects in this scattering amplitude are similar to the isotropic case, except now there are  also expansions in $a_sW^\xi$ in addition to the expansions in $a_sW$.  Thus, as expected, if all the anisotropic effects are removed, only the isotropic limit remains.  The only new symmetry breaking effect is the $w_1$ term which is not a hypercubic invariant term.  In other words, all of the hypercubic breaking due to anisotropy at this order is contained in this term.  However, it's effects in $\D T_{aniso}$ appear as just another contribution to the linear combination of the LECs in front of the term $a_sW^\xi$.  Therefore, it is difficult to determine the effect of the the hypercubic breaking term alone from $\D T_{aniso}$  since its effects will be mixed in with the other anisotropic LECs\footnote{The total hypercubic breaking effect would be more visible from the differences of $\fpt$ and $\fps$}.   The resulting artifacts for the anisotropic Wilson lattice are the isotropic artifacts, $\D(k\cot \d_0)_{iso}$, and the anisotropic artifacts, $\D(k\cot \d_0)_{aniso}$.  Therefore, the total effect of the lattice artifacts due to lattice spacings are 

\begin{equation}
\D(k\cot \d_0) = \D(k\cot \d_0)_{iso} + \D(k\cot \d_0)_{aniso}.
\end{equation}
The anisotropic lattice artifacts are given by

\begin{align}\label{aniso_k}
\D(k\cot \d_0)_{aniso} \approx \frac{\mp}{2\pi} &\Bigg\{ \bigg(w_{\pi\pi}^{\xi a}(\mu)\frac{a_sW^\xi}{\mp^2}+\hat{w}_{\pi\pi}^{ \xi a}(\mu)\frac{(a_sW^\xi)^2}{\mp^4}+ \bar{w}_{\pi\pi}^{\xi a}(\mu) \frac{(a_sW)(a_sW^\xi)}{\mp^4}\bigg) \nonumber\\
&-\frac{1}{2} \bigg(w_{\pi\pi}^{\xi r}(\mu)\frac{a_sW^\xi}{\mp^2}+7w_{\pi\pi}^{\prime \xi a}(\mu)\frac{(a_sW^\xi)^2}{\mp^4}+ 7\bar{w}_{\pi\pi}^{\xi a}(\mu) \frac{(a_sW)(a_sW^\xi)}{\mp^4}\bigg)\frac{k^2}{\mp^2}\nn\\
&+ \cdots \Bigg\},
\end{align}
where

\begin{align}\label{eq:def_w}
    w_{\pi\pi}^{\xi a}(\mu) =& -8(4\pi)^2\big(w_3^{\xi R}(\mu) - w_{eff}^{\xi R}(\mu)+w_1^{\xi R}(\mu) -2\ell_3^R(\mu)+\ell_4^R(\mu)\big) ,\nonumber\\
    \hat{w}_{\pi\pi}^{ \xi a}(\mu) =&-16(4\pi)^2\big(w_3^{\prime R}(\mu)-w_3^R(\mu)+\ell_3^R(\mu)\big),\nonumber\\
    \bar{w}_{\pi\pi}^{\xi a}(\mu) =&-16(4\pi)^2\big(\bar{w}_3^{\xi R}(\mu)-w_3^R(\mu)-w_3^{\xi R}(\mu)+2\ell_3^R(\mu)\big), \nonumber\\
    w_{\pi\pi}^{\xi r}(\mu)=& -8(4\pi)^2\big(7w_3^{\xi R}(\mu)+9w_{eff}^{\xi R}(\mu)+w_1^{\xi R}(\mu)-14\ell_3^R(\mu)-9\ell_4^R(\mu)\big).
\end{align}
  
As a result, the ultimate effects of the anisotropic lattice on $\D(k\cot \d_0)$ are more terms that will require variation of $a_s$ and $a_t$ independently to fit.  Three more terms require fitting to correct the constant term in the expansion and one more term requires fitting for the $\mc O(k^2/\mp^2)$ term in order to remove it's effects.  As mentioned with the isotropic correction, no physical information is gained by picking off the individual anisotropic LECs. Analogous to the isotropic case, the anisotropic Wilson action first has lattice spacing dependence at the NLO counter-terms. The aim here is to determine these linear combinations and remove their effects from the lattice measurements.

%
%
\section{Discussion \label{sec:concl}}

In this work, I=2 $\pi\pi$ scattering was calculated from the \CPT for isotropic and anisotropic lattice spacings.  Also, connections between these \CPT calculations and the $k \cot \d_0$ value measured from lattice calculations were illustrated.  When $\D(k \cot \d_0)$ is given in terms of the lattice-physical parameters, these lattice spacing effects first appear at the NLO LECs and can be removed from the result of the lattice calculation.  However,  $\D(k \cot \d_0)$ has numerous undetermined linear combinations of LECs, which would need to be determined (by fitting several different lattice spacings) in order to successfully remove it from the lattice result. Therefore, as more lattice calculations of $\pi\pi$ scattering are completed (for both isotropic and anisotropic lattice spacings), these combinations of LECs can be determined better, which will result in a more accurate result after these lattice artifacts are removed.   


\renewcommand{\thechapter}{4}

\chapter{$\pi\pi$ Scattering in Twisted Mass Chiral Perturbation Theory}\label{ch:TM_pi_pi}

\section{Overview}

Calculations of hadron interactions from lattice QCD have received significant attention in recent years~\cite{Beane:2008dv}, with dynamical calculations of two-meson systems~\cite{Yamazaki:2004qb,Beane:2005rj,Beane:2007xs,Beane:2006gj,Beane:2007uh}, two-baryon systems~\cite{Beane:2006mx,Beane:2006gf} and systems of up to 12 pions~\cite{Beane:2007es,Detmold:2008fn} and kaons~\cite{Detmold:2008yn}.  Additionally, lattice methods are currently being applied to the low-energy effective field theory of multinucleon interactions~\cite{Seki:2005ns,Chen:2003vy,Chen:2004rq,Lee:2004si,Lee:2004qd,Borasoy:2006qn,Borasoy:2007vi,Borasoy:2007vk}, for which there exists a nice review~\cite{Lee:2008fa}.  The majority of the dynamical lattice QCD calculations of hadron interactions to date, have either been performed with Wilson fermions or a mixed lattice action~\cite{Renner:2004ck,Edwards:2005kw} of domain-wall valence fermions~\cite{Kaplan:1992bt,Shamir:1993zy,Furman:1994ky} and the Asqtad improved~\cite{Orginos:1998ue,Orginos:1999cr} rooted staggered MILC configurations~\cite{Bernard:2001av,Aubin:2004wf}.  Twisted mass lattice QCD~\cite{Frezzotti:1999vv,Frezzotti:2000nk} has recently emerged as a viable fermion discretization method for lattice calculations with two flavors of light quarks, ($up$ and $down$) ~\cite{Boucaud:2007uk,Blossier:2007vv,Boucaud:2008xu} and hopeful prospects of $2+1+1$ ($up$, $down$, $strange$ and $charm$) flavors of dynamical sea fermions~\cite{Chiarappa:2006ae} in the chiral regime.  Therefore, it is natural that hadron interactions will be computed with the twisted mass fermion discretization method, and initial twisted mass results for I=2 $\pi\pi$ interactions have been recently published \cite{Feng:2009ij, Feng:2009ck}\footnote{These references made use of the work developed in this chapter}.

To truly understand the twisted mass results of I=2 $\pi\pi$ interactions, the systematic effects due to lattice artifacts must first be understood.  Lattice discretization effects can be incorporated into the chiral Lagrangian through a two-step process first detailed in Ref.~\cite{Sharpe:1998xm}.   One first constructs the effective continuum Symanzik Lagrangian~\cite{Symanzik:1983dc,Symanzik:1983gh} for a given lattice action.  One then builds the low energy chiral Lagrangian from the Symanzik theory, giving rise to new unphysical operators with their own LECs.  These new operators capture the discretization effects for a given lattice action.  This process is spelled out for the Wilson action in Chapter \ref{ch:Intro}. 

In this chapter, we briefly review the construction of the twisted mass chiral Lagrangian in Sec.~\ref{sec:tmLag}.
We then determine the lattice spacing $(b)$ corrections to low-energy $\pi\pi$ scattering specific to the twisted mass lattice action.  We work through $\mc{O}(b^2)\sim\mc{O}(bm_\pi^2)$.

\section{Twisted mass lattice QCD and the continuum effective action\label{sec:tmLag}}

The twisted mass chiral Lagrangian was determined previously in Refs.~\cite{Munster:2003ba,Scorzato:2004da,Sharpe:2004ps,Aoki:2004ta,Sharpe:2004ny}, and for baryons in Ref.~\cite{WalkerLoud:2005bt}.  In this work, we focus on twisted mass lattice QCD with degenerate light flavors given by the lattice action%
%
\begin{align}
S =&\ \sum_{x} \bar{\psi}(x) \bigg[ \frac{1}{2} \sum_\nu \g_\nu (\nabla^*_\nu + \nabla_\nu)
	-\frac{r}{2} \sum_\nu \nabla^*_\nu \nabla_\nu +m_0 + i\g_5 \t_3 \mu_0 \bigg] \psi(x)\, ,
\end{align}
where $\psi$ and $\bar{\psi}$ are the dimensionless lattice fermion fields, $\nabla_\nu(\nabla_\nu^*)$ are the covariant forward (backward) lattice derivatives in the $\nu$ direction, $m_0$ is the dimensionless bare quark mass and $\mu_0$ is the dimensionless bare twisted quark mass.  The fermion fields are flavor doublets, $\tau_3$ is the third Pauli-spin matrix and the bare mass term is implicitly accompanied by a flavor identity matrix.  Our twisted mass $\chi$PT analysis also holds for dynamical lattice calculations with $2+1+1$ flavors, the only difference being the numerical values of the LECs determined when fitting the extrapolation formula to the calculation results.  This action is beneficial  in several ways.  While it has a term that breaks parity, it contains a parity $\bigotimes$ flavor symmetry that prevents unwanted radiative corrections.  However, this term, which is referred to as the twisted mass term, is a mass term that multiplicatively renormalized akin to how the quark mass term is in continuum QCD.  This prevents issues involving exceptional configurations.  Additionally, for certain choices of $m_0$ and $\mu_0$ parameters, known as maximal twist, this action can additionally be  $\mc O(a)$ improved without including a Sheikholeslami-Wohlert term.  

The continuum chiral Lagrangian, supplemented by discretization effects is determined with the two step procedure of Ref.~\cite{Sharpe:1998xm}.  This was done for the twisted mass lattice action in Ref.~\cite{Sharpe:2004ps}, to NLO in which a power counting $m_q \sim b \Lambda_{QCD}^2$ was used and which we shall adopt.  The resulting effective  Lagrangian is
\begin{align}\label{eq:Leff}
\mc{L}_{eff} =&\ \mc{L}_{glue}
	+\bar{q} (\Dslash + m + i \g_5 \t_3 \mu ) q+c_{SW} b\ \bar{q}\, i \s_{\mu\nu} F_{\mu\nu} q
	+\mc{O}(b^2, bm_q, m_q^2)\, ,
\end{align}
where $\mc{L}_{glue}$ is the Yang-Mills Lagrangian.  The quark fields are an isodoublet, $q^T = (q_u, q_d)$ and the quark masses are given by
\begin{align}
	m &= Z_m (m_0 - m_c) / b\, ,
\nonumber\\
	\mu &= Z_\mu \mu_0 / b\, .
\end{align}
The symmetry properties of the twisted mass lattice action protect the twisted mass from additive mass renormalization.  With Eq.~\eqref{eq:Leff}, one can construct the two flavor chiral Lagrangian.  This is the Gasser-Leutwyler Lagrangian~\cite{Gasser:1983yg} supplemented by chiral and flavor symmetry breaking terms proportional to the lattice spacing.  The Lagrangian through NLO relevant to our work takes the form~\cite{Sharpe:2004ps,Sharpe:2004ny},

\begin{align}\label{eq:Lag}
\mc{L}_\chi^{tw} =&\ \frac{f^2}{8} \tr (\partial_\mu \S \partial_\mu \S^\dagger )
	-\frac{f^2}{8}\tr ( \chi^{\prime \dagger} \S + \S^\dagger \chi^\prime )
	-\frac{l_1}{4} \tr (\partial_\mu \S \partial_\mu \S^\dagger )^2
	-\frac{l_2}{4} \tr (\partial_\mu \S \partial_\nu \S^\dagger ) 
		\tr (\partial_\mu \S \partial_\nu \S^\dagger )
\nonumber\\&
	-\frac{l_3+l_4}{16} \Big[ \tr ( \chi^{\prime \dagger}\S + \S^\dagger \chi^\prime) \Big]^2
	+\frac{l_4}{8} \tr (\partial_\mu \S \partial_\mu \S^\dagger )
		\tr ( \chi^{\prime \dagger}\S + \S^\dagger \chi^\prime)
\nonumber\\&
	+\tilde{W} \tr (\partial_\mu \S \partial_\mu \S^\dagger )
		\tr (\hat{A}^\dagger \S + \S^\dagger \hat{A} )
	-W \tr (\chi^{\prime \dagger}\S + \S^\dagger \chi^\prime)
		\tr (\hat{A}^\dagger \S + \S^\dagger \hat{A} )
		\nonumber\\&
	-W^\prime \Big[ \tr (\hat{A}^\dagger \S + \S^\dagger \hat{A} ) \Big]^2\, ,
\end{align}

where the LECs, $l_1$--$l_4$ are the $SU(2)$ Gasser-Leutwyler coefficients and the coefficients $\tilde{W}$, $W$ and $W^\prime$ are unphysical LECs arising from the explicit chiral symmetry breaking of the twisted mass lattice action.  The spurion fields are defined as
\begin{align}
	&\chi^\prime = 2B_0(m + i\t_3 \mu) + 2W_0 b \equiv \hat{m} + i\t_3 \hat{\mu} + \hat{b}&
\nonumber\\
	&\hat{A} = 2W_0 b \equiv \hat{b}\, .&
\end{align}
As discussed in Ref.~\cite{Sharpe:2004ny}, the vacuum of the theory as written is not aligned with the flavor identity but is given at LO by
\begin{equation}
\S_0 \equiv \langle 0 | \S | 0 \rangle
	= \frac{\hat{m} + \hat{b} + i\t_3 \hat{\mu}}{M^\prime}
	= \textrm{exp}(i \w_0 \t_3)\, ,
\end{equation}
with
\begin{equation}
	M^\prime = \sqrt{(\hat{m}+\hat{b})^2 +\hat{\mu}^2}\, .
\end{equation}
Therefore, to determine the Feynman rules which leave the interactions of the theory the most transparent, one expands the Lagrangian around the physical vacuum.    Extending this analysis to NLO, one finds the vacuum angle shifts to $\w = \w_0 + \e$ where one can determine $\e$ either by finding the minimum of the potential, as was done in Ref.~\cite{Sharpe:2004ny} or by requiring the single pion vertices to vanish,
\begin{equation}\label{eq:deltaw}
	\e(\w_0) = -\frac{32}{f^2}\hat{b} \sin \w_0 \left[ W + 2W^\prime \cos \w_0 \frac{\hat{b}}{M^\prime} \right]\, .
\end{equation}
One can expand about the physical vacuum by making the replacement
\begin{align}
	&\S = \xi_m\, \S_{ph}\, \xi_m,&
	&\textrm{with}&
	&\xi_m = \textrm{exp} (i \w \t_3 / 2),&
\end{align}
and
\begin{align}
&\S_{ph} = \textrm{exp} \left( \frac{2i \phi}{f} \right),& 
&\phi = \begin{pmatrix} \frac{\pi^0}{\sqrt{2}} & \pi^+ \\ \pi^- & -\frac{\pi^0}{\sqrt{2}}
	\end{pmatrix}\, .&
\end{align}
One then finds the Lagrangian is given by
\begin{align}
\mc{L} =&\  \mc{L}_{cont.} +
	\tilde{W} \hat{b} \cos \w\, \tr ( \partial_\mu \S_{ph} \partial_\mu \S_{ph}^\dagger) 
		\tr (\S_{ph} + \S_{ph}^\dagger)
	-\hat{b} \cos \w \left( W M^\prime +W^\prime \hat{b} \cos \w \right)
		\left[ \tr (\S_{ph} + \S_{ph}^\dagger) \right]^2
\nonumber\\&
	+\tilde{W} \hat{b}\sin \w\, \tr ( \partial_\mu \S_{ph} \partial_\mu \S_{ph}^\dagger) 
		\tr ( i\t_3 (\S_{ph} - \S_{ph}^\dagger ) )
	-W^\prime \hat{b}^2 \sin^2 \w\, 
		\left[ \tr ( i\t_3 (\S_{ph} - \S_{ph}^\dagger ) ) \right]^2
\nonumber\\&
	-\tr ( i\t_3 (\S_{ph} - \S_{ph}^\dagger ) ) \left[
		\e(\w)\, \frac{M^\prime f^2}{8}
		+\hat{b} \sin \w\, \left( W M^\prime + 2 W^\prime \hat{b} \cos \w\, \right)
			\tr (\S_{ph} + \S_{ph}^\dagger) \right]\, ,
\end{align}
where $\mc{L}_{cont.}$ is the continuum $SU(2)$ chiral Lagrangian to NLO.  
Of particular interest to us are the new two, three and four pion interactions which result from the discretization errors in the twisted mass Lagrangian.
We find, in agreement with Ref.~\cite{Sharpe:2004ny}
\begin{equation}
\mc{L} = \mc{L}_{cont.} + \D\mc{L}_{2\phi} + \D\mc{L}_{3\phi} + \D\mc{L}_{4\phi}\, ,
\end{equation}
where
\begin{align}\label{eq:L2phi}
\D\mc{L}_{2\phi} = &\ 
	\cos \w \frac{16\tilde{W}\hat{b}}{f^2} \tr (\partial_\mu \phi \partial_\mu \phi )
	+\frac{1}{2}\D M^\prime(\w)\, \tr( \phi^2 ) 
	+ \frac{1}{2} \D M^\prime_0(\w)\, \left[ \tr \left( \frac{\t_3 \phi}{\sqrt{2}} \right) \right]^2\, ,
\end{align} 
\begin{align}\label{eq:L3phi}
\D\mc{L}_{3\phi} =&
	-\sin \w\, \frac{16 \tilde{W} \hat{b}}{f^3} \tr(\t_3 \phi) \tr( \partial_\mu \phi \partial_\mu \phi)
	+\frac{\e(\w) M^\prime}{2f} \tr( \t_3 \phi) \tr(\phi^2)\, ,
\end{align}
\begin{align}\label{eq:L4phi}
\D\mc{L}_{4\phi} =&
	-\frac{\D M^\prime(\w)}{3 f^2} [\tr ( \phi^2 ) ]^2
	-\frac{\D M^\prime_0(\w)}{3f^2} [\tr (\frac{\t_3 \phi}{\sqrt{2}} )]^2 \tr (\phi^2 )
\nonumber\\&
	+\cos \w\, \frac{32 \tilde{W} \hat{b}}{3f^4} \tr ( \phi\, \partial_\mu \phi [ \phi, \partial_\mu \phi] )
	-\cos \w\, \frac{16 \tilde{W} \hat{b}}{f^4} \tr(\partial_\mu \phi \partial_\mu \phi ) \tr ( \phi^2)\, ,
\end{align}
and the mass corrections are given by
\begin{align}
	&\D M^\prime(\w) = \cos \w \frac{64 \hat{b}}{f^2} \left( W M^\prime + \cos \w\, W^\prime \hat{b} \right),&
\nonumber\\ \label{eq:Mprime0}
	&\D M^\prime_0(\w) = -\sin^2 \w \frac{64 W^\prime \hat{b}^2}{f^2}\, .&
\end{align}
From this Lagrangian, one can determine the pion masses, decay constants and wave-function corrections.  One finds the masses are (using the modified \textit{dimensional regularization} of Ref.~\cite{Gasser:1983yg})
\begin{align}
m_{\pi^\pm}^2 =&\ M^\prime \left[
	1 + \frac{M^\prime}{(4\pi f)^2}\ln \left( \frac{M^\prime}{\mu^2} \right) + l_3^r(\mu)\frac{4 M^\prime}{f^2} \right]
\nonumber\\&
	+\D M^\prime(\w) - \cos\w \frac{32\tilde{W}\hat{b}M^\prime}{f^2}\, ,
\\ \label{eq:mpi0}
m_{\pi^0}^2 =&\ m_{\pi^\pm}^2 + \D M^\prime_0(\w)\, ,
\end{align}
the decay constants are%
\footnote{There is an exact Ward identity one can exploit to compute the charged pion decay constant and avoid issues of the axial current renormalization discussed for example in Ref.~\cite{Aoki:2007es}.} 
\begin{align}
f_{\pi} =&\ f \bigg[ 
	1 - \frac{2M^\prime}{(4\pi f)^2} \ln \left(\frac{M^\prime}{\mu^2} \right)
	+l_4^r(\mu)\frac{2 M^\prime}{f^2} 
	+\cos\w \frac{16\tilde{W}\hat{b}}{f^2} \bigg]\, ,
\end{align}
and the wave-function correction is
\begin{align}
\d\mc{Z}_\pi = &\ \frac{4M^\prime}{3(4\pi f)^2} \ln \left( \frac{M^\prime}{\mu^2} \right)
	-l_4^r(\mu) \frac{4M^\prime}{f^2}
	-\cos\w \frac{32\tilde{W}\hat{b}}{f^2}
\end{align}
These expressions will be needed to express the scattering in terms of the lattice-physical parameters (by lattice-physical, we mean the renormalized mass and decay constant as measured from the correlation functions, and not extrapolated to the continuum or infinite volume limit).  As we discuss in the next section, these interactions lead to three types of new contributions to $\pi\pi$ scattering states: there are discretization corrections to the scattering parameters, the scattering lengths, effective ranges, etc., which appear in a mild manner as those from the Wilson chiral Lagrangian~\cite{Buchoff:2008ve}.  There are corrections which can potentially significantly modify the chiral behavior, arising from the three-pion interactions, and there are new corrections which mix different scattering channels, for example, the $I=2, I_3=0$ and the $I=0$ scattering states.

%
\section{$\pi\pi$ Scattering in Twisted Mass $\chi$PT}
In this section we calculate corrections to the two-pion scattering channels.  We begin with the maximally stretched $I=2$ states, which have the simplest corrections.

\subsection{$I=2, I_3=\pm 2$ Channels}
There are two types of discretization corrections which modify the $I=2, I_3=\pm2$ scattering, those which are similar to the corrections for the Wilson lattice action~\cite{Buchoff:2008ve,Aoki:2008gy} and those which arise from the three-pion interactions, Eq.~\eqref{eq:L3phi} and give rise to new Feynman diagrams.  We will express the scattering parameters in terms of the lattice-physical pion mass and decay constant.  As was shown in detail, this has dramatic consequences on the formula for the scattering parameters in both partially quenched and mixed action $\chi$PT~\cite{Chen:2005ab,Chen:2006wf,Chen:2007ug}, such that the extrapolation formulae were free of unphysical counterterms through NLO.  There is a second benefit to expressing the scattering parameters in lattice-physical parameters.  This allows one to perform a chiral extrapolation in terms of the ratio $m_\pi / f_\pi$, and thus avoid the need for scale setting.  This was crucial in allowing the NPLQCD Collaboration to make a precision prediction of the $I=2$ scattering length~\cite{Beane:2005rj,Beane:2007xs}.

\begin{figure}[t]
\center
\begin{tabular}{ccc}
\includegraphics[width=0.25\columnwidth]{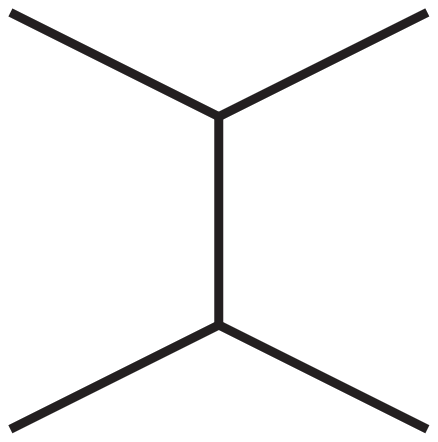}
&\phantom{space}&
\includegraphics[width=0.25\columnwidth]{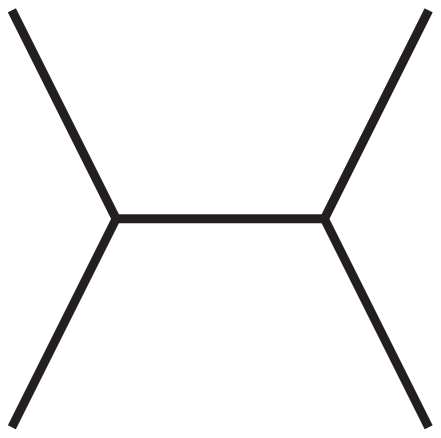}
\\
$(a)$ &&$(b)$
\end{tabular}
\caption{\label{fig:3phi_graphs} New unphysical graphs from twisted mass interactions in the $t$($u$)-channel $(a)$ and $s$-channel $(b)$.  Fig.~$(b)$ can only contribute to $I_3=0$ scattering.}
\end{figure}

The simple corrections to the scattering amplitude are determined from $\D\mc{L}_{4\phi}$, Eq.~\eqref{eq:L4phi}.  The three-pion interactions from Eq.~\eqref{eq:L3phi} lead to new graphs in the scattering amplitude, which we depict in Fig.~\ref{fig:3phi_graphs}.  The $I=2, I_3=\pm2$ scattering channels receive corrections from Fig.~\ref{fig:3phi_graphs}$(a)$ and its $u$-channel counterpart.
The internal propagating pion is a $\pi^0$, which, for present lattice actions, is known to be lighter than the charged pions in dynamical twisted mass lattice calculations with degenerate light quark masses~\cite{Boucaud:2007uk,Boucaud:2008xu}.

Putting all the corrections together, one finds the scattering amplitude, which we express in Minkowski-space, is given by
\begin{equation}
\mc{T}^{2,\pm2} = \mc{T}^{2,\pm2}_{cont.} + \D\mc{T}^{2,\pm2}\, ,
\end{equation}
where the discretization corrections are 
\begin{align}
\D \mc{T}^{2,\pm2}(\w) &= 4\frac{\D M^\prime(\w)}{f_\pi^2}
	+\cos\w \frac{64\tilde{W} \hat{b}}{f_\pi^4}\, s
	+\frac{2\e^2(\w)m_\pi^4}{f_\pi^2} \left( \frac{1}{m_{\pi^0}^2 - t} + \frac{1}{m_{\pi^0}^2 - u} \right)\, .
\end{align}
The first two terms arise from Eq.~\eqref{eq:L4phi} as well as from the conversion to the lattice-physical parameters.  The second two terms arise from Fig.~\ref{fig:3phi_graphs}$(a)$.  These terms are formally NNLO.  However, depending upon the precision with which the twist angle is tuned, these terms may become large and require promotion to lower order.  Expanding the NLO contribution to the twist angle, Eq.~\eqref{eq:deltaw}, one finds
\begin{equation}
\e^2(\w) m_\pi^4 = \left(\frac{64 W^\prime \hat{b}^2}{f_\pi^2}\right)^2 \sin^2 \w \cos^2 \w
	+\mc{O}(m_\pi^2)\, .
\end{equation}
We can then determine the corrections to the $I=2, I_3=\pm2$ scattering lengths, for which we find
\begin{align}
\D m_\pi a_{\pi\pi}^{I=2,\pm2}(\w) =&\
	\frac{\D M^\prime(\w)}{8\pi f_\pi^2}
	-\cos(\w)\frac{8 \tilde{W} \hat{b} m_\pi^2}{\pi f_\pi^4}
	+\frac{(32W^\prime)^2}{2\pi} \frac{\sin^2 \w \cos^2 \w}{m_{\pi^0}^2/f_\pi^2} \frac{\hat{b}^4}{f_\pi^8}\, .
\end{align}
The first observation we make is that at maximal twist, $\w=\pi/2$, these leading discretization errors exactly cancel through NLO (this is true of the corrections to the scattering amplitude and not just the scattering length)%
\footnote{We have assumed that a suitable definition of the maximal twist angle has been used in the numerical lattice computations such that one is not restricted to the regime $m_q >> b^2 \Lambda_{QCD}^3$, but rather one is allowed $m_q \gtrsim b \L_{QCD}^2$~\cite{Frezzotti:2003ni,Aoki:2004ta,Sharpe:2004ny,Sharpe:2005rq}.} 
\begin{equation}
	\D m_\pi a_{\pi\pi}^{I=2,\pm2}(\pi/2) = 0\, .
\end{equation}
This is independent of the use of lattice-physical parameters, and holds also for the scattering length expressed in bare parameters, or any combination of bare and physical.  At zero twist, $\w=0$, our expressions reduce to those of Ref.~\cite{Buchoff:2008ve}.  Converting $f_\pi \rightarrow f$, our answer agrees with that in Ref.~\cite{Aoki:2008gy}.  The scattering length at maximal twist is simply given by the continuum formula
\begin{equation}
m_\pi a_{\pi\pi}^{I=2} = -2\pi \left(\frac{m_\pi}{4\pi f_\pi}\right)^2 \bigg\{
	1 + \left(\frac{m_\pi}{4\pi f_\pi}\right)^2 \left[ 
		3 \ln \left( \frac{m_\pi^2}{\mu^2} \right)
		-1 -l_{\pi\pi}^{I=2}(\mu) \right]
	\bigg\}\, ,
\end{equation}
where the combination of Gasser-Leutwyler coefficients is~\cite{Bijnens:1995yn,Bijnens:1997vq}
\begin{equation}
	l_{\pi\pi}^{I=2} = 4(4\pi)^2 (4l_1^r +4l_2^r + l_3^r - l_4^r )\, .
\end{equation}
Furthermore, the discretization errors only enter at tree level at this order (when the expression is expressed in lattice-physical parameters), and thus at arbitrary twist, the exponentially suppressed finite volume corrections to L\"{u}scher's method are also given by those determined in continuum finite volume $\chi$PT~\cite{Bedaque:2006yi}.

Returning to the new graphs arising from the three-pion interactions, we can estimate the size of the corrections to the scattering amplitude using the known mass splitting between the charged and neutral pions~\cite{Boucaud:2007uk,Boucaud:2008xu}.  Estimating the splitting with the leading correction, Eq.~\eqref{eq:mpi0}, and solving for $W^\prime$ from Eq.~\eqref{eq:Mprime0}, we can estimate the corrections to the $I=2, I_3=\pm2$ scattering length near maximal twist.  As a ratio to the LO prediction for the scattering length, one finds
\begin{equation}
\frac{\D m_\pi a_{\pi\pi}^{I=2,\pm2}(\w)}{m_\pi^2 / 8\pi f_\pi^2} \simeq
	\frac{\cot^2\w \left(\D M^\prime_0(\w) / m_\pi^2 \right)^2}{1 + \D M^\prime_0(\w) / m_\pi^2}\, .
\end{equation}
At the lightest mass point calculated in Refs.~\cite{Boucaud:2007uk,Boucaud:2008xu}, which corresponds to $m_\pi \simeq 300$~MeV, the pion mass splitting is
\begin{equation}\label{eq:pionSplitting}
	\frac{\D M^\prime_0(\w\sim \pi/2)}{m_\pi^2} \simeq - 0.33\, ,
\end{equation}
and therefore one must have $\cot\w \geq 0.3$ for this term to make more than a 1\% correction.  Therefore, for current twisted mass lattice calculations, corrections to the $I=2, I_3=\pm2$ scattering length (and other parameters) should be negligible provided higher order corrections are as small as expected.

\subsection{$I_3=0$ scattering channels\label{sec:I3=0}}
There are several features which make scattering in the $I_3=0$ channels more complicated than in the $I=2,I_3=\pm2$ channels, most of which stem from the fact that the twisted mass lattice action explicitly breaks the full $SU(2)$ symmetry down to $U(1)$, the conserved $I_3$ symmetry.  The first technical complication is not specific to twisted mass calculations, but is simply the need to compute quark disconnected diagrams.  The second complication stems from the mass splitting of the charged and neutral pions.  Generally, one determines the scattering phase shift for two particles with the L\"{u}scher method~\cite{Maiani:1990ca,Hamber:1983vu,Luscher:1986pf,Luscher:1990ux}, by determining the interaction energy
\begin{equation}
	\D E_{\pi\pi} = 2\sqrt{p^2 + m_\pi^2} - 2m_\pi\, .
\end{equation}
In the isospin limit, the $|2,0\rangle$ and $|0,0\rangle$ states (in the $|I,I_3\rangle$ basis) are given by
\begin{align}\label{eq:pipiStates}
|2,0\rangle &= 
	\frac{1}{\sqrt{6}}\left( |\pi^+ \pi^-\rangle + |\pi^- \pi^+ \rangle - 2|\pi^0\pi^0\rangle \right)\, ,
\nonumber\\
|0,0\rangle &= 
	\frac{1}{\sqrt{3}}\left( |\pi^+ \pi^-\rangle + |\pi^- \pi^+ \rangle + |\pi^0\pi^0\rangle \right)\, .
\end{align}
However, given the relatively large mass splitting in current twisted mass lattice calculations, Eq.~\eqref{eq:pionSplitting}, the propagating eigenstates will be arbitrarily shifted from the physical states, perhaps shifting nearly to the $\{|\pi^+ \pi^-\rangle, |\pi^- \pi^+\rangle, |\pi^0 \pi^0\rangle \}$ basis.  This would have to be disentangled numerically.  Even ignoring this issue, which we deem the most serious, and working with the continuum $\{|2,0\rangle, |0,0\rangle \}$ basis, there is a mixing of these states, which first appears at NLO as the second operator in Eq.~\eqref{eq:L4phi}.  Working with the states
\begin{equation}
| I,0 \rangle = 
	\begin{pmatrix}
	| 2, 0 \rangle \\
	| 0, 0 \rangle
	\end{pmatrix}
\end{equation}
one finds
\begin{equation}
\D \mc{T}^{2,0;0,0}_{\D\mc{L}_{4\phi}} = \frac{8\D M^\prime_0(\w)}{9f^2}
	\begin{pmatrix} 
		 4 & -\frac{7}{\sqrt{2}} \\
		-\frac{7}{\sqrt{2}} & 5
	\end{pmatrix}\, .
\end{equation}
Given the correction $\D M^\prime_0(\w)$, one sees this mixing is in fact maximal at maximal twist.  This is nominally a NLO effect, thus possibly leaving the states mostly unmixed.  However, a comparison of this term with the LO amplitude of the $I=2, I_3=\pm2$ scattering, one finds close to maximal twist
\begin{equation}
\frac{\D \mc{T}^{2,0;0,0}_{\D\mc{L}_{4\phi}} / (32\pi)}{-2\pi (m_\pi / 4\pi f_\pi)^2} 
	\simeq \frac{1}{m_\pi^2 / f_\pi^2}
		\begin{pmatrix} 
		 1.11 & -1.37 \\
		-1.37 & 1.38
		\end{pmatrix}\, .
\end{equation}
For $m_\pi / f_\pi =2$, all terms in this scattering matrix are approximately $1/3$ the size of the LO amplitude.  Since we now know that the physical NLO corrections to the $I=2$ scattering length for example, only provide a few percent deviation from the LO term~\cite{Beane:2005rj,Beane:2007xs,Chen:2005ab,Chen:2006wf}, we conclude that this NLO operator in fact provides a relatively large contribution to the scattering amplitude, and furthermore provides a large mixing term, and thus cannot be neglected.  This, combined with the problem we mentioned previously, means a coupled channel version of L\"{u}scher's method of determining the scattering parameters would be needed to explore the $I_3=0$ scattering channels with twisted mass lattice QCD.
\begin{figure}[t]
\center
\begin{tabular}{c}
\includegraphics[width=0.75 \columnwidth]{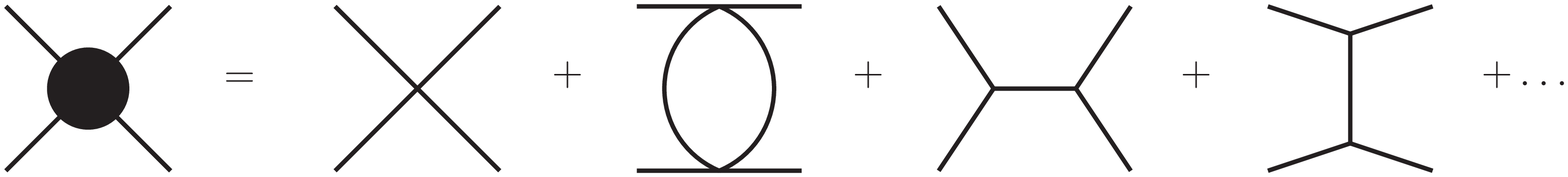}
\\
$(a)$ \\ \\
\includegraphics[width=0.75 \columnwidth]{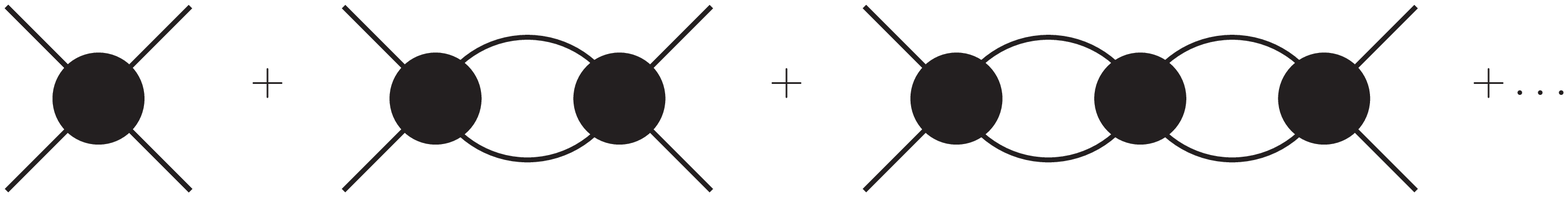}
\\
$(b)$
\end{tabular}
\caption{\label{fig:vertex} Modified four-point function, $(a)$ consisting of all off-shell graphs.  These vertices can then be iterated and summed $(b)$, to determine the $\pi\pi$ interactions.  This summation gives rise to the L\"{u}scher relation, valid below the inelastic threshold.}
\end{figure}

The $\D\mc{L}_{4\phi}$ Lagrangian is not the only source of mixing.  The three-pion interactions, depicted in Fig.~\ref{fig:3phi_graphs}, in the $s$, $t$ and $u$ channels, will also lead to a mixing of the $|2,0\rangle$ and $|0,0\rangle$ states, as one can check with an explicit calculation.  One may be concerned that the new $s$ channel graph will invalidate L\"{u}scher's method.  This is not the case however, as the internal pion propagator is always off-shell, and thus these diagrams do not contribute to the power-law volume dependence of the two-particle energy levels.  An alternative way to understand this is diagrammatically.  One can define a modified (momentum dependent) four-point function, which is order by order all the diagrams which do not go on-shell below the inelastic threshold.  We depict this modified vertex in Fig.~\ref{fig:vertex}$(a)$.  These $2PI$ diagrams can then be resummed to all orders to produce the scattering matrix, Fig.~\ref{fig:vertex}$(b)$.  It is this resummation that produces the L\"{u}scher relation, relating the finite volume scattering to the infinite volume scattering parameters~\cite{Maiani:1990ca,Hamber:1983vu,Luscher:1986pf,Luscher:1990ux}.  In this way, one can see that the new interactions will not lead to a modification of the structure of the L\"{u}scher relation.

Our last note of caution regards the construction of the interpolating fields.  In the physical basis, the states which become those of definite isospin in the continuum limit are given by Eq.~\eqref{eq:pipiStates}.
However, the interpolating fields are generally constructed with quark fields in the twisted basis, with a known definite \textit{twist} from the physical basis fields.  While this is also true of the $\pi^+\pi^+$ scattering channel, the phase is trivial since there is only one term contributing to the $|2,2\rangle$ state.  Thus if one were to undertake a calculation of these coupled scattering channels, care should be taken in constructing the correct interpolating fields.

\section{Conclusions}
In this report, we have detailed $\pi\pi$ interactions in twisted mass $\chi$PT.  We have shown that through NLO, at maximal twist the corrections to the $I=2, I_3=\pm2$ scattering parameters from discretization errors are identically zero.  However, near maximal twist there are corrections which can modify the expected chiral behavior which we demonstrated by an explicit calculation of the correction to the scattering length.  We found however, that for the dynamical twisted mass lattice configurations which exist today, the expected corrections are negligible.

The $I_3=0$ scattering channels proved to have more significant discretization corrections, most notably a mixing term between the $|2,0\rangle$ and $|0,0\rangle$ states which is relatively large.  In fact, these mixing terms combined with the need for computing quark disconnected diagrams and the expected nonperturbative shift of the twisted mass eigenstates, as discussed in Sec.~\ref{sec:I3=0}, may make a calculation of these $I_3=0$ scattering channels prohibitively complicated.


\renewcommand{\thechapter}{5}

\chapter{Meson-Baryon Scattering Parameters from Lattice QCD with an Isospin Chemical Potential}\label{Iso_Chem}

\section{Overview}

As stated in Chapter \ref{ch:TM_pi_pi}, the determination of hadronic interactions via lattice QCD has shown substantial progress in recent years.  The main focus on this chapter will involve meson-baryon scattering.  Several meson-bayon scattering channels are numerically feasible, and have been simulated in Ref.~\cite{Torok:2009dg}.    
The computation of a different class of hadronic interactions, 
however, 
is hindered by the existence of annihilation diagrams, 
which, 
using current lattice methods, 
require all-to-all propagators.  
While some progress has been made in the computation of these propagators~%
\cite{Peardon:2009gh}, 
the scaling to larger lattices is still prohibitively expensive for practical calculations of hadronic interactions. 
As a result, 
information on pion-nucleon scattering, 
and 
$\ol K N$ scattering, 
a process thought to be important for kaon condensation in neutron stars~%
\cite{Kaplan:1986yq,Nelson:1987dg,Brown:1987hj,Brown:1992ib,Brown:2007ara}, 
has remained elusive to a first principles lattice calculation.

In this work, 
we propose a novel method for extracting the low-energy constants for scattering processes with or without annihilation diagrams by employing an isospin chemical potential.  In terms of lattice calculations, 
the addition of an isospin chemical potential results in a positive definite determinant, 
which avoids the fermion sign problem that plagues other finite density calculations\footnote{For more information and examples, see~%
\cite{Stephanov:1996ki,Barbour:1997ej,Alford:1998sd,deForcrand:1999cy,Cox:1999nt}.}.  
An additional feature of an isospin chemical potential intrinsic to our analysis is the formation of a pion condensate when the chemical potential reaches a critical value~%
\cite{Son:2000xc,Son:2000by}.  
This feature has been observed in previous lattice calculations~%
\cite{Kogut:2002tm,Kogut:2002zg,Kogut:2002cm,Kogut:2003ju,Kogut:2004zg,deForcrand:2007uz}.  
In the presence of this dynamically generated pion condensate, 
propagating baryons have their masses shifted by an amount depending on the value of the condensate, 
chemical potential, 
and a linear combination of LECs, 
with a dependence that can be calculated in heavy baryon chiral perturbation theory.   
As a result, 
precision measurement of baryon masses for several values of the chemical potential coupled with the HB\CPT\ analysis gives an avenue for extracting scattering parameters, even for processes with annihilation diagrams.

The organization of this chapter is as follows.  
In Sec.~\ref{s:Iso}, 
the continuum QCD action with an isospin chemical potential is detailed along with the 
leading-order values for the condensates.    
In Sec.~\ref{s:baryonsSU2}, 
we present the two-flavor HB\CPT\ analysis of nucleon and hyperon masses in the presence of the pion condensate. 
Our results show the linear combinations of LECs that can be extracted from doing several measurements at different chemical potentials above the critical point.  
In Sec.~\ref{s:Strange}, 
the results of three-flavor HB\CPT\ are derived by matching conditions to the two-flavor results.

\section{Isospin Chemical Potential}       %
\label{s:Iso}                                              %

The primary focus of the next two section will be QCD with two light-quark flavors with a finite isospin chemical potential.  
Using a continuum Minkowski space action, 
the matter part of the QCD Lagrangian is 
\begin{equation} \label{eq:L}
\cL 
=
\ol \psi 
\left[ 
i 
\left(
\Dslash 
+
i
\mu_I
\gamma_0
\frac{\tau^3}{2}
\right)
-
m_q
+
i 
\epsilon
\gamma_5
\frac{\tau^2}{2}
\right]
\psi
,\end{equation}
where 
$\Dslash \,$ 
is the 
$SU(3)_c$ 
gauge covariant derivative,  
$m_q = \diag (m, m)$
is the quark mass matrix in the isospin limit, 
and 
$\psi = ( u, d )^T$ an isodoublet of quarks. 
The isospin chemical potential, 
$\mu_I$,
appears as the time-component of a uniform gauge field. 
The term proportional to 
$\epsilon$
is an explicit isospin breaking term. 
It is included here because the spontaneous  breaking of isospin is an essential part of our analysis. 
Because symmetries do not break spontaneously at finite volumes, 
a small explicit breaking is necessary.  
This explicit breaking should go to zero as the size of the lattice goes to infinity.%
\footnote{
In Euclidean space, 
the action can be written in the form 
$S_E = \ol \psi ( \cD + m_q ) \psi$,
where
\begin{equation}
\cD = \Dslash + \mu_I \gamma_4 \frac{\tau^3}{2} + i \epsilon \gamma_5 \frac{\tau^2}{2} 
.\end{equation} 
The Dirac operator 
$\cD$ 
satisfies the condition
$\tau^1 \gamma_5 \, \cD \, \tau^1 \gamma_5 = \cD^\dagger$, 
which ensures that the product of $u$ and $d$ fermionic determinants is positive.
}

For vanishing quark masses, 
$m=0$,
and vanishing external sources,
$\mu_I = \epsilon = 0$, 
the QCD action has an 
$SU(2)_L \otimes SU(2)_R$
symmetry that is spontaneously 
broken down to 
$SU(2)_V$. 
For non-vanishing chemical potential, 
the QCD action has only a
$U(1)_L \otimes U(1)_R$
symmetry associated with $I_3$. 
This will be broken down to
$U(1)_V$
by both spontaneous chiral symmetry breaking 
and the explicit symmetry breaking introduced by 
the quark mass. 
At finite 
$\mu_I$ 
and finite
$\epsilon$, 
the only continuous symmetry of 
$\cL$
is the 
$U(1)_B$
associated with baryon number.
\footnote{
With both 
$\mu_I$ 
and 
$\epsilon$
non-vanishing there is no symmetry that prevents their renormalization. 
These parameters are only multiplicatively renormalized because, for example, 
the isospin chemical potential is the only term in the action that breaks 
$C \otimes \tau^3$, 
and the explicit isospin breaking term is the only term in the action that breaks parity.   
Furthermore
because we are interested in the limit of small 
$\epsilon$,
the action will be approximately symmetric under the 
$U(1)_V$ 
associated with $\tau^3$. 
Consequently 
$\mu_I$ 
cannot be appreciably renormalized. 
}
We can write down an effective field theory of QCD
that takes into account this pattern of spontaneous and explicit symmetry 
breaking. 
This effective theory is \CPT, 
and at leading order has the form
\begin{equation} \label{eq:CPT}
\cL
=
\frac{f^2}{8}
\left[
<
D_\mu U D^\mu U^\dagger
>
+
2 \lambda 
<
M^\dagger U + U^\dagger M
>
\right]
.\end{equation}
Here the angled brackets denote flavor traces, 
and the external sources have been included in the terms
\begin{eqnarray}
M &=& s + i p ,  
\notag
\\
D_\mu U 
&=&
\partial_\mu U
+ i [ \mathbb{V}_\mu, U]
,\end{eqnarray}
where the external vector potential is given by: 
$\mathbb{V}_\mu = \mu_I \frac{\tau^3}{2} \delta_{\mu,0}$,  
the pseusdoscalar source is given by:
$p = \epsilon \frac{\tau^2}{2}$,
and the scalar source is just the quark mass,
$s = m_q$.

The isospin chemical potential favors the presence of up quarks as opposed to down quarks. When it is larger than about 
$m_\pi$, 
it becomes energetically favorable for the ground state to contain positive pions. 
Assuming a uniform value $U_0$ 
for the vacuum expectation value of $U(x)$, 
we can minimize the effective potential to determine
the vacuum (ground) state. 
Using the standard parametrization of 
$SU(2)$, 
namely
$U_0 = \exp [ i \alpha \, \bm{n} \cdot \bm{\tau} ]$, 
with $\bm{n} \cdot \bm{n} = 1$,
we find
\begin{equation}
U_0
=
\begin{cases}
1, & | \mu_I | < m_\pi, 
\\
\exp [ i \alpha \, \tau^2 ], &
|\mu_I| > m_\pi
\end{cases}
.\end{equation}
The angle 
$\a$
is determined, at lowest order in the chiral expansion, by the transcendental equation
\begin{equation} \label{eq:alpha}
\cos \alpha 
= 
\frac{m_\pi^2}{\mu_I^2}
- 
\frac{\lambda \epsilon}{\mu_I^2} \cot \a
,\end{equation}
where the pion mass satisfies the Gell-Mann--Oakes--Renner relation,
$m_\pi^2 = 2 \lambda m$.
For small values of 
$\epsilon$,
one can determine 
$\a$
using a perturbative expansion of Eq.~\eqref{eq:alpha}.
This expansion, however, breaks down as one nears 
the critical value of the chemical potential from above. 
In the condensed phase, 
the condensates have values
\begin{eqnarray}
\langle \ol \psi \psi \rangle 
&=&
f^2 \lambda \cos \a,
\notag
\\
i \langle \ol \psi \tau^2 \gamma_5 \psi \rangle
&=&
f^2 \lambda \sin \a
.\end{eqnarray}

\section{Baryons in an isospin chemical potential: $SU(2)$}  %
\label{s:baryonsSU2}                                                               %

When a hadron propagates in the background of a pion condensate its properties are changed due to the interactions of the hadron with the pions in the condensate. Their mass, for instance, is shifted by an amount closely related to the forward scattering amplitude of pions on the hadron. Unfortunately, the pions in the condensate are not exactly on-shell and no model independent relation can be found between pion-hadron scattering amplitudes and the mass shift. However, the chiral expansion of these two quantities are related in the way described in this section.

\subsection{Nucleons}      %

We can address effects of the isospin chemical potential 
on nucleons by using the heavy nucleon chiral Lagrangian~\cite{Bernard:1992qa}. 
This effective theory accounts for spontaneous and explicit 
chiral symmetry breaking in the nucleon sector. The theory is 
written in terms of static nucleon fields, and thereby the 
nucleon mass can be treated in accordance with the chiral symmetry
breaking scale.

The leading-order heavy nucleon chiral Lagrangian is written 
in terms of the nucleon doublet,
$N = (p, n)^T$,
and has the form
\begin{eqnarray} \label{eq:nuke}
\cL^{(1)}
&=&
N^\dagger \left( i v \cdot D + 2 g_A S \cdot A \right) N
.\end{eqnarray}
The vector $v_\mu$ is the nucleon four-velocity, 
while the vector, 
$V_\mu$,  
and axial-vector,
$A_\mu$,
fields of pions
are defined by 
\begin{eqnarray}
V_\mu 
&=&
\frac{1}{2} 
\left(  
\xi^\dagger_L D^L_\mu \xi_L
+
\xi_R D^R_\mu \xi_R^\dagger
\right)
\notag
,\\
A_\mu
&=& 
\frac{i}{2} 
\left(  
\xi^\dagger_L D^L_\mu \xi_L
-
\xi_R D^R_\mu \xi_R^\dagger
\right)
.\end{eqnarray}
The former appears in the chirally covariant derivative
$D_\mu$, 
which has the following action on the nucleon field
\begin{equation}
(D_\mu N)_i
=
\partial_\mu N_i
+ 
(V_\mu)_{i} {}^j N_j
.\end{equation} 
Appearing in the vector and axial-vector pion fields are 
$\xi_L$, 
and 
$\xi_R$  
which arise from the definition
$U = \xi_L \xi_R$. 
Under a chiral transformation,
$(L,R) \in SU(2)_L \otimes SU(2)_R$, 
the pion fields transform as:
$\xi_L(x) \to L \, \xi_L(x) \mathcal{U}(x)$,
and
$\xi_R(x) \to \mathcal{U}^\dagger(x) \xi_R(x) R^\dagger$, 
where $\mathcal{U}(x)$ is the matrix entering the transformation of the nucleon field, 
$N_i \to \mathcal{U}(x)_i {}^j N_j$. 
The left-handed and right-handed derivatives act according to the rules
\begin{eqnarray}
D^L_\mu \xi_L
&=&
\partial_\mu \xi_L
+ i \mathbb{L}_\mu \xi_L
\notag
,\\
D^R_\mu \xi_R
&=&
\partial_\mu \xi_R
-
i \xi_R \mathbb{R}_\mu 
.\end{eqnarray}
For the case of an external vector field, 
the left- and right-handed vector sources coincide:
$\mathbb{L}_\mu = \mathbb{R}_\mu = \mathbb{V}_\mu$. 
In an isospin chemical potential, 
the spurions are given the final values
\begin{eqnarray}
\xi_L(x) &=& \xi_0 \,  \xi(x),
\notag
\\
\xi_R(x) &=& \xi(x) \xi_0,
\end{eqnarray}
so that the vacuum value is 
$U_0 = \xi_0^2$.

At  leading-order using Lagrangian Eq.~\eqref{eq:nuke}, 
the shift in nucleon mass due to the chemical potential is the trivial result
\begin{equation}
M_N 
= 
M_{N}^{(0)}  -  \mu_I \cos \a \frac{\tau^3}{2}
,\end{equation}
where 
$M_N^{(0)}$ 
is the chiral limit value. 
Beyond leading-order, 
the nucleon mass receives corrections that depend on low-energy constants. 
These constants are the coefficients of terms in the second-order nucleon chiral Lagnrangian. 
Including relativistic corrections with fixed coefficients, 
the complete
\footnote{
When one considers external sources with non-vanishing field strength tensors, 
additional terms are present. 
The external sources we consider, 
however, 
are uniform in spacetime. 
} 
 second-order Lagrangian is 
\begin{eqnarray} 
\cL^{(2)}
&=&
N^\dagger 
\left[
- 
\frac{D_\perp^2}{2 M_{N}^{(0)}} 
+
\frac{1}{2 M}
[S^\mu, S^\nu] \, [ D_\mu, D_\nu]
- 
\frac{i g_A}{M_N^{(0)}}
\Big\{
v \cdot A, S \cdot D
\Big\}
+
4 
c_1 
< \cM_+ >
\right.
\notag \\
&& \phantom{s}
\left.
+
4 
\left(c_2
- 
\frac{g_A^2}{8 M_N^{(0)}}
\right)
(v \cdot A)^2
+
4 c_3 A^2
+ 
4 c_4 [S^\mu, S^\nu] A_\mu A_\nu
+
4 c_5 \tilde{\cM}_+
\right] N,\nn\\
\label{eq:secondnuke}
\end{eqnarray}
where 
$\tilde{\cM}_+ = \cM_+ - \frac{1}{2}  < \cM_+ >$, 
and 
$\cM_+ = \frac{1}{2} \lambda (\xi^\dagger_L M \xi^\dagger_R + \xi_R M^\dagger \xi_L)$. 
The last term with coefficient $c_5$ only contributes in the presence of strong isospin breaking.

Utilizing the second-order nucleon Lagrangian, 
we arrive at the nucleon mass
\begin{equation}
M_N 
=
M_N^{(0)} 
- 
\mu_I \cos \a \frac{\tau^3}{2}
+
4 c_1 \left( m_\pi^2 \cos \a + \lambda \epsilon \sin \a \right)
+
\left( c_2 - \frac{g_A^2}{8 M}  + c_3 \right) \mu_I^2 \sin^2 \a
\end{equation}
The second-order correction is entirely isoscalar,
and allows access to the low-energy constant
$c_1$, 
and 
the combination 
$c_2 + c_3$. 
One can vary the quark mass and isospin chemical potential 
to isolate these coefficients from the observed behavior of the nucleon mass in lattice QCD.
\footnote{
One can go further and impose isospin twisted boundary conditions 
on the quark fields, 
e.g.~$\psi(x + L \hat{z}) = \exp [ i \theta \tau^3 ] \psi(x)$.
This has the advantage of introducing non-vanishing spatial components of the vacuum value of 
$A_\mu$,
and leads to the ability to isolate more low-energy constants.
For the example mentioned, 
the 
$c_3$ 
coefficient can be determined from the mixing angle between nucleons. 
}

\subsection{Hyperons}                         %
\label{s:hyper}                                      %

Recently there has been interest in treating hyperons using 
$SU(2)$ \CPT, effectively integrating out the kaon interactions and grouping interactions by strangeness ~\cite{Beane:2003yx,Frink:2002ht,Tiburzi:2008bk,Jiang:2009sf,Mai:2009ce,Tiburzi:2009cf}. 
We consider hyperons in an isospin chemical potential, 
and determine which low-energy constants of $SU(2)$ 
\CPT\ 
can be determined from their masses.
This is a natural approach because the isospin chemical potential 
produces only pion-hyperon interactions.

In the stangeness $S=1$ sector are the $\L$ and $\S$ baryons. 
The 
$\L$ 
baryon is an isosinglet, 
while the 
$\S$ 
baryons form an isotriplet. 
The leading-order Lagrangian in the $S=1$ sector has the form
\begin{equation}
\cL^{(1)}
=
\Lambda^\dagger i v \cdot \partial \Lambda 
+
< 
\Sigma^\dagger \left( i v \cdot D - \D_{\L\S}  \right) \Sigma 
>
.\end{equation}
Here we have packaged the $\S$ fields into an adjoint matrix
\begin{equation}
\S = 
\begin{pmatrix}
\frac{1}{\sqrt{2}} \Sigma^0 & \Sigma^+ \\
\Sigma^- & - \frac{1}{\sqrt{2}} \Sigma^0
\end{pmatrix}
,\end{equation}
and use angled brackets to denote flavor traces. 
The covariant derivative acts as 
$D_\mu \Sigma = \partial_\mu \S + [ V_\mu, \S]$. 
The parameter 
$\D_{\L\S}$ 
is the 
$SU(2)$ 
chiral limit value of the mass splitting,
$\D_{\L\S} = M^{(0)}_\S - M^{(0)}_\L$. 
In writing the leading-order Lagrangian, 
we have omitted the axial couplings to pions 
as these terms are not relevant for our computation. 
Similarly
we write down only the contributing terms from the 
second-order 
$S=1$ 
Lagrangian
\begin{eqnarray}
\cL 
&=&
\L^\dagger \L  
\left[
4 c_1^\L
 < \cM_+ >
+
2 c_2^\L
< v \cdot A^2  >
+ 
2 c_3^\L
<  A^2 >
\right]
\notag \\
&&
+ 
< 
\Sigma^\dagger \Sigma
>
\left[
4 c_1^\S
< \cM_+ >
+
2
c_2^\S
< v \cdot A^2 >
+
2
c_3^\S
< A^2 >
\right]
\notag \\
&& +
4 c_6
< \S^\dagger  v \cdot A> <  v \cdot A \, \S >
+
4 c_7 
< \S^\dagger A_\mu > < A^\mu \S >
.\end{eqnarray}

Using the first- and second-order Lagrangians, 
we find the mass of the 
$\L$ 
baryon
\begin{equation}
M_\L 
= 
M_\L^{(0)}
+
4 c_1^\L 
\left( 
m_\pi^2 \cos \a 
+ 
\lambda \epsilon \sin \a 
\right)
+
(c_2^\L + c_3^\L)
\mu_I^2 \sin^2 \a
,\end{equation}
and see that the low-energy constant 
$c_1^\L$, 
and the combination 
$c_2^\L + c_3^\L$
can be accessed. 
For the 
$\S^0$ 
baryon,
we find the mass
\begin{eqnarray}
M_{\S^0}
&=&
M_{\S}^{(0)}
+
4 c_1^\S 
\left( 
m_\pi^2 \cos \a 
+ 
\lambda \epsilon \sin \a 
\right)
+
(c_2^\S + c_3^\S) 
\mu_I^2 \sin^2 \a
.\end{eqnarray}
At this order in the chiral expansion, 
the charged 
$\S$ 
baryons mix. 
We find the mass eigenvalues
\begin{eqnarray}
M_\pm
&=&
M_{\S^0}
+
( c_6^\S + c_7^\S)
\mu_I^2 \sin^2 \a
\pm 
\mu_I
\sqrt{
\cos^2 \a
+
(c_6^\S + c_7^\S)^2 
\mu_I^2 \sin^4 \a
}
.\end{eqnarray}
The corresponding eigenstates are
\begin{eqnarray}
| + \rangle
&=& 
\cos \frac{\Theta}{2} 
| \S^+ \rangle
-
\sin \frac{\Theta}{2} \,
| \S^- \rangle
\notag
\\
| - \rangle
&=&
\sin \frac{\Theta}{2} \,
| \S^+ \rangle
-
\cos \frac{\Theta}{2} \,
| \S^- \rangle 
,\end{eqnarray}
where the mixing angle can be found from
\begin{equation}
\cos \Theta
= 
- \frac{\cos \a}
{\sqrt{ 
\cos^2 \a
+
(c_6^\S + c_7^\S)^2 
\mu_I^2 \sin^4 \a
}}
.\end{equation}
Finally because the 
$c_j^\S$ 
coefficients should scale with the inverse of the chiral symmetry breaking scale, 
the mixing between 
$\S$ 
baryons should be small for moderately sized 
$\mu_I$.

In the strangeness $S=2$ sector are the cascades. 
The case of the cascade is similar to the nucleon
as the cascades transform as an isospin doublet, 
$\Xi = ( \Xi^0, \Xi^- )^T$. 
Consequently the first- and second-order
Lagrangians have the same form as those for the 
nucleon, Eqs.~\eqref{eq:nuke} and \eqref{eq:secondnuke}.
One merely replaces the chiral limit mass of the nucleon, 
$M_N^{(0)}$,
with the chiral limit mass of the cascade, 
$M_{\X}^{(0)}$, 
the nucleon axial charge, 
$g_A$
with the cascade axial charge, 
$g_{\X\X}$. 
Finally each of the low-energy constants
$c_j$ 
have different values for the cascades;
these we denote by $c^{\X}_j$.
The cascade masses are thus given by
\begin{equation}
M_\X
=
M^{(0)}_\X 
- 
\mu_I \cos \a \frac{\tau^3}{2}
+
4 c^\X_1 \left( m_\pi^2 \cos \a + \lambda \epsilon \sin \a \right)
+
\left( c^\X_2 - \frac{g_{\X\X}^2}{8 M_\X^{(0)}}  + c^\X_3 \right) \mu_I^2 \sin^2 \a
.\end{equation}
Varying the quark mass and isospin chemical potential allows one 
to access 
$c_1^\X$, 
and the combination 
$c_2^\X + c_3^\X$.

\section{Strange Quark and $SU(3)$}                              %
\label{s:Strange}                                                               %

We can repeat the above analysis including a third light quark.
The isospin chemical potential is added proportional to the 
$\l^3$ Gell-Mann matrix, 
while the explicit isospin breaking term we choose in the direction of 
$\l^2$. 
For a massless strange quark, 
the QCD action with 
$\mu_I$ 
and 
$\epsilon$ 
both non-vanishing possesses a 
$U(1)_L \otimes U(1)_R$ 
symmetry associated with strangeness. 
The strange quark mass breaks this symmetry 
down to 
$U(1)_V$. 
This symmetry precludes tilting the vacuum value of 
$U$
in any strange direction. 
In 
$SU(3)$,  
we must have
\begin{equation}
U_0 
= 
\begin{pmatrix}
e^{ i \a \tau^2 } & 0 \\
0 & 1
\end{pmatrix} 
.\end{equation}
Consequently the non-vanishing values of the condensates are 
\begin{eqnarray}
\frac{1}{3} \langle \ol \psi (1 - \sqrt{3} \l^8) \psi \rangle
&=&
\lambda f^2,
\notag
\\
\frac{1}{3} \langle \ol \psi (2 + \sqrt{3} \l^8 ) \psi \rangle 
&=& 
\lambda f^2 \cos \a
\notag
,\\
i \langle \ol \psi \l^2 \gamma_5 \psi \rangle 
&=& 
\l f^2 \sin \a
.\end{eqnarray}

It is now straightforward to determine the effects of the isospin chemical potential on the octet baryons. 
The octet baryons can be packaged into an $SU(3)$ adjoint matrix 
$B$, 
given by
\begin{equation}
B
= 
\begin{pmatrix}
\frac{1}{\sqrt{2}} \S^0 + \frac{1}{\sqrt{6}} \L & \S^+ & p \\
\S^- & - \frac{1}{\sqrt{2}} \S^0 + \frac{1}{\sqrt{6}} \L & n \\
\X^- & \X^0 & - \frac{2}{\sqrt{6}} \L
\end{pmatrix}
.\end{equation}
The relevant term of the leading-order 
$SU(3)$
Lagrangian is
\begin{equation}
\cL^{(1)} 
=
< B^\dagger [ i v \cdot D, B] > 
,\end{equation}
with 
$D_\mu B = \partial_\mu B + [ V_\mu, B]$. 
The kinetic term reproduces the 
isospin-dependent mass shifts 
for the octet baryons found above.

The contributing terms of the second-order 
$SU(3)$ 
chiral Lagrangian are
\begin{eqnarray}
\cL^{(2)}
&=&
2 b_F
< B^\dagger [\cM_+ , B] >
+
2 b_D
< B^\dagger \{ \cM_+, B \} >
+ 
2 b_0
<B^\dagger B> < \cM_+>
\notag \\
&&+
b_1 
< B^\dagger [ A_\mu, [ A^\mu, B] ] >
+
b_2
< B^\dagger [ A_\mu, \{ A^\mu, B \} ] >
+ 
b_3
< B^\dagger \{ A_\mu, \{ A^\mu, B \} \} >
\notag \\
&&+
b_4
< B^\dagger [ v \cdot A, [ v \cdot A, B ] ] > 
+ 
b_5
< B^\dagger [ v \cdot A, \{ v \cdot A, B \} ] >
\notag \\
&&+
b_6
< B^\dagger \{ v \cdot A, \{ v \cdot A, B \} \} >
+ 
b_7
< B^\dagger B > < A^2 > 
+ 
b_8 
< B^\dagger B> < v \cdot A^2 >
.
\notag \\
\end{eqnarray}
Evaluating the masses of the octet baryons, 
we arrive at results identical to the 
$SU(2)$
case but with  
$SU(3)$ 
relations between parameters.
The matching of parameters for the baryon masses 
have been given in~\cite{Tiburzi:2008bk}, 
while those for pion-baryon 
scattering have been given in~\cite{Mai:2009ce}. 
Terms proportional to the strange quark mass
give rise to shifts of the 
$SU(3)$ chiral limit value of the baryon mass.
Such terms are the leading contributions in 
$SU(3)$
to the 
$SU(2)$ 
chiral limit values of the baryon masses employed above. 
As our interest is with the 
$\mu_I$ 
dependence, 
we do not duplicate the matching relations between 
the chiral limit baryon masses here.
Matching of 
$\cM_+$ 
operators in the upper 
$2 \times 2$ 
block yields the relations~\cite{Tiburzi:2008bk}
\begin{eqnarray}
4 c_1 &=& 
b_D + b_F + 2 b_0,
\notag
\\
4 c_1^\L &=& 
\frac{2}{3} b_D + 2 b_0,
\notag
\\
4 c_1^\S &=&
2  b_D + 2 b_0,
\notag
\\
4 c_1^\X &=&
b_D - b_F + 2 b_0, 
.\end{eqnarray}
Matching of operators involving the product of two axial-vector pion fields gives the relations~\cite{Mai:2009ce}
\begin{eqnarray}
4(c_2 + c_3)
&=&
(b_1 + b_4) +  (b_2  + b_5) + ( b_3 + b_6)  + 2 ( b_7 + b_8)
\notag
,\\
4 (c_2^\L + c_3^\L)
&=&
\frac{4}{3} ( b_3 +  b_6 ) +   2 ( b_7 + b_8)
\notag
,\\
4( c_2^\S + c_3^\S )
&=&
4 ( b_1 + b_4)  + 2 ( b_7 + b_8 )
\notag
,\\
4( c_6^\S + c_7^\S )
&=&
- 2 ( b_1 + b_4)  + 2  ( b_3  + b_6)
\notag
,\\
4( c_2^\X + c_3^\X )
&=&
( b_1  + b_4) -  ( b_2 + b_5)  +  (b_3 + b_6)  + 2 (b_7 + b_8)
.\end{eqnarray}
From these matching conditions, 
we see that all three of the LECs for the quark mass dependent
operators can be determined, 
along with four combinations 
of the eight double axial-insertion operators. 
The overdetermination of these parameters, 
moreover, 
serves as a useful check on 
$SU(3)$ 
symmetry.

\section{Discussion}                                                         %
\label{s:End}                                                                    %

Above, 
we show what aspects of meson-baryon scattering can be determined
from simulating lattice QCD with an isospin chemical potential.
Our approach allows for the determination of the relevant LECs from first principles; 
and,
specifically for pion-nucleon processes,
avoids the need to evaluate all-to-all propagators. 
Several attempts at extracting  the non-derivative pion-nucleon couplings from analysis of experimental data have been attempted in the past. 
In \cite{Rentmeester:2003mf}, 
nucleon-nucleon scattering data was used to determine the pion-nucleon constants through their effect on two-pion exchange contributions to the nuclear force. Pion-nucleon scattering data has also been used for this purpose, for instance \cite{Fettes:1998ud}. 
Significant uncertainties and discrepancies between different extractions remain. 
Our method can help resolve this situation.   
The values of the antikaon-nucleon couplings are considerably more uncertain, 
although attempts of extracting them from experimental data have been made, 
see, for example,~\cite{Oller:2000fj,Meissner:2004jr,Borasoy:2005ie,Oller:2005ig}.
Our framework can shed light on this issue by testing 
$SU(3)$ 
symmetry predictions.

Of particular interest is the coefficient 
$c_1$, 
related to the nucleon 
$\sigma$-term, 
which is important for addressing dark matter interactions with nuclei, 
and has been the subject of numerous studies.
The value estimated in \cite{Gasser:1990ce} has been challenged~
\cite{Buettiker:1999ap}
. 
The pion-nucleon constants also affect the dependence of the nucleon mass on the quark mass at fourth order in the chiral expansion,  
and this dependence has been used to extract them from lattice QCD calculations at different quark masses (for a recent extraction, see \cite{Ohki:2009mt}). 
These extractions are complicated, however, by the fact that the quark mass dependence of nucleon mass is barely compatible with chiral perturbation theory predictions at the current values of quark masses \cite{WalkerLoud:2008bp,WalkerLoud:2008pj}. 
The method proposed here adds another tool that is somewhat complementary to the quark mass variation of the nucleon mass. In fact, in order to extract the pion-nucleon couplings one needs to know the nucleon mass for several values of the quark mass small enough so that chiral perturbation theory is trustable. But, apparently, this range might be small \cite{WalkerLoud:2008pj}. The dependence of the nucleon mass on the isospin chemical potential adds another handle that can be used for this purpose. The computational costs of changing the quark mass or the chemical potential are comparable.  Of course, gauge configurations at different quark masses are also useful for other purposes besides nucleon mass extraction. The dependence on $\mu_I$, moreover, can also help in the extraction of other observables, making gauge configurations with $\mu_I$ more generally useful, and our proposal a realistic option. 
\renewcommand{\thechapter}{6}

\chapter{Search for Chiral Fermion Actions on Non-Orthogonal Lattices}\label{ch:Non_orth}

\section{Overview}
One obstacle in lattice QCD is fermion doubling, where the continuum limit describes multiple fermions despite only one fermion being attached to each node.  While more complicated discretizations can reduce the amount of doubling, this often comes at the expense of exact chiral invariance as shown in the Nielsen-Ninomiya ``no-go'' theorem~\cite{Nielsen:1980rz,Nielsen:1981xu,Nielsen:1981hk}.   In the past, attempts were made by Karsten~\cite{Karsten:1981gd} and Wilczek~\cite{Wilczek:1987kw} to minimize the doubling (only two fermions) allowed by the Nielsen-Ninomiya ``no-go'' theorem while preserving chiral symmetry.  However, these actions broke additional symmetries leading to the generation of non-physical dimension three and four operators in the continuum limit.

Recently, Creutz~\cite{Creutz:2007af} and, shortly after, Bori\c{c}i~\cite{Borici:2007kz} introduced a non-orthogonal graphene-inspired action in four-dimensions in order to achieve both minimal doubling and chiral symmetry.  Unfortunately, as shown in Ref.~\cite{Bedaque:2008xs}, this new action suffers from similar symmetry breaking issues that plagued the actions proposed by Karsten and Wilczek.   However, an intriguing realization upon studying the Bori\c{c}i-Creutz action is that if a four-dimension action contains the minimum $\mathbb{Z}_5$ permutation symmetry, the generation of the divergent dimension 3 operators could be prevented in the continuum limit.  As a result, Ref.~\cite{Bedaque:2008jm}, explores several non-orthogonal lattices in order to achieve this symmetry, but these lattices also lead to other undesireable features.  These actions will be discussed throughout this note along with the Bori\c{c}i-Creutz action.

\section{Graphene}
Graphene is a single layer of graphite, where the carbon atoms form a two-dimensional hexagonal (or honeycomb) lattice.  The purpose of this section is to describe some of the most basic results of hexagonal lattices that have attracted a great deal of attention throughout the condenced matter community. Hexagonal lattices have the intriguing property that massless fermions on the sites lead to exact Dirac fermions.

\begin{figure}[ht]
\center
\begin{tabular}{cc}
\includegraphics[width=0.5\columnwidth]{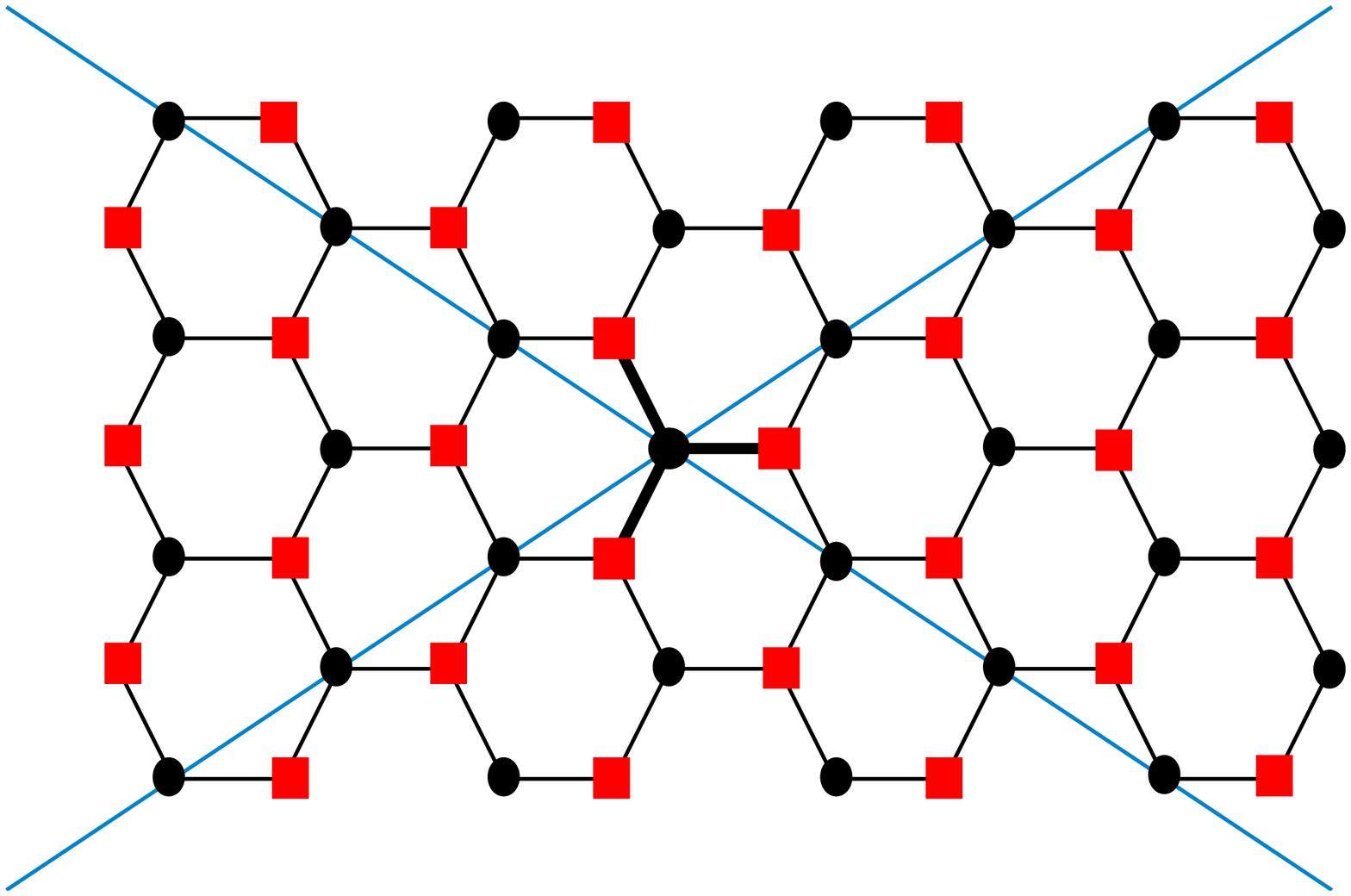}
\includegraphics[width=0.5\columnwidth]{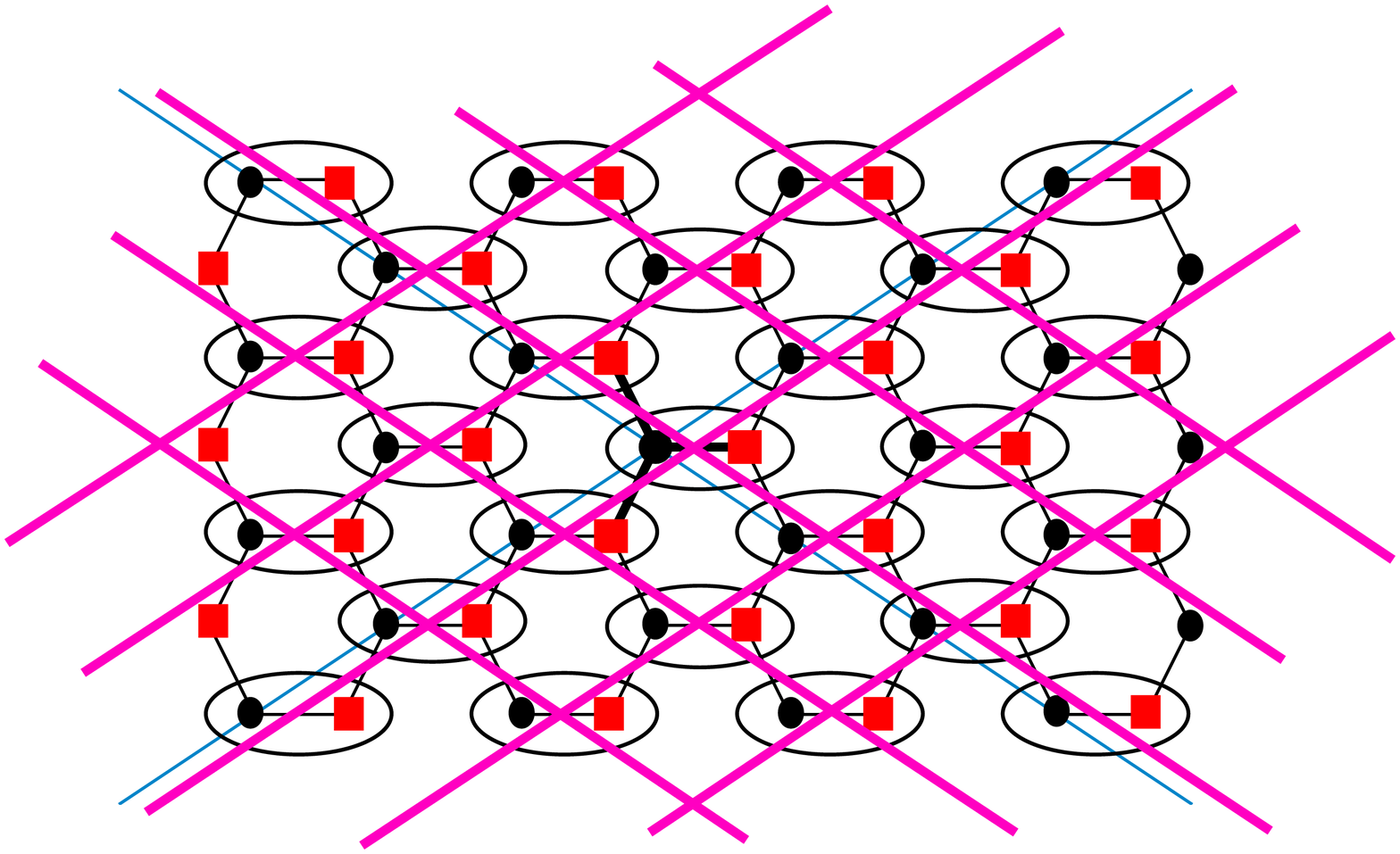}
\end{tabular}
\caption{(left) Graphene lattice composed of two sub-lattices.  The ``a-site'' lattice is represented by black circles and the ``b-site'' lattice is represented by red squares. (right) Illustrates the proposal by Creutz \cite{Creutz:2007af} to group horizontal neighbors into two-atom sites, given by the ellipses.  With these sites, a grid can be constructed, which is given by the pink lines.}
\label{Grid}
\end{figure}
These lattices are better analyzed by a two sub-lattice description.  As shown in Fig.~\ref{Grid}, one sub-lattice (often referred to as the ``a-sites'' or ``l-sites'') only communicates with its three nearest neighbors which are members of the other sub-lattice (often referred to as the ``b-sites'' or ``r-sites'').  Another important feature is that these three nearest neighbors possess a $\mathbb{Z}_3$ permutation symmetry, which means that these lattices are invariant under $120^\circ$ rotations.  As a result, the addition of the three primary vectors, $\ee^\alpha$, created by the three connections to the neareast neighbors yields zero ($\sum_\alpha \ee^\alpha = 0$).  Without going into detail, these features of a $1+1$ dimensional graphene lattice, along with two component Dirac spinors,  result in chiral Dirac fermions in the massless limit.  The question that is of interest to this chapter is how to extend this construction to four-dimensions.

\section{Graphene-Inspired Lattice Actions}

Several years ago, Creutz in Ref.~\cite{Creutz:2007af} proposed a viable method of extending this construction to four dimensional lattice gauge theory.  The main idea behind this method was to combine the horizontally connected ``a'' and ``b'' sites into an individual two-atom ``site'' (as shown in Fig.~\ref{Grid} as the ellipses around these two-atom sites).  Through this process, the connecting segments between these two atom sides form a non-orthogonal grid, which is illustrated by the pink lines imposed on the lattice in Fig.~\ref{Grid}.  With this formulation,  Creutz generalized this non-orthogonal grid to four-dimensions while maintaining a construction similar to graphene.  

 \section{Bori\c{c}i-Creutz Action}
The action of this graphene-inspired, non-orthogonal grid in four dimensions proposed by Creutz \cite{Creutz:2007af}, and later Bori\c{c}i \cite{Borici:2007kz} is given by 
\begin{eqnarray} 
S_{BC} 
&=& 
\frac{1}{2}\sum_{x}
\Bigg[
\sum_{\mu=1}^4
\left(\ol \psi_{x- \mu}
\, \ee^\mu \cdot \Gamma \,
\psi_{x}
-
\ol \psi_{x+\mu}
\,
\ee^\mu \cdot \Gamma^\dagger
\,
\psi_{x}
\right)
+
\ol \psi_x
\,
\ee^5 \cdot \Gamma
\,
\psi_x-\ol \psi_x
\,
\ee^5 \cdot \Gamma^\dagger
\,
\psi_x
\Bigg],\nn\\
\label{eq:CreutzZ5}
\end{eqnarray}
where 
$\Gamma_\mu = (\vec{\gamma}, i \gamma_4)$ and the four-vectors $\ee^\alpha$ are defined
in terms of two parameters $B$ and $C$ as
\bea\label{eq:es2}
\ee^1 &=& \phantom{-}(\phantom{+}1,\phantom{+}1,\phantom{+}1,\phantom{4}B\ ),\nn\quad 
\ee^2 = \phantom{-}(\phantom{+}1,-1,-1,\phantom{4}B\ ),\nn\quad
\ee^3 = \phantom{-}(-1,-1,\phantom{+}1,\phantom{4}B\ ),\nn\\
\ee^4 &=& \phantom{-}(-1,\phantom{+}1,-1,\phantom{4}B\ ),\nn\quad
\ee^5 = -(\phantom{+}0,\phantom{+}0,\phantom{+}0,4BC).
\eea 
Note, each $\ee^\alpha$ above is a four vector and the $\alpha$ does not refer to the individual components of this vector.  These $\ee^\alpha$ vectors are generated from this non-orthogonal grid\footnote{See Ref.~\cite{Bedaque:2008jm} for more details.}, which are related to the diagonal pink lines in Fig.~\ref{Grid}.  While written in a way that is similar to the na\"ive fermion lattice action, there are several key differences in this action.  First, the definition of $\Gamma_\mu$ contains an $i \gamma_4$, as opposed to the na\"ive fermion action that does not contain this factor of $i$.  This leads to an important sign difference in the fourth component of these vectors.  Second, these $\ee^\alpha$ vectors contain the non-orthogonal structure of this lattice action, which lead to non-trivial linear combinations of Dirac matricies.    To understand the behavior of this lattice action, it is beneficial to write this action in momentum space, which is given by
\begin{equation}
\label{BC_Mom }
S_{BC}=
\int_p 
\ol\psi_p \left[ 
	i\sum_{\mu=1}^4 \bigg(\sin(p_\mu) \vec{\ee}^\mu \cdot \vec{\gamma}
	+B\gamma_4(\cos(p_\mu)-C)\bigg)
\right] \psi_p.
\end{equation}
When written in momentum space, one can immediately confirm that this action has exact chiral symmetry.  In addition, for $B \neq 0$ and $0<C<1$, this action has two poles at 
\begin{eqnarray}
p_\mu^{(1)} &=& \tilde{p}(\phantom{+}1,\phantom{+}1,\phantom{+}1,\phantom{4}1\ )\nn\quad\quad
p_\mu^{(2)} = -\tilde{p}(\phantom{+}1,\phantom{+}1,\phantom{+}1,\phantom{4}1\ ),
\end{eqnarray}
where $\cos(\tilde{p}) = C$.  The existence of only two poles results in the action having minimal fermion doubling under the Nielsen-Ninomiya ``no-go'' theorem~\cite{Nielsen:1980rz,Nielsen:1981xu,Nielsen:1981hk}.  

While the features of exact chiral symmetry and minimal doubling are both convenient and attractive features in a lattice action, there are undesirable consequences as well.   These consequences stem from the action breaking additional discrete symmetries, namely parity ($\psi(\vec{p},p_4) \rightarrow \gamma_4 \psi(-\vec{p},p_4)$), charge conjugation ($\psi(\vec{p},p_4) \rightarrow C \bar{\psi}^T(\vec{p},p_4)$), and time reversal ($\psi(\vec{p},p_4) \rightarrow \gamma_5\gamma_4 \psi(\vec{p},-p_4)$).  These broken symmetries will lead to new unwanted operators that are a result of radiative corrections.

\subsection{Radiative Corrections}
The presence of additional broken symmetries in lattice actions allow for new operators to be generated through radiative corrections.  The new unphysical operators due to this lattice discretization fall into three categories; relevant, marginal, or irrelevant operators.  Irrelevant operators, which are operators of mass dimension five or higher, are proportional to positive powers of the lattice spacing $a$ and would vanish in the continuum limit, $a\rightarrow 0$.  The other two categories, relevant and marginal, remain in the continuum limit and add unphysical terms to the final results that require fine tuning to eliminate.  Marginal operators, dimension four, can have logarithmic lattice spacing dependence. Relevant operators, dimension three or less, are particularly bad since they are proportional to negative powers of $a$ and diverge in the continuum limit.  

The exact chirality of the Bori\c{c}i-Creutz action prevents the relevant operator from the Wilson action proportional to $a^{-1}\ol \psi \psi$, but the breaking of parity and time reversal allow to two additional relevant operators, $a^{-1}\sum_j c_3^{(j)} \mc O_3^{(j)}$, where
\begin{eqnarray}
\mc O_3^{(1)} &=& 4iB\ol \psi \gamma_4 \psi =i\sum_{\mu=1}^4 \ol \psi (\ee^\mu \cdot \gamma) \psi, \nn\quad\quad
\mc O_3^{(2)} = 4iB\ol \psi \gamma_4 \gamma_5\psi = i\sum_{\mu=1}^4 \ol \psi (\ee^\mu \cdot \gamma)\gamma_5 \psi.
\end{eqnarray}
An emphasis should be placed on the fact that $\mu = 1,2,3,4$ and does not include $e^5$ in this sum.   This illustrates the $S_4$ permutation symmetry the Bori\c{c}i-Creutz action possesses, which is apparent in this form since these relevant operators are invariant under any exchange of the four $\mu$ values.  In addition to these two relevant operators, there are additionally ten marginal operators\footnote{See Ref.~\cite{Bedaque:2008xs} for more detail on the marginal operators.}, which we will not focus on in this note.  All these operators will be generated unless the action possesses an additional symmetry. 

\subsection{Additional Symmetry?}
In physical two-dimensional graphene, there exists an additional $\mathbb{Z}_3$ rotational symmetry.  The question that now remains is whether an analogous symmetry exists for the graphene-inspired Bori\c{c}i-Creutz action.  Analysis on the subject \cite{Bedaque:2008jm} shows that if an action in four dimensions  
poscesses the minimal $\mathbb{Z}_5$ permutation symmetry (or the larger two-operation $A_5$ or $S_5$ permutation symmetry), then for $\alpha = 1,2,3,4,5$, 
\begin{equation}
\label{sum_ea}
\sum_{\alpha=1}^5 \ee^{\alpha} = 0,
\end{equation}
which results in the relevant operators   
\begin{eqnarray}
\mc O_3^{(1)} &=&i\sum_{\alpha=1}^5 \ol \psi (\ee^\alpha \cdot \gamma) \psi = 0, \nn\quad\quad
\mc O_3^{(2)} =i\sum_{\alpha=1}^5 \ol \psi (\ee^\alpha \cdot \gamma) \gamma_5\psi = 0.
\end{eqnarray}
Thus, a lattice action with this minimal symmetry would not have any relevant operators.  

\begin{figure}[ht]
\center
\begin{tabular}{cc}
\includegraphics[width=0.41\columnwidth]{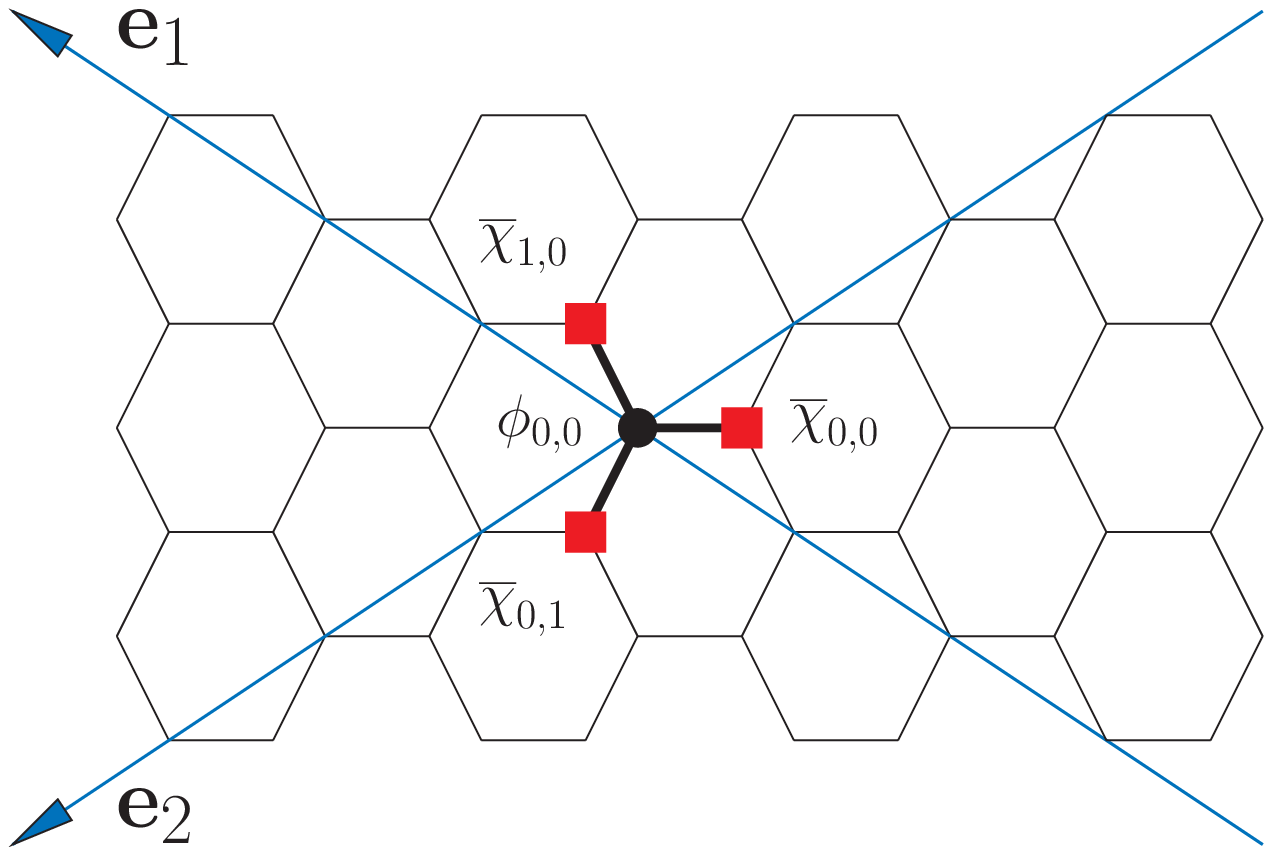} &
\includegraphics[width=0.41\columnwidth]{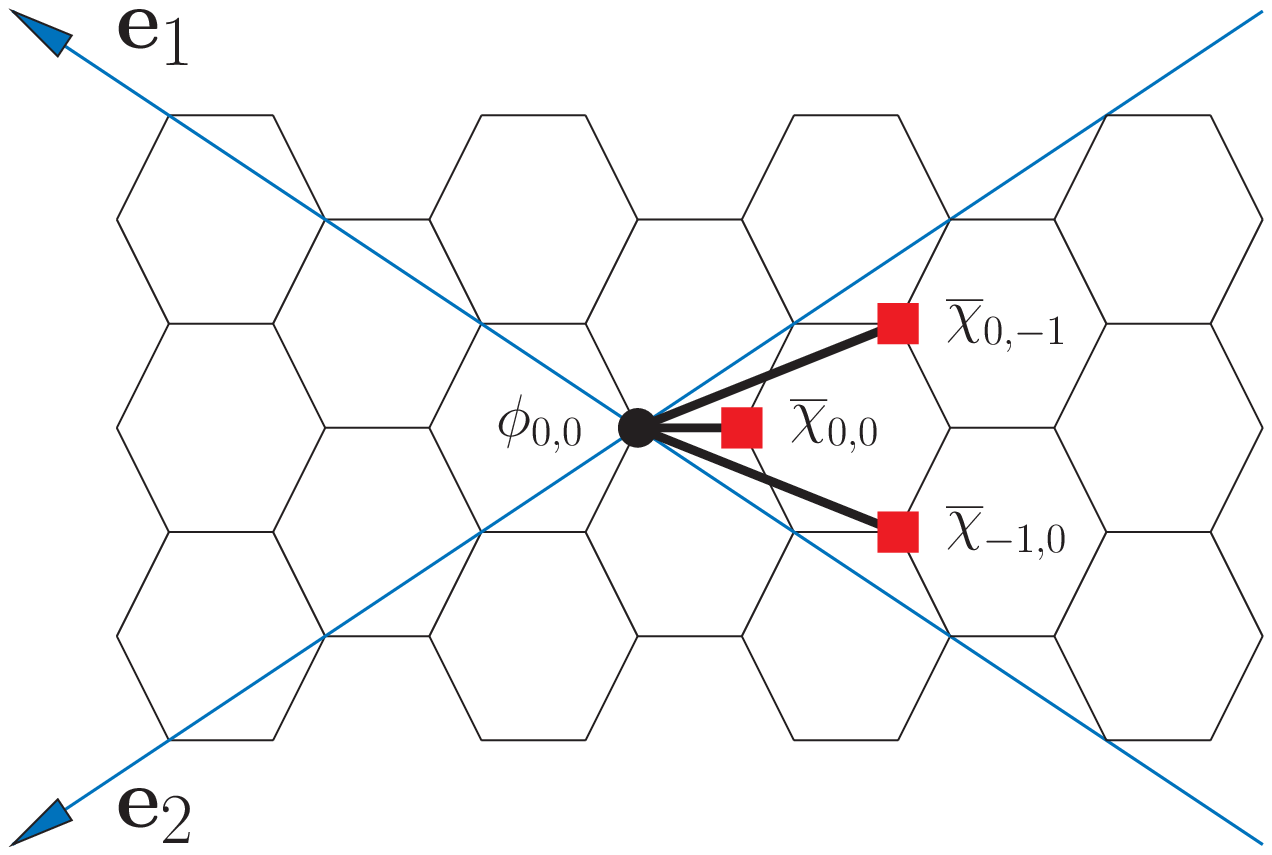}
\end{tabular}
\caption{Two-dimensional projection of  Bori\c{c}i-Creutz action with the blue duel basis vectors $\ee_\mu = \ee^\mu-\ee^5$. (left) Projection of the first line in of the two-component action shows nearest neighbor interactions leading to $\mathbb{Z}_3$ rotational symmetry. (right) Projection of the second line shows next-to-nearest neighbor interactions breaking $\mathbb{Z}_3$.} 
\label{GB}
\end{figure}

So does the Bori\c{c}i-Creutz action have this minimal $\mathbb{Z}_5$ permutation symmetry?  To explore this question, it is useful to write the action in terms of two two-component spinors, one which acts as the ``a-site'' and one which acts as the ``b-site'' as illustrated in Fig.~\ref{Grid}.  The action written in this way is 
\bea\label{eq:BC_twocomponent}
S_{BC} 
&=& 
\frac{1}{2}\sum\limits_x 
\Bigg[
\sum\limits_{\mu=1}^4\left(
\ol\phi_{x-\mu} \, \Sigma \cdot \ee^\mu \, \chi_{x}
 -\ol\chi_{x+\mu} \, \Sigma \cdot \ee^\mu \, \phi_x  \right)
 +\ol\phi_{x} \, \Sigma \cdot \ee^5 \, \chi_{x} 
 -\ol\chi_{x} \, \Sigma \cdot \ee^5 \, \phi_x
  \nn\\
&& \phantom{spaci} + \sum\limits_{\mu=1}^4
 \left(\ol\chi_{x-\mu} \, \ol\Sigma \cdot \ee^\mu \, \phi_{x}
 -\ol\phi_{x+\mu} \, \ol\Sigma \cdot \ee^\mu \, \chi_x \right)
  +\ol\chi_{x} \, \ol\Sigma \cdot \ee^5 \, \phi_x
-\ol\phi_{x} \, \ol\Sigma \cdot \ee^5 \, \chi_{x} 
\Bigg],\nn\\
\eea with $\Sigma=(\vec{\sigma}, -1)$ and $\ol\Sigma=(\vec{\sigma}, 1)$ and
 \beq
\psi_p = \begin{pmatrix}
\phi_p\\
\chi_p
\end{pmatrix},
\quad
\text{and}
 \quad
\ol \psi_p  
= 
\left( \ol \phi_p , \ol \chi_p  \right).\nn
\eeq 
With this two-component form of the action, examining whether or not this action has this symmetry is most easily accomplished by looking at a two-dimensional projection of this action and using the graphene picture.  In order for any of the terms to display this $\mathbb{Z}_5$ permutation symmetry, the choice of $B=1/\sqrt{5}$ and $C=1$ is required.  The first line of Eq.~\eqref{eq:BC_twocomponent} is given by the left figure in Fig.~\ref{GB} and shows the desired $\mathbb{Z}_3$, nearest neighbor behavior in this two dimensional projection (generalizes to the $\mathbb{Z}_5$ permutation symmetry in four dimensions).  However, the next line leads to the right figure in Fig.~\ref{GB}, which violates $\mathbb{Z}_3$ with next-to-nearest neighbor interactions (thus, violates the desired $\mathbb{Z}_5$ permutation symmetry in four dimensions).  Therefore, this action does not have the minimal symmetry required to eliminate the relevant operators.

\section{Other Non-Orthogonal Actions}
With the motivation of finding an action with exact chiral symmetry and this minimal $\mathbb{Z}_5$ permutation symmetry needed to eliminate the relevant operators, Ref.~\cite{Bedaque:2008jm} explored several non-orthogonal actions which will be reviewed here.
\subsection{Modified Bori\c{c}i-Creutz Action}
With $B=1/\sqrt{5}$ and $C=1$, the first line of Eq.~\eqref{eq:BC_twocomponent} obeys the $\mathbb{Z}_5$ symmetry and the second line breaks this symmetry. A logical extension is to simply delete this last line.  Therefore, using the same notation as Eq.~\eqref{eq:BC_twocomponent}, this action with $B=1/\sqrt{5}$ and $C=1$ is given by
\bea
S_{MBC} 
&=& 
\frac{1}{2}\sum\limits_x 
\Bigg[
\sum\limits_{\mu=1}^4\left(
\ol\phi_{x-\mu} \, \Sigma \cdot \ee^\mu \, \chi_{x}
 -\ol\chi_{x+\mu} \, \Sigma \cdot \ee^\mu \, \phi_x  \right)
 +\ol\phi_{x} \, \Sigma \cdot \ee^5 \, \chi_{x} 
 -\ol\chi_{x} \, \Sigma \cdot \ee^5 \, \phi_x\Bigg].\nn\\
\eea
However, the analysis of this action about the pole $p_\mu \simeq 0$ yields the behavior $\ol \psi (i\vec{\gamma}\cdot \vec{k} + \gamma_5\gamma_4 k_4) \psi$.   This behavior, refered to as mutilated fermions \cite{Celmaster:1982ht,Celmaster:1983jq}, is not the desired Dirac structure.

\subsection{``Hyperdiamond'' Action}
A further modification to the modified Bori\c{c}i-Creutz action to ensure the correct Dirac structure in the continuum limit and preserve at least the minimal $\mathbb{Z}_5$ permutation symmetry led to the ``hyperdiamond'' action, which is given by
\bea\label{eq:Z5}
S &=&  \sum_x \Bigg[\sum\limits_{\mu=1}^4\left(  \,
 \ol\phi_{x-\mu} \, \sigma \cdot \ee^\mu \, \chi_{x}
 -\ol\chi_{x+\mu} \, \ol\sigma \cdot \ee^\mu \, \phi_x \right)
 +\ol\phi_{x} \, \sigma \cdot \ee^5 \, \chi_{x} 
 -\ol\chi_{x} \, \ol\sigma \cdot \ee^5 \, \phi_x\Bigg],\nn\\
\eea 
where 
\bea
\label{eq:es}
\ee^1 &=& \frac{1}{4} (\phantom{+}\sqrt{5},\phantom{+}\sqrt{5},\phantom{+}\sqrt{5},\ 1),\nn\quad
\ee^2 = \frac{1}{4} (\phantom{+}\sqrt{5},-\sqrt{5},-\sqrt{5},\ 1),\nn\\
\ee^3 &=& \frac{1}{4} (-\sqrt{5},-\sqrt{5},\phantom{+}\sqrt{5},\ 1),\quad
\ee^4 = \frac{1}{4} (-\sqrt{5},\phantom{+}\sqrt{5},-\sqrt{5},\ 1),\nn\\
\ee^5 &=& -  ( \phantom{++}0,\phantom{++}0,\phantom{++}0,\,\ 1) .\nn
\eea 
and $\ol\sigma=(\vec{\sigma}, -i)$, $\sigma=(\vec{\sigma}, i)$.  The key difference between this action and the modified Bori\c{c}i-Creutz action is the imaginary fourth component of the sigma four-vectors.  This alteration allows for the action to correctly reproduce the Dirac equation about $p_\mu \simeq 0$ (which is clear in Eq.~\eqref{eq:hyper_mom}).  In addition, the vectors $\ee^\alpha$ satisfy the desired behavior needed for  $\mathbb{Z}_5$ (or $A_5$ or $S_5$) permutation symmetry, $\sum_\alpha \ee^\alpha = 0$ and 
\bea\label{eq:e_properties}
\ee^\alpha \cdot \ee^\beta &=& \left\{ \begin{matrix}
                               \phantom{--}1 ,\ \  {\rm for} \ \  \alpha=\beta\\
                             -1/4,\ \  {\rm for} \ \  \alpha\neq\beta
                        \end{matrix} \right.
.\eea 
While the exact details of the transformation won't be covered here, one can show that this action has an $A_5$ permutation symmetry\footnote{See Ref.~\cite{Bedaque:2008xs} for details.} invariant under pairs of permutations, which is enough symmetry to prevent the relevant operators from being generated.  In momentum space, the action is given by 
\beq \label{eq:hyper_mom}
S=
\int_p 
\ol\psi_p \left[ 
	i\sum_{\mu=1}^4 \sin(p_\mu) \ee^\mu \cdot \gamma
	-\Big(\sum_{\mu=1}^4 \cos(p_\mu) \ee^\mu+\ee^5\Big) \cdot \gamma\, \gamma_5
\right] \psi_p
,\eeq 
with
\beq
\psi_p = \begin{pmatrix}
\phi_p\\
\chi_p
\end{pmatrix},
\quad 
\ol \psi_p  
= 
\left( \ol \phi_p , \ol \chi_p  \right),
\quad 
\text{and}
\quad 
\gamma_\mu
=
\begin{pmatrix}
0 & \sigma_\mu\\
 \ol\sigma_\mu  & 0 
 \end{pmatrix}
.\nn
\eeq 
From this form of the action, it is clear that it maintains exact chiral symmetry and correctly reproduces the Dirac equation in the continuum limit.  Additionally, due to the $A_5$ permutation symmetry, relevant operators will not be generated.  Unfortunately, this action yields excessive fermion doubling, not unlike the na\"ive fermion action.  An example of six poles in addition to the one at $p_\mu = 0$ is  $p_1 = -p_2 = -p_3 = p_4 = \cos^{-1}(-2/3)$. 

\section{The Wilczek Action} \label{Wilczek}

Karsten~\cite{Karsten:1981gd} and Wilczek~\cite{Wilczek:1987kw} pointed out, over twenty years ago, the possibility of minimal 
doubling with an explicit construction.
The Wilczek action has the form
\begin{eqnarray}
S_W
&=& 
\frac{1}{2}
\sum_{x,\mu}
\left[
\ol \psi (x) 
\gamma_\mu
\psi (x+\mu)
- 
\ol \psi (x)
\gamma_\mu
\psi (x-\mu)
\right]
+
3 \lambda
\sum_{x}
\ol \psi (x)
i \gamma_4 
\psi(x)
\notag \\
&& \phantom{space}
-
\frac{\lambda}{2}
\sum_{x,j}
\left[
\ol \psi(x) i \gamma_4 \psi (x + j)
+ 
\ol \psi(x) i \gamma_4 \psi (x - j)
\right]
,\end{eqnarray}
with 
$\lambda >\frac{1}{2}$
as a parameter.
The action is diagonal in momentum space
\begin{equation}
S_W 
= 
\int_{-\pi}^{\pi} \frac{d^4 k}{(2 \pi)^4}
\ol \psit (k) 
D^{-1}(k)
\psi(k)
,\end{equation}
with the propagator 
$D(k)$ 
given by
\begin{equation}
D(k) 
=
\left[
\sum_\mu i \gamma_\mu \sin k_\mu
- 
i \gamma_4 \lambda
\sum_{j}
( \cos k_j - 1)
\right]^{-1}
.\end{equation}
There are exactly two poles of the propagator, 
namely at the values
$k_\mu^{(1)} = (0,0,0,0)$,
and 
$k_\mu^{(2)} = (0,0,0,\pi)$.
As this action has been analyzed 
in detail previously~\cite{Pernici:1994yj},
we shall not pursue the decomposition
of the fields in terms of flavor and chirality.
Instead we will focus on the discrete
symmetries of the action:
\begin{enumerate}
\item
Cubic invariance
\item $CT$ invariance
\item $P$ invariance
\item $\Theta$ invariance,
\end{enumerate}
where 
$\Theta$ 
is the time link-reflection symmetry operator. 
For the Bori\c{c}i and Creutz actions, 
$\Theta$ is not a symmetry.

Given these discrete symmetries, it is straightforward to write
down the relevant and marginal terms in the Symanzik action. 
There is one relevant operator
\begin{equation}
\cO_3 
= 
\ol \psi 
i \gamma_4
\psi
\notag
,\end{equation}
and several marginal operators
\begin{align*}
\cO_4^{(1)} &=  F_{\mu\nu} F_{\mu\nu} \\
\cO_4^{(2)} &=  F_{4\mu} F_{4\mu} \\
\cO_4^{(3)} &= 
\bar{\psi}\, D_4 \gamma_4\, \psi \\
\cO_4^{(4)} &=  \bar \psi D_\mu \gamma_\mu  \psi
\end{align*}
As with anisotropic lattices, the breaking of hypercubic symmetry down to
cubic symmetry requires a speed of light tuning.
It has been argued in Ref.~\cite{Pernici:1994yj} that the relevant operator can be absorbed by a field redefinition, 
namely
\begin{equation} \label{eq:fieldredef}
\psi 
\longrightarrow
\exp \left(- \frac{i c_3 x_4}{a ( 1 + c_4^{(3)})} \right) \psi
,\end{equation}
where 
$c_3$
is the non-perturbatively determined 
coefficient of the operator 
$\cO_3$ 
in the Symanzik action, 
while $c_4^{(3)}$ that of $\cO_4^{(3)}$. 
The field redefinition, however, modifies the 
boundary condition and just moves the fine tuning problem from the bulk of the action 
to the boundary. 
The physical reality of this term can be ascertained by noticing that it plays the role of  
an imaginary chemical potential.
The relevant operator thus cannot 
be avoided in the case of the Wilczek action.

\section{Conclusion}
Non-orthogonal lattice actions can be used to enforce desirable features from a lattice action.  As shown for the Bori\c{c}i-Creutz action, clever discretizations can enforce exact chiral symmetry and minimal doubling.  However, upon gaining these benefits, additional symmetries are broken, which can (and often will) lead to the generation of relevant or marginal operators from radiative corrections.  An intricate balance of symmetry is needed in a discrete lattice action in order to have exact chiral symmetry, minimal doubling, and no relevant operators.  At this time, no non-orthogonal lattice action has accomplished this task.  However, this is by no means a proof that such a lattice action does not exist.  The possibility still stands that an action with just enough symmetry to rule out these relevant operators exists and if found, it would be a very efficient, cheap way to simulate chiral fermions.  

\renewcommand{\thechapter}{7}

\chapter{Discussion and Outlook}

Effective field theory provides a formalism not only to account for lattice spacing effects from lattice QCD, but also can be used extract information indirectly from lattice calculations or account for potential issues using new lattice actions.  In this work, examples from all three of these categories have been presented.

Lattice QCD is now entering a truly interesting time as a result of increased computational power and improved  algorithms.    In fact, several groups are currently performing simulations at the physical pion mass \cite{Aoki:2009ix,Lellouch:2009fg}.   For the first time, physically accurate non-perturbative calculations involving nuclear interactions, hadronic spectra, and QCD thermodynamics realistically appear to be on the horizon.  

In one sense,  extrapolation formula to the physical point from effective field theory are not necessary if the calculations are at physical pion masses.  While both finite volume and lattice spacing effects are just as, if not more, significant as a systematic error, increased computing power can reduce these effects substantially.  However, from the point of view of the effective field theory, there are several interesting questions that can finally be addressed.  For example, the pion mass is thought of as a small expansion parameter in hadronic effective field theories; is this mass small enough to capture all the chiral dynamics?  

An additional question of interest is how to extract parameters from effective field theory at the physical point.  For example, the sigma parameter in HB\CPT is one of great interest restricting dark matter models and the $KN$ coupling is important for kaon condensation scenarios in neutron stars.  One suggestion proposed in this work in Chapter \ref{Iso_Chem}, is to employ an isospin chemical potential as an additional parameter (or knob) that can be varied to help extract these parameters.  

Ultimately, effective field theory and lattice QCD are subjects often intertwined.  Effective field theory can be used to make clever proposals for lattice QCD measurements and alternatively, lattice QCD results can have interesting, unexpected effects that can alter the underlying assumptions of an effective field theory.  For this reason, the two fields will remain intertwined as we enter this new era of hadronic physics.



%
\nocite{*}   
\bibliographystyle{jhep}
\bibliography{EFT_Lattice}

\providecommand{\href}[2]{#2}\begingroup\raggedright\begin{thebibliography}{10%
0}

\bibitem{DeGrand:2006zz}
T.~DeGrand and C.~E. Detar, {\it {Lattice methods for quantum chromodynamics}},
  . New Jersey, USA: World Scientific (2006) 345 p.

\bibitem{Sharpe:1993wt}
S.~R. Sharpe, {\it {Introduction to lattice field theory}}, . Prepared for
  Uehling Summer School on Phenomenology and Lattice QCD, Seattle, WA, 21 Jun -
  2 Jul 1993.

\bibitem{Golterman:2000hr}
M.~Golterman, {\it {Lattice chiral gauge theories}},  {\em Nucl. Phys. Proc.
  Suppl.} {\bf 94} (2001) 189--203,
  [\href{http://xxx.lanl.gov/abs/hep-lat/0011027}{{\tt hep-lat/0011027}}].

\bibitem{Luscher:2000hn}
M.~Luscher, {\it {Chiral gauge theories revisited}},
  \href{http://xxx.lanl.gov/abs/hep-th/0102028}{{\tt hep-th/0102028}}.

\bibitem{Kaplan:2009yg}
D.~B. Kaplan, {\it {Chiral Symmetry and Lattice Fermions}},
  \href{http://xxx.lanl.gov/abs/0912.2560}{{\tt arXiv:0912.2560}}.

\bibitem{Poppitz:2010lq}
E.~Poppitz and Y.~Shang, {\it Chiral lattice gauge theories via mirror-fermion
  decoupling: A mission (im)possible?},
  \href{http://xxx.lanl.gov/abs/1003.5896}{{\tt arXiv:1003.5896}}.

\bibitem{Sharatchandra:1981si}
H.~S. Sharatchandra, H.~J. Thun, and P.~Weisz, {\it {Susskind Fermions on a
  Euclidean Lattice}},  {\em Nucl. Phys.} {\bf B192} (1981) 205.

\bibitem{Ginsparg:1981bj}
P.~H. Ginsparg and K.~G. Wilson, {\it {A Remnant of Chiral Symmetry on the
  Lattice}},  {\em Phys. Rev.} {\bf D25} (1982) 2649.

\bibitem{Kaplan:1992bt}
D.~B. Kaplan, {\it {A Method for simulating chiral fermions on the lattice}},
  {\em Phys. Lett.} {\bf B288} (1992) 342--347,
  [\href{http://xxx.lanl.gov/abs/hep-lat/9206013}{{\tt hep-lat/9206013}}].

\bibitem{Shamir:1993zy}
Y.~Shamir, {\it {Chiral fermions from lattice boundaries}},  {\em Nucl. Phys.}
  {\bf B406} (1993) 90--106,
  [\href{http://xxx.lanl.gov/abs/hep-lat/9303005}{{\tt hep-lat/9303005}}].

\bibitem{Furman:1994ky}
V.~Furman and Y.~Shamir, {\it {Axial symmetries in lattice QCD with Kaplan
  fermions}},  {\em Nucl. Phys.} {\bf B439} (1995) 54--78,
  [\href{http://xxx.lanl.gov/abs/hep-lat/9405004}{{\tt hep-lat/9405004}}].

\bibitem{Narayanan:1992wx}
R.~Narayanan and H.~Neuberger, {\it {Infinitely many regulator fields for
  chiral fermions}},  {\em Phys. Lett.} {\bf B302} (1993) 62--69,
  [\href{http://xxx.lanl.gov/abs/hep-lat/9212019}{{\tt hep-lat/9212019}}].

\bibitem{Narayanan:1994gw}
R.~Narayanan and H.~Neuberger, {\it {A Construction of lattice chiral gauge
  theories}},  {\em Nucl. Phys.} {\bf B443} (1995) 305--385,
  [\href{http://xxx.lanl.gov/abs/hep-th/9411108}{{\tt hep-th/9411108}}].

\bibitem{Neuberger:1997fp}
H.~Neuberger, {\it {Exactly massless quarks on the lattice}},  {\em Phys.
  Lett.} {\bf B417} (1998) 141--144,
  [\href{http://xxx.lanl.gov/abs/hep-lat/9707022}{{\tt hep-lat/9707022}}].

\bibitem{Neuberger:1997bg}
H.~Neuberger, {\it {Vector like gauge theories with almost massless fermions on
  the lattice}},  {\em Phys. Rev.} {\bf D57} (1998) 5417--5433,
  [\href{http://xxx.lanl.gov/abs/hep-lat/9710089}{{\tt hep-lat/9710089}}].

\bibitem{Kikukawa:1999sy}
Y.~Kikukawa and T.~Noguchi, {\it {Low energy effective action of domain-wall
  fermion and the Ginsparg-Wilson relation}},
  \href{http://xxx.lanl.gov/abs/hep-lat/9902022}{{\tt hep-lat/9902022}}.

\bibitem{Gasser:1983yg}
J.~Gasser and H.~Leutwyler, {\it {Chiral Perturbation Theory to One Loop}},
  {\em Ann. Phys.} {\bf 158} (1984) 142.

\bibitem{Symanzik:1983dc}
K.~Symanzik, {\it {Continuum Limit and Improved Action in Lattice Theories. 1.
  Principles and phi**4 Theory}},  {\em Nucl. Phys.} {\bf B226} (1983) 187.

\bibitem{Symanzik:1983gh}
K.~Symanzik, {\it {Continuum Limit and Improved Action in Lattice Theories. 2.
  O(N) Nonlinear Sigma Model in Perturbation Theory}},  {\em Nucl. Phys.} {\bf
  B226} (1983) 205.

\bibitem{Sheikholeslami:1985ij}
B.~Sheikholeslami and R.~Wohlert, {\it {Improved Continuum Limit Lattice Action
  for QCD with Wilson Fermions}},  {\em Nucl. Phys.} {\bf B259} (1985) 572.

\bibitem{Luscher:1996sc}
M.~Luscher, S.~Sint, R.~Sommer, and P.~Weisz, {\it {Chiral symmetry and O(a)
  improvement in lattice QCD}},  {\em Nucl. Phys.} {\bf B478} (1996) 365--400,
  [\href{http://xxx.lanl.gov/abs/hep-lat/9605038}{{\tt hep-lat/9605038}}].

\bibitem{Bar:2003mh}
O.~Bar, G.~Rupak, and N.~Shoresh, {\it {Chiral perturbation theory at O(a**2)
  for lattice QCD}},  {\em Phys. Rev.} {\bf D70} (2004) 034508,
  [\href{http://xxx.lanl.gov/abs/hep-lat/0306021}{{\tt hep-lat/0306021}}].

\bibitem{Lee:1999zxa}
W.-J. Lee and S.~R. Sharpe, {\it {Partial Flavor Symmetry Restoration for
  Chiral Staggered Fermions}},  {\em Phys. Rev.} {\bf D60} (1999) 114503,
  [\href{http://xxx.lanl.gov/abs/hep-lat/9905023}{{\tt hep-lat/9905023}}].

\bibitem{Bernard:2001yj}
{\bf MILC} Collaboration, C.~Bernard, {\it {Chiral Logs in the Presence of
  Staggered Flavor Symmetry Breaking}},  {\em Phys. Rev.} {\bf D65} (2002)
  054031, [\href{http://xxx.lanl.gov/abs/hep-lat/0111051}{{\tt
  hep-lat/0111051}}].

\bibitem{Aubin:2003mg}
C.~Aubin and C.~Bernard, {\it {Pion and Kaon masses in Staggered Chiral
  Perturbation Theory}},  {\em Phys. Rev.} {\bf D68} (2003) 034014,
  [\href{http://xxx.lanl.gov/abs/hep-lat/0304014}{{\tt hep-lat/0304014}}].

\bibitem{Sharpe:1998xm}
S.~R. Sharpe and R.~L. Singleton, Jr, {\it {Spontaneous flavor and parity
  breaking with Wilson fermions}},  {\em Phys. Rev.} {\bf D58} (1998) 074501,
  [\href{http://xxx.lanl.gov/abs/hep-lat/9804028}{{\tt hep-lat/9804028}}].

\bibitem{Rupak:2002sm}
G.~Rupak and N.~Shoresh, {\it {Chiral perturbation theory for the Wilson
  lattice action}},  {\em Phys. Rev.} {\bf D66} (2002) 054503,
  [\href{http://xxx.lanl.gov/abs/hep-lat/0201019}{{\tt hep-lat/0201019}}].

\bibitem{Bedaque:2007xg}
P.~F. Bedaque, M.~I. Buchoff, and A.~Walker-Loud, {\it {Effective Field Theory
  for the Anisotropic Wilson Lattice Action}},  {\em Phys. Rev.} {\bf D77}
  (2008) 074501, [\href{http://xxx.lanl.gov/abs/0708.2254}{{\tt
  arXiv:0708.2254}}].

\bibitem{Buchoff:2008ve}
M.~I. Buchoff, {\it {Isotropic and Anisotropic Lattice Spacing Corrections for
  I=2 pi-pi Scattering from Effective Field Theory}},  {\em Phys. Rev.} {\bf
  D77} (2008) 114502, [\href{http://xxx.lanl.gov/abs/0802.2931}{{\tt
  arXiv:0802.2931}}].

\bibitem{Buchoff:2008hh}
M.~I. Buchoff, J.-W. Chen, and A.~Walker-Loud, {\it {pi-pi Scattering in
  Twisted Mass Chiral Perturbation Theory}},  {\em Phys. Rev.} {\bf D79} (2009)
  074503, [\href{http://xxx.lanl.gov/abs/0810.2464}{{\tt arXiv:0810.2464}}].

\bibitem{Feng:2009ij}
X.~Feng, K.~Jansen, and D.~B. Renner, {\it {The pi+ pi+ scattering length from
  maximally twisted mass lattice QCD}},  {\em Phys. Lett.} {\bf B684} (2010)
  268--274, [\href{http://xxx.lanl.gov/abs/0909.3255}{{\tt arXiv:0909.3255}}].

\bibitem{Feng:2009ck}
X.~Feng, K.~Jansen, and D.~B. Renner, {\it {Scattering from finite size methods
  in lattice QCD}},  \href{http://xxx.lanl.gov/abs/0910.4871}{{\tt
  arXiv:0910.4871}}.

\bibitem{Bedaque:2009yh}
P.~F. Bedaque, M.~I. Buchoff, and B.~C. Tiburzi, {\it {Meson-Baryon Scattering
  Parameters from Lattice QCD with an Isospin Chemical Potential}},  {\em Phys.
  Rev.} {\bf D80} (2009) 114501, [\href{http://xxx.lanl.gov/abs/0910.4595}{{\tt
  arXiv:0910.4595}}].

\bibitem{Buchoff:2008ei}
M.~I. Buchoff, {\it {Search for Chiral Fermion Actions on Non-Orthogonal
  Lattices}},  {\em PoS} {\bf LATTICE2008} (2008) 068,
  [\href{http://xxx.lanl.gov/abs/0809.3943}{{\tt arXiv:0809.3943}}].

\bibitem{Bedaque:2008xs}
P.~F. Bedaque, M.~I. Buchoff, B.~C. Tiburzi, and A.~Walker-Loud, {\it {Broken
  Symmetries from Minimally Doubled Fermions}},  {\em Phys. Lett.} {\bf B662}
  (2008) 449--455, [\href{http://xxx.lanl.gov/abs/0801.3361}{{\tt
  arXiv:0801.3361}}].

\bibitem{Bedaque:2008jm}
P.~F. Bedaque, M.~I. Buchoff, B.~C. Tiburzi, and A.~Walker-Loud, {\it {Search
  for Fermion Actions on Hyperdiamond Lattices}},  {\em Phys. Rev.} {\bf D78}
  (2008) 017502, [\href{http://xxx.lanl.gov/abs/0804.1145}{{\tt
  arXiv:0804.1145}}].

\bibitem{Creutz:2007af}
M.~Creutz, {\it {Four-dimensional graphene and chiral fermions}},  {\em JHEP}
  {\bf 04} (2008) 017, [\href{http://xxx.lanl.gov/abs/0712.1201}{{\tt
  arXiv:0712.1201}}].

\bibitem{Creutz:1996bg}
M.~Creutz, {\it {Wilson fermions at finite temperature}},
  \href{http://xxx.lanl.gov/abs/hep-lat/9608024}{{\tt hep-lat/9608024}}.

\bibitem{Beane:2003xv}
S.~R. Beane and M.~J. Savage, {\it {Nucleons properties at finite lattice
  spacing in chiral perturbation theory}},  {\em Phys. Rev.} {\bf D68} (2003)
  114502, [\href{http://xxx.lanl.gov/abs/hep-lat/0306036}{{\tt
  hep-lat/0306036}}].

\bibitem{Tiburzi:2005vy}
B.~C. Tiburzi, {\it {Baryon masses at O(a**2) in chiral perturbation theory}},
  {\em Nucl. Phys.} {\bf A761} (2005) 232--258,
  [\href{http://xxx.lanl.gov/abs/hep-lat/0501020}{{\tt hep-lat/0501020}}].

\bibitem{Bedaque:2007pe}
P.~F. Bedaque and A.~Walker-Loud, {\it {Restless pions: orbifold boundary
  conditions and noise suppression in lattice QCD}},  {\em Phys. Lett.} {\bf
  B660} (2008) 369--375, [\href{http://xxx.lanl.gov/abs/0708.0207}{{\tt
  arXiv:0708.0207}}].

\bibitem{Alford:1996nx}
M.~G. Alford, T.~R. Klassen, and G.~P. Lepage, {\it {Improving lattice quark
  actions}},  {\em Nucl. Phys.} {\bf B496} (1997) 377--407,
  [\href{http://xxx.lanl.gov/abs/hep-lat/9611010}{{\tt hep-lat/9611010}}].

\bibitem{Klassen:1998ua}
T.~R. Klassen, {\it {The anisotropic Wilson gauge action}},  {\em Nucl. Phys.}
  {\bf B533} (1998) 557--575,
  [\href{http://xxx.lanl.gov/abs/hep-lat/9803010}{{\tt hep-lat/9803010}}].

\bibitem{Morningstar:1997ff}
C.~J. Morningstar and M.~J. Peardon, {\it {Efficient glueball simulations on
  anisotropic lattices}},  {\em Phys. Rev.} {\bf D56} (1997) 4043--4061,
  [\href{http://xxx.lanl.gov/abs/hep-lat/9704011}{{\tt hep-lat/9704011}}].

\bibitem{Morningstar:1999rf}
C.~J. Morningstar and M.~J. Peardon, {\it {The glueball spectrum from an
  anisotropic lattice study}},  {\em Phys. Rev.} {\bf D60} (1999) 034509,
  [\href{http://xxx.lanl.gov/abs/hep-lat/9901004}{{\tt hep-lat/9901004}}].

\bibitem{Basak:2006ww}
S.~Basak {\em et.~al.}, {\it {Lattice QCD determination of states with spin 5/2
  or higher in the spectrum of nucleons}},
  \href{http://xxx.lanl.gov/abs/hep-lat/0609052}{{\tt hep-lat/0609052}}.

\bibitem{Fukugita:1994ve}
M.~Fukugita, Y.~Kuramashi, M.~Okawa, H.~Mino, and A.~Ukawa, {\it {Hadron
  scattering lengths in lattice QCD}},  {\em Phys. Rev.} {\bf D52} (1995)
  3003--3023, [\href{http://xxx.lanl.gov/abs/hep-lat/9501024}{{\tt
  hep-lat/9501024}}].

\bibitem{Beane:2006mx}
S.~R. Beane, P.~F. Bedaque, K.~Orginos, and M.~J. Savage, {\it {Nucleon nucleon
  scattering from fully-dynamical lattice QCD}},  {\em Phys. Rev. Lett.} {\bf
  97} (2006) 012001, [\href{http://xxx.lanl.gov/abs/hep-lat/0602010}{{\tt
  hep-lat/0602010}}].

\bibitem{Beane:2006gf}
{\bf NPLQCD} Collaboration, S.~R. Beane {\em et.~al.}, {\it {Hyperon nucleon
  scattering from fully-dynamical lattice QCD}},  {\em Nucl. Phys.} {\bf A794}
  (2007) 62--72, [\href{http://xxx.lanl.gov/abs/hep-lat/0612026}{{\tt
  hep-lat/0612026}}].

\bibitem{Lin:2009zr}
H.-W. Lin, S.~D. Cohen, J.~Dudek, R.~G. Edwards, B.~Jo{\'o}, D.~G. Richards,
  J.~Bulava, J.~Foley, C.~Morningstar, E.~Engelson, S.~Wallace, K.~J. Juge,
  N.~Mathur, M.~J. Peardon, and S.~M. Ryan, {\it First results from 2+1
  dynamical quark flavors on an anisotropic lattice: light-hadron spectroscopy
  and setting the strange-quark mass},  {\em Phys.Rev.D} {\bf 79} (2009)
  034502, [\href{http://xxx.lanl.gov/abs/0810.3588}{{\tt arXiv:0810.3588}}].

\bibitem{Beane:2009ys}
S.~R. Beane, W.~Detmold, T.~C. Luu, K.~Orginos, A.~Parreno, M.~J. Savage,
  A.~Torok, and A.~Walker-Loud, {\it High statistics analysis using anisotropic
  clover lattices: (ii) three-baryon systems},  {\em Phys.Rev.D} {\bf 80}
  (2009) 074501, [\href{http://xxx.lanl.gov/abs/0905.0466}{{\tt
  arXiv:0905.0466}}].

\bibitem{Beane:2009rt}
S.~R. Beane, W.~Detmold, H.-W. Lin, T.~C. Luu, K.~Orginos, M.~J. Savage,
  A.~Torok, and A.~Walker-Loud, {\it High statistics analysis using anisotropic
  clover lattices: (iii) baryon-baryon interactions},
  \href{http://xxx.lanl.gov/abs/0912.4243}{{\tt arXiv:0912.4243}}.

\bibitem{Beane:2009fr}
S.~R. Beane, W.~Detmold, T.~C. Luu, K.~Orginos, A.~Parreno, M.~J. Savage,
  A.~Torok, and A.~Walker-Loud, {\it High statistics analysis using anisotropic
  clover lattices: (i) single hadron correlation functions},  {\em Phys.Rev.D}
  {\bf 79} (2009) 114502, [\href{http://xxx.lanl.gov/abs/0903.2990}{{\tt
  arXiv:0903.2990}}].

\bibitem{Wilson:1974sk}
K.~G. Wilson, {\it {Confinement of quarks}},  {\em Phys. Rev.} {\bf D10} (1974)
  2445--2459.

\bibitem{Weinberg:1978kz}
S.~Weinberg, {\it {Phenomenological Lagrangians}},  {\em Physica} {\bf A96}
  (1979) 327.

\bibitem{Gasser:1984gg}
J.~Gasser and H.~Leutwyler, {\it {Chiral Perturbation Theory: Expansions in the
  Mass of the Strange Quark}},  {\em Nucl. Phys.} {\bf B250} (1985) 465.

\bibitem{Aoki:1983qi}
S.~Aoki, {\it {New Phase Structure for Lattice QCD with Wilson Fermions}},
  {\em Phys. Rev.} {\bf D30} (1984) 2653.

\bibitem{Klassen:1998fh}
T.~R. Klassen, {\it {Non-perturbative improvement of the anisotropic Wilson QCD
  action}},  {\em Nucl. Phys. Proc. Suppl.} {\bf 73} (1999) 918--920,
  [\href{http://xxx.lanl.gov/abs/hep-lat/9809174}{{\tt hep-lat/9809174}}].

\bibitem{Chen:2000ej}
P.~Chen, {\it {Heavy quarks on anisotropic lattices: The charmonium spectrum}},
   {\em Phys. Rev.} {\bf D64} (2001) 034509,
  [\href{http://xxx.lanl.gov/abs/hep-lat/0006019}{{\tt hep-lat/0006019}}].

\bibitem{Liao:2001yh}
X.~Liao and T.~Manke, {\it {Relativistic bottomonium spectrum from anisotropic
  lattices}},  {\em Phys. Rev.} {\bf D65} (2002) 074508,
  [\href{http://xxx.lanl.gov/abs/hep-lat/0111049}{{\tt hep-lat/0111049}}].

\bibitem{GellMann:1968rz}
M.~Gell-Mann, R.~J. Oakes, and B.~Renner, {\it {Behavior of current divergences
  under SU(3) x SU(3)}},  {\em Phys. Rev.} {\bf 175} (1968) 2195--2199.

\bibitem{Antonio:2008zz}
{\bf RBC} Collaboration, D.~J. Antonio {\em et.~al.}, {\it {Localization and
  chiral symmetry in 3 flavor domain wall QCD}},  {\em Phys. Rev.} {\bf D77}
  (2008) 014509, [\href{http://xxx.lanl.gov/abs/0705.2340}{{\tt
  arXiv:0705.2340}}].

\bibitem{Jenkins:1990jv}
E.~E. Jenkins and A.~V. Manohar, {\it {Baryon chiral perturbation theory using
  a heavy fermion Lagrangian}},  {\em Phys. Lett.} {\bf B255} (1991) 558--562.

\bibitem{Jenkins:1991es}
E.~E. Jenkins and A.~V. Manohar, {\it {Chiral corrections to the baryon axial
  currents}},  {\em Phys. Lett.} {\bf B259} (1991) 353--358.

\bibitem{Frezzotti:2000nk}
{\bf Alpha} Collaboration, R.~Frezzotti, P.~A. Grassi, S.~Sint, and P.~Weisz,
  {\it {Lattice QCD with a chirally twisted mass term}},  {\em JHEP} {\bf 08}
  (2001) 058, [\href{http://xxx.lanl.gov/abs/hep-lat/0101001}{{\tt
  hep-lat/0101001}}].

\bibitem{Kogut:1974ag}
J.~B. Kogut and L.~Susskind, {\it {Hamiltonian Formulation of Wilson's Lattice
  Gauge Theories}},  {\em Phys. Rev.} {\bf D11} (1975) 395.

\bibitem{Susskind:1976jm}
L.~Susskind, {\it {Lattice Fermions}},  {\em Phys. Rev.} {\bf D16} (1977)
  3031--3039.

\bibitem{WalkerLoud:2005bt}
A.~Walker-Loud and J.~M.~S. Wu, {\it {Nucleon and Delta masses in twisted mass
  chiral perturbation theory}},  {\em Phys. Rev.} {\bf D72} (2005) 014506,
  [\href{http://xxx.lanl.gov/abs/hep-lat/0504001}{{\tt hep-lat/0504001}}].

\bibitem{Bailey:2006zn}
J.~A. Bailey, {\it {Staggered baryon operators with flavor SU(3) quantum
  numbers}},  {\em Phys. Rev.} {\bf D75} (2007) 114505,
  [\href{http://xxx.lanl.gov/abs/hep-lat/0611023}{{\tt hep-lat/0611023}}].

\bibitem{Bailey:2007iq}
J.~A. Bailey, {\it {Staggered heavy baryon chiral perturbation theory}},  {\em
  Phys. Rev.} {\bf D77} (2008) 054504,
  [\href{http://xxx.lanl.gov/abs/0704.1490}{{\tt arXiv:0704.1490}}].

\bibitem{Georgi:1990um}
H.~Georgi, {\it {An effective field theory for heavy quarks at low-energies}},
  {\em Phys. Lett.} {\bf B240} (1990) 447--450.

\bibitem{Manohar:2000dt}
A.~V. Manohar and M.~B. Wise, {\it {Heavy quark physics}},  {\em Camb. Monogr.
  Part. Phys. Nucl. Phys. Cosmol.} {\bf 10} (2000) 1--191.

\bibitem{Hemmert:1997ye}
T.~R. Hemmert, B.~R. Holstein, and J.~Kambor, {\it {Chiral Lagrangians and
  Delta(1232) interactions: Formalism}},  {\em J. Phys.} {\bf G24} (1998)
  1831--1859, [\href{http://xxx.lanl.gov/abs/hep-ph/9712496}{{\tt
  hep-ph/9712496}}].

\bibitem{WalkerLoud:2004hf}
A.~Walker-Loud, {\it {Octet baryon masses in partially quenched chiral
  perturbation theory}},  {\em Nucl. Phys.} {\bf A747} (2005) 476--507,
  [\href{http://xxx.lanl.gov/abs/hep-lat/0405007}{{\tt hep-lat/0405007}}].

\bibitem{Tiburzi:2004rh}
B.~C. Tiburzi and A.~Walker-Loud, {\it {Decuplet baryon masses in partially
  quenched chiral perturbation theory}},  {\em Nucl. Phys.} {\bf A748} (2005)
  513--536, [\href{http://xxx.lanl.gov/abs/hep-lat/0407030}{{\tt
  hep-lat/0407030}}].

\bibitem{Tiburzi:2005na}
B.~C. Tiburzi and A.~Walker-Loud, {\it {Strong isospin breaking in the nucleon
  and Delta masses}},  {\em Nucl. Phys.} {\bf A764} (2006) 274--302,
  [\href{http://xxx.lanl.gov/abs/hep-lat/0501018}{{\tt hep-lat/0501018}}].

\bibitem{WalkerLoud:2006sa}
A.~Walker-Loud, {\it {Topics in effective field theory for lattice QCD}},
  \href{http://xxx.lanl.gov/abs/hep-lat/0608010}{{\tt hep-lat/0608010}}.

\bibitem{Luke:1992cs}
M.~E. Luke and A.~V. Manohar, {\it {Reparametrization invariance constraints on
  heavy particle effective field theories}},  {\em Phys. Lett.} {\bf B286}
  (1992) 348--354, [\href{http://xxx.lanl.gov/abs/hep-ph/9205228}{{\tt
  hep-ph/9205228}}].

\bibitem{Chen:2005ab}
J.-W. Chen, D.~O'Connell, R.~S. Van~de Water, and A.~Walker-Loud, {\it
  {Ginsparg-Wilson pions scattering on a staggered sea}},  {\em Phys. Rev.}
  {\bf D73} (2006) 074510, [\href{http://xxx.lanl.gov/abs/hep-lat/0510024}{{\tt
  hep-lat/0510024}}].

\bibitem{Chen:2006wf}
J.-W. Chen, D.~O'Connell, and A.~Walker-Loud, {\it {Two meson systems with
  Ginsparg-Wilson valence quarks}},  {\em Phys. Rev.} {\bf D75} (2007) 054501,
  [\href{http://xxx.lanl.gov/abs/hep-lat/0611003}{{\tt hep-lat/0611003}}].

\bibitem{Orginos:2007tw}
K.~Orginos and A.~Walker-Loud, {\it {Mixed meson masses with domain-wall
  valence and staggered sea fermions}},  {\em Phys. Rev.} {\bf D77} (2008)
  094505, [\href{http://xxx.lanl.gov/abs/0705.0572}{{\tt arXiv:0705.0572}}].

\bibitem{Chen:2007ug}
J.-W. Chen, D.~O'Connell, and A.~Walker-Loud, {\it {Universality of Mixed
  Action Extrapolation Formulae}},  {\em JHEP} {\bf 04} (2009) 090,
  [\href{http://xxx.lanl.gov/abs/0706.0035}{{\tt arXiv:0706.0035}}].

\bibitem{Huang:1957im}
K.~Huang and C.~N. Yang, {\it {Quantum-mechanical many-body problem with
  hard-sphere interaction}},  {\em Phys. Rev.} {\bf 105} (1957) 767--775.

\bibitem{Hamber:1983vu}
H.~W. Hamber, E.~Marinari, G.~Parisi, and C.~Rebbi, {\it {Considerations on
  numerical analysis of QCD}},  {\em Nucl. Phys.} {\bf B225} (1983) 475.

\bibitem{Luscher:1990ux}
M.~Luscher, {\it {Two particle states on a torus and their relation to the
  scattering matrix}},  {\em Nucl. Phys.} {\bf B354} (1991) 531--578.

\bibitem{Luscher:1986pf}
M.~Luscher, {\it {Volume Dependence of the Energy Spectrum in Massive Quantum
  Field Theories. 2. Scattering States}},  {\em Commun. Math. Phys.} {\bf 105}
  (1986) 153--188.

\bibitem{Weinberg:1966kf}
S.~Weinberg, {\it {Pion scattering lengths}},  {\em Phys. Rev. Lett.} {\bf 17}
  (1966) 616--621.

\bibitem{Knecht:1995tr}
M.~Knecht, B.~Moussallam, J.~Stern, and N.~H. Fuchs, {\it {The Low-energy pi pi
  amplitude to one and two loops}},  {\em Nucl. Phys.} {\bf B457} (1995)
  513--576, [\href{http://xxx.lanl.gov/abs/hep-ph/9507319}{{\tt
  hep-ph/9507319}}].

\bibitem{Bijnens:1995yn}
J.~Bijnens, G.~Colangelo, G.~Ecker, J.~Gasser, and M.~E. Sainio, {\it {Elastic
  $\pi\pi$ scattering to two loops}},  {\em Phys. Lett.} {\bf B374} (1996)
  210--216, [\href{http://xxx.lanl.gov/abs/hep-ph/9511397}{{\tt
  hep-ph/9511397}}].

\bibitem{Bijnens:1997vq}
J.~Bijnens, G.~Colangelo, G.~Ecker, J.~Gasser, and M.~E. Sainio, {\it {Pion
  pion scattering at low energy}},  {\em Nucl. Phys.} {\bf B508} (1997)
  263--310, [\href{http://xxx.lanl.gov/abs/hep-ph/9707291}{{\tt
  hep-ph/9707291}}].

\bibitem{Gupta:1993rn}
R.~Gupta, A.~Patel, and S.~R. Sharpe, {\it {I = 2 pion scattering amplitude
  with Wilson fermions}},  {\em Phys. Rev.} {\bf D48} (1993) 388--396,
  [\href{http://xxx.lanl.gov/abs/hep-lat/9301016}{{\tt hep-lat/9301016}}].

\bibitem{Fukugita:1994na}
M.~Fukugita, Y.~Kuramashi, H.~Mino, M.~Okawa, and A.~Ukawa, {\it {An
  Exploratory study of nucleon-nucleon scattering lengths in lattice QCD}},
  {\em Phys. Rev. Lett.} {\bf 73} (1994) 2176--2179,
  [\href{http://xxx.lanl.gov/abs/hep-lat/9407012}{{\tt hep-lat/9407012}}].

\bibitem{Fiebig:1999hs}
H.~R. Fiebig, K.~Rabitsch, H.~Markum, and A.~Mihaly, {\it {Exploring the pi+
  pi+ interaction in lattice QCD}},  {\em Few Body Syst.} {\bf 29} (2000)
  95--120, [\href{http://xxx.lanl.gov/abs/hep-lat/9906002}{{\tt
  hep-lat/9906002}}].

\bibitem{Aoki:1999pt}
{\bf JLQCD} Collaboration, S.~Aoki {\em et.~al.}, {\it {I = 2 pion scattering
  length with Wilson fermions}},  {\em Nucl. Phys. Proc. Suppl.} {\bf 83}
  (2000) 241--243, [\href{http://xxx.lanl.gov/abs/hep-lat/9911025}{{\tt
  hep-lat/9911025}}].

\bibitem{Liu:2001ss}
C.~Liu, J.-h. Zhang, Y.~Chen, and J.~P. Ma, {\it {Calculating the I = 2 pion
  scattering length using tadpole improved clover Wilson action on coarse
  anisotropic lattices}},  {\em Nucl. Phys.} {\bf B624} (2002) 360--376,
  [\href{http://xxx.lanl.gov/abs/hep-lat/0109020}{{\tt hep-lat/0109020}}].

\bibitem{Aoki:2001hc}
{\bf CP-PACS} Collaboration, S.~Aoki {\em et.~al.}, {\it {I = 2 pion scattering
  length and phase shift with Wilson fermions}},  {\em Nucl. Phys. Proc.
  Suppl.} {\bf 106} (2002) 230--232,
  [\href{http://xxx.lanl.gov/abs/hep-lat/0110151}{{\tt hep-lat/0110151}}].

\bibitem{Aoki:2002in}
{\bf JLQCD} Collaboration, S.~Aoki {\em et.~al.}, {\it {I=2 Pion Scattering
  Length with the Wilson Fermion}},  {\em Phys. Rev.} {\bf D66} (2002) 077501,
  [\href{http://xxx.lanl.gov/abs/hep-lat/0206011}{{\tt hep-lat/0206011}}].

\bibitem{Aoki:2002sg}
{\bf CP-PACS} Collaboration, S.~Aoki {\em et.~al.}, {\it {I = 2 pion scattering
  phase shift with Wilson fermions}},  {\em Nucl. Phys. Proc. Suppl.} {\bf 119}
  (2003) 311--313, [\href{http://xxx.lanl.gov/abs/hep-lat/0209056}{{\tt
  hep-lat/0209056}}].

\bibitem{Aoki:2002ny}
{\bf CP-PACS} Collaboration, S.~Aoki {\em et.~al.}, {\it {I=2 Pion Scattering
  Phase Shift with Wilson Fermions}},  {\em Phys. Rev.} {\bf D67} (2003)
  014502, [\href{http://xxx.lanl.gov/abs/hep-lat/0209124}{{\tt
  hep-lat/0209124}}].

\bibitem{Ishizuka:2003nb}
N.~Ishizuka and T.~Yamazaki, {\it {I = 2 pion scattering length from two-pion
  wave function}},  {\em Nucl. Phys. Proc. Suppl.} {\bf 129} (2004) 233--235,
  [\href{http://xxx.lanl.gov/abs/hep-lat/0309168}{{\tt hep-lat/0309168}}].

\bibitem{Yamazaki:2004qb}
{\bf CP-PACS} Collaboration, T.~Yamazaki {\em et.~al.}, {\it {I = 2 pi pi
  scattering phase shift with two flavors of O(a) improved dynamical quarks}},
  {\em Phys. Rev.} {\bf D70} (2004) 074513,
  [\href{http://xxx.lanl.gov/abs/hep-lat/0402025}{{\tt hep-lat/0402025}}].

\bibitem{Du:2004ib}
X.~Du, G.-w. Meng, C.~Miao, and C.~Liu, {\it {I = 2 pion scattering length with
  improved actions on anisotropic lattices}},  {\em Int. J. Mod. Phys.} {\bf
  A19} (2004) 5609--5614, [\href{http://xxx.lanl.gov/abs/hep-lat/0404017}{{\tt
  hep-lat/0404017}}].

\bibitem{Aoki:2004wq}
{\bf CP-PACS} Collaboration, S.~Aoki {\em et.~al.}, {\it {I = 2 pion scattering
  length from two-pion wave function}},  {\em Nucl. Phys. Proc. Suppl.} {\bf
  140} (2005) 305--307, [\href{http://xxx.lanl.gov/abs/hep-lat/0409063}{{\tt
  hep-lat/0409063}}].

\bibitem{Aoki:2005uf}
{\bf CP-PACS} Collaboration, S.~Aoki {\em et.~al.}, {\it {I = 2 pion scattering
  length from two-pion wave functions}},  {\em Phys. Rev.} {\bf D71} (2005)
  094504, [\href{http://xxx.lanl.gov/abs/hep-lat/0503025}{{\tt
  hep-lat/0503025}}].

\bibitem{Li:2007ey}
{\bf CLQCD} Collaboration, X.~Li {\em et.~al.}, {\it {Hadron Scattering in an
  Asymmetric Box}},  {\em JHEP} {\bf 06} (2007) 053,
  [\href{http://xxx.lanl.gov/abs/hep-lat/0703015}{{\tt hep-lat/0703015}}].

\bibitem{Sharpe:1992pp}
S.~R. Sharpe, R.~Gupta, and G.~W. Kilcup, {\it {Lattice calculation of I = 2
  pion scattering length}},  {\em Nucl. Phys.} {\bf B383} (1992) 309--356.

\bibitem{Kuramashi:1993ka}
Y.~Kuramashi, M.~Fukugita, H.~Mino, M.~Okawa, and A.~Ukawa, {\it {Lattice QCD
  calculation of full pion scattering lengths}},  {\em Phys. Rev. Lett.} {\bf
  71} (1993) 2387--2390.

\bibitem{Juge:2003mr}
{\bf BGR} Collaboration, K.~J. Juge, {\it {I = 2 pion scattering length with
  the parametrized fixed point action}},  {\em Nucl. Phys. Proc. Suppl.} {\bf
  129} (2004) 194--196, [\href{http://xxx.lanl.gov/abs/hep-lat/0309075}{{\tt
  hep-lat/0309075}}].

\bibitem{Feng:2009rc}
X.~Feng, K.~Jansen, and D.~B. Renner, {\it The pi+ pi+ scattering length from
  maximally twisted mass lattice qcd},
  \href{http://xxx.lanl.gov/abs/0909.3255}{{\tt arXiv:0909.3255}}.

\bibitem{Beane:2005rj}
{\bf NPLQCD} Collaboration, S.~R. Beane, P.~F. Bedaque, K.~Orginos, and M.~J.
  Savage, {\it {I = 2 pi pi scattering from fully-dynamical mixed-action
  lattice QCD}},  {\em Phys. Rev.} {\bf D73} (2006) 054503,
  [\href{http://xxx.lanl.gov/abs/hep-lat/0506013}{{\tt hep-lat/0506013}}].

\bibitem{Beane:2007xs}
S.~R. Beane {\em et.~al.}, {\it {Precise Determination of the I=2 pipi
  Scattering Length from Mixed-Action Lattice QCD}},  {\em Phys. Rev.} {\bf
  D77} (2008) 014505, [\href{http://xxx.lanl.gov/abs/0706.3026}{{\tt
  arXiv:0706.3026}}].

\bibitem{Tiburzi:2005is}
B.~C. Tiburzi, {\it {Baryons with Ginsparg-Wilson quarks in a staggered sea}},
  {\em Phys. Rev.} {\bf D72} (2005) 094501,
  [\href{http://xxx.lanl.gov/abs/hep-lat/0508019}{{\tt hep-lat/0508019}}].

\bibitem{Bedaque:2006yi}
P.~F. Bedaque, I.~Sato, and A.~Walker-Loud, {\it {Finite volume corrections to
  pi pi scattering}},  {\em Phys. Rev.} {\bf D73} (2006) 074501,
  [\href{http://xxx.lanl.gov/abs/hep-lat/0601033}{{\tt hep-lat/0601033}}].

\bibitem{Colangelo:2001df}
G.~Colangelo, J.~Gasser, and H.~Leutwyler, {\it {pi pi scattering}},  {\em
  Nucl. Phys.} {\bf B603} (2001) 125--179,
  [\href{http://xxx.lanl.gov/abs/hep-ph/0103088}{{\tt hep-ph/0103088}}].

\bibitem{Maiani:1990ca}
L.~Maiani and M.~Testa, {\it {Final state interactions from Euclidean
  correlation functions}},  {\em Phys. Lett.} {\bf B245} (1990) 585--590.

\bibitem{Beane:2003yx}
S.~R. Beane, P.~F. Bedaque, A.~Parreno, and M.~J. Savage, {\it {Exploring
  hyperons and hypernuclei with lattice QCD}},  {\em Nucl. Phys.} {\bf A747}
  (2005) 55--74, [\href{http://xxx.lanl.gov/abs/nucl-th/0311027}{{\tt
  nucl-th/0311027}}].

\bibitem{Beane:2003da}
S.~R. Beane, P.~F. Bedaque, A.~Parreno, and M.~J. Savage, {\it {Two Nucleons on
  a Lattice}},  {\em Phys. Lett.} {\bf B585} (2004) 106--114,
  [\href{http://xxx.lanl.gov/abs/hep-lat/0312004}{{\tt hep-lat/0312004}}].

\bibitem{Bernard:2006zp}
C.~Bernard {\em et.~al.}, {\it {Low energy constants from the MILC
  Collaboration}},  \href{http://xxx.lanl.gov/abs/hep-lat/0611024}{{\tt
  hep-lat/0611024}}.

\bibitem{Boucaud:2007uk}
{\bf ETM} Collaboration, P.~Boucaud {\em et.~al.}, {\it {Dynamical twisted mass
  fermions with light quarks}},  {\em Phys. Lett.} {\bf B650} (2007) 304--311,
  [\href{http://xxx.lanl.gov/abs/hep-lat/0701012}{{\tt hep-lat/0701012}}].

\bibitem{DelDebbio:2006cn}
L.~Del~Debbio, L.~Giusti, M.~Luscher, R.~Petronzio, and N.~Tantalo, {\it {QCD
  with light Wilson quarks on fine lattices (I): first experiences and physics
  results}},  {\em JHEP} {\bf 02} (2007) 056,
  [\href{http://xxx.lanl.gov/abs/hep-lat/0610059}{{\tt hep-lat/0610059}}].

\bibitem{DelDebbio:2007pz}
L.~Del~Debbio, L.~Giusti, M.~Luscher, R.~Petronzio, and N.~Tantalo, {\it {QCD
  with light Wilson quarks on fine lattices. II: DD-HMC simulations and data
  analysis}},  {\em JHEP} {\bf 02} (2007) 082,
  [\href{http://xxx.lanl.gov/abs/hep-lat/0701009}{{\tt hep-lat/0701009}}].

\bibitem{Leutwyler:2007ae}
H.~Leutwyler, {\it {Insights and puzzles in light quark physics}},
  \href{http://xxx.lanl.gov/abs/0706.3138}{{\tt arXiv:0706.3138}}.

\bibitem{Aoki:2007es}
S.~Aoki and O.~Bar, {\it {The vector and axial vector current in Wilson ChPT}},
   {\em PoS} {\bf LAT2007} (2007) 062,
  [\href{http://xxx.lanl.gov/abs/0710.0072}{{\tt arXiv:0710.0072}}].

\bibitem{Sharpe:2004ny}
S.~R. Sharpe and J.~M.~S. Wu, {\it {Twisted mass chiral perturbation theory at
  next-to-leading order}},  {\em Phys. Rev.} {\bf D71} (2005) 074501,
  [\href{http://xxx.lanl.gov/abs/hep-lat/0411021}{{\tt hep-lat/0411021}}].

\bibitem{Umeda:2003pj}
{\bf CP-PACS} Collaboration, T.~Umeda {\em et.~al.}, {\it {Two flavors of
  dynamical quarks on anisotropic lattices}},  {\em Phys. Rev.} {\bf D68}
  (2003) 034503, [\href{http://xxx.lanl.gov/abs/hep-lat/0302024}{{\tt
  hep-lat/0302024}}].

\bibitem{Beane:2008dv}
S.~R. Beane, K.~Orginos, and M.~J. Savage, {\it {Hadronic Interactions from
  Lattice QCD}},  {\em Int. J. Mod. Phys.} {\bf E17} (2008) 1157--1218,
  [\href{http://xxx.lanl.gov/abs/0805.4629}{{\tt arXiv:0805.4629}}].

\bibitem{Beane:2006gj}
S.~R. Beane {\em et.~al.}, {\it {Pi-K Scattering in Full QCD with Domain-Wall
  Valence Quarks}},  {\em Phys. Rev.} {\bf D74} (2006) 114503,
  [\href{http://xxx.lanl.gov/abs/hep-lat/0607036}{{\tt hep-lat/0607036}}].

\bibitem{Beane:2007uh}
{\bf NPLQCD} Collaboration, S.~R. Beane {\em et.~al.}, {\it {The K+K+
  Scattering Length from Lattice QCD}},  {\em Phys. Rev.} {\bf D77} (2008)
  094507, [\href{http://xxx.lanl.gov/abs/0709.1169}{{\tt arXiv:0709.1169}}].

\bibitem{Beane:2007es}
S.~R. Beane {\em et.~al.}, {\it {Multi-Pion Systems in Lattice QCD and the
  Three-Pion Interaction}},  {\em Phys. Rev. Lett.} {\bf 100} (2008) 082004,
  [\href{http://xxx.lanl.gov/abs/0710.1827}{{\tt arXiv:0710.1827}}].

\bibitem{Detmold:2008fn}
W.~Detmold {\em et.~al.}, {\it {Multi-Pion States in Lattice QCD and the
  Charged-Pion Condensate}},  {\em Phys. Rev.} {\bf D78} (2008) 014507,
  [\href{http://xxx.lanl.gov/abs/0803.2728}{{\tt arXiv:0803.2728}}].

\bibitem{Detmold:2008yn}
W.~Detmold, K.~Orginos, M.~J. Savage, and A.~Walker-Loud, {\it {Kaon
  Condensation with Lattice QCD}},  {\em Phys. Rev.} {\bf D78} (2008) 054514,
  [\href{http://xxx.lanl.gov/abs/0807.1856}{{\tt arXiv:0807.1856}}].

\bibitem{Seki:2005ns}
R.~Seki and U.~van Kolck, {\it {Effective field theory of nucleon nucleon
  scattering on large discrete lattices}},  {\em Phys. Rev.} {\bf C73} (2006)
  044006, [\href{http://xxx.lanl.gov/abs/nucl-th/0509094}{{\tt
  nucl-th/0509094}}].

\bibitem{Chen:2003vy}
J.-W. Chen and D.~B. Kaplan, {\it {A lattice theory for low energy fermions at
  finite chemical potential}},  {\em Phys. Rev. Lett.} {\bf 92} (2004) 257002,
  [\href{http://xxx.lanl.gov/abs/hep-lat/0308016}{{\tt hep-lat/0308016}}].

\bibitem{Chen:2004rq}
J.-W. Chen, D.~Lee, and T.~Schafer, {\it {Inequalities for Light Nuclei in the
  Wigner Symmetry Limit}},  {\em Phys. Rev. Lett.} {\bf 93} (2004) 242302,
  [\href{http://xxx.lanl.gov/abs/nucl-th/0408043}{{\tt nucl-th/0408043}}].

\bibitem{Lee:2004si}
D.~Lee, B.~Borasoy, and T.~Schafer, {\it {Nuclear lattice simulations with
  chiral effective field theory}},  {\em Phys. Rev.} {\bf C70} (2004) 014007,
  [\href{http://xxx.lanl.gov/abs/nucl-th/0402072}{{\tt nucl-th/0402072}}].

\bibitem{Lee:2004qd}
D.~Lee and T.~Schafer, {\it {Neutron matter on the lattice with pionless
  effective field theory}},  {\em Phys. Rev.} {\bf C72} (2005) 024006,
  [\href{http://xxx.lanl.gov/abs/nucl-th/0412002}{{\tt nucl-th/0412002}}].

\bibitem{Borasoy:2006qn}
B.~Borasoy, E.~Epelbaum, H.~Krebs, D.~Lee, and U.-G. Meissner, {\it {Lattice
  simulations for light nuclei: Chiral effective field theory at leading
  order}},  {\em Eur. Phys. J.} {\bf A31} (2007) 105--123,
  [\href{http://xxx.lanl.gov/abs/nucl-th/0611087}{{\tt nucl-th/0611087}}].

\bibitem{Borasoy:2007vi}
B.~Borasoy, E.~Epelbaum, H.~Krebs, D.~Lee, and U.-G. Meissner, {\it {Chiral
  effective field theory on the lattice at next-to- leading order}},  {\em Eur.
  Phys. J.} {\bf A35} (2008) 343--355,
  [\href{http://xxx.lanl.gov/abs/0712.2990}{{\tt arXiv:0712.2990}}].

\bibitem{Borasoy:2007vk}
B.~Borasoy, E.~Epelbaum, H.~Krebs, D.~Lee, and U.-G. Meissner, {\it {Dilute
  neutron matter on the lattice at next-to-leading order in chiral effective
  field theory}},  {\em Eur. Phys. J.} {\bf A35} (2008) 357--367,
  [\href{http://xxx.lanl.gov/abs/0712.2993}{{\tt arXiv:0712.2993}}].

\bibitem{Lee:2008fa}
D.~Lee, {\it {Lattice simulations for few- and many-body systems}},  {\em Prog.
  Part. Nucl. Phys.} {\bf 63} (2009) 117--154,
  [\href{http://xxx.lanl.gov/abs/0804.3501}{{\tt arXiv:0804.3501}}].

\bibitem{Renner:2004ck}
{\bf LHP} Collaboration, D.~B. Renner {\em et.~al.}, {\it {Hadronic physics
  with domain-wall valence and improved staggered sea quarks}},  {\em Nucl.
  Phys. Proc. Suppl.} {\bf 140} (2005) 255--260,
  [\href{http://xxx.lanl.gov/abs/hep-lat/0409130}{{\tt hep-lat/0409130}}].

\bibitem{Edwards:2005kw}
{\bf LHPC} Collaboration, R.~G. Edwards {\em et.~al.}, {\it {Hadron structure
  with light dynamical quarks}},  {\em PoS} {\bf LAT2005} (2006) 056,
  [\href{http://xxx.lanl.gov/abs/hep-lat/0509185}{{\tt hep-lat/0509185}}].

\bibitem{Orginos:1998ue}
{\bf MILC} Collaboration, K.~Orginos and D.~Toussaint, {\it {Testing improved
  actions for dynamical Kogut-Susskind quarks}},  {\em Phys. Rev.} {\bf D59}
  (1999) 014501, [\href{http://xxx.lanl.gov/abs/hep-lat/9805009}{{\tt
  hep-lat/9805009}}].

\bibitem{Orginos:1999cr}
{\bf MILC} Collaboration, K.~Orginos, D.~Toussaint, and R.~L. Sugar, {\it
  {Variants of fattening and flavor symmetry restoration}},  {\em Phys. Rev.}
  {\bf D60} (1999) 054503, [\href{http://xxx.lanl.gov/abs/hep-lat/9903032}{{\tt
  hep-lat/9903032}}].

\bibitem{Bernard:2001av}
C.~W. Bernard {\em et.~al.}, {\it {The QCD spectrum with three quark flavors}},
   {\em Phys. Rev.} {\bf D64} (2001) 054506,
  [\href{http://xxx.lanl.gov/abs/hep-lat/0104002}{{\tt hep-lat/0104002}}].

\bibitem{Aubin:2004wf}
C.~Aubin {\em et.~al.}, {\it {Light hadrons with improved staggered quarks:
  Approaching the continuum limit}},  {\em Phys. Rev.} {\bf D70} (2004) 094505,
  [\href{http://xxx.lanl.gov/abs/hep-lat/0402030}{{\tt hep-lat/0402030}}].

\bibitem{Frezzotti:1999vv}
R.~Frezzotti, P.~A. Grassi, S.~Sint, and P.~Weisz, {\it {A local formulation of
  lattice QCD without unphysical fermion zero modes}},  {\em Nucl. Phys. Proc.
  Suppl.} {\bf 83} (2000) 941--946,
  [\href{http://xxx.lanl.gov/abs/hep-lat/9909003}{{\tt hep-lat/9909003}}].

\bibitem{Blossier:2007vv}
{\bf European Twisted Mass} Collaboration, B.~Blossier {\em et.~al.}, {\it
  {Light quark masses and pseudoscalar decay constants from Nf=2 Lattice QCD
  with twisted mass fermions}},  {\em JHEP} {\bf 04} (2008) 020,
  [\href{http://xxx.lanl.gov/abs/0709.4574}{{\tt arXiv:0709.4574}}].

\bibitem{Boucaud:2008xu}
{\bf ETM} Collaboration, P.~Boucaud {\em et.~al.}, {\it {Dynamical Twisted Mass
  Fermions with Light Quarks: Simulation and Analysis Details}},  {\em Comput.
  Phys. Commun.} {\bf 179} (2008) 695--715,
  [\href{http://xxx.lanl.gov/abs/0803.0224}{{\tt arXiv:0803.0224}}].

\bibitem{Chiarappa:2006ae}
T.~Chiarappa {\em et.~al.}, {\it {Numerical simulation of QCD with u, d, s and
  c quarks in the twisted-mass Wilson formulation}},  {\em Eur. Phys. J.} {\bf
  C50} (2007) 373--383, [\href{http://xxx.lanl.gov/abs/hep-lat/0606011}{{\tt
  hep-lat/0606011}}].

\bibitem{Munster:2003ba}
G.~Munster and C.~Schmidt, {\it {Chiral perturbation theory for lattice QCD
  with a twisted mass term}},  {\em Europhys. Lett.} {\bf 66} (2004) 652--656,
  [\href{http://xxx.lanl.gov/abs/hep-lat/0311032}{{\tt hep-lat/0311032}}].

\bibitem{Scorzato:2004da}
L.~Scorzato, {\it {Pion mass splitting and phase structure in twisted mass
  QCD}},  {\em Eur. Phys. J.} {\bf C37} (2004) 445--455,
  [\href{http://xxx.lanl.gov/abs/hep-lat/0407023}{{\tt hep-lat/0407023}}].

\bibitem{Sharpe:2004ps}
S.~R. Sharpe and J.~M.~S. Wu, {\it {The phase diagram of twisted mass lattice
  QCD}},  {\em Phys. Rev.} {\bf D70} (2004) 094029,
  [\href{http://xxx.lanl.gov/abs/hep-lat/0407025}{{\tt hep-lat/0407025}}].

\bibitem{Aoki:2004ta}
S.~Aoki and O.~Bar, {\it {Twisted-mass QCD, O(a) improvement and Wilson chiral
  perturbation theory}},  {\em Phys. Rev.} {\bf D70} (2004) 116011,
  [\href{http://xxx.lanl.gov/abs/hep-lat/0409006}{{\tt hep-lat/0409006}}].

\bibitem{Aoki:2008gy}
S.~Aoki, O.~Bar, and B.~Biedermann, {\it {Pion scattering in Wilson ChPT}},
  {\em Phys. Rev.} {\bf D78} (2008) 114501,
  [\href{http://xxx.lanl.gov/abs/0806.4863}{{\tt arXiv:0806.4863}}].

\bibitem{Frezzotti:2003ni}
R.~Frezzotti and G.~C. Rossi, {\it {Chirally improving Wilson fermions. I: O(a)
  improvement}},  {\em JHEP} {\bf 08} (2004) 007,
  [\href{http://xxx.lanl.gov/abs/hep-lat/0306014}{{\tt hep-lat/0306014}}].

\bibitem{Sharpe:2005rq}
S.~R. Sharpe, {\it {Observations on discretization errors in twisted-mass
  lattice QCD}},  {\em Phys. Rev.} {\bf D72} (2005) 074510,
  [\href{http://xxx.lanl.gov/abs/hep-lat/0509009}{{\tt hep-lat/0509009}}].

\bibitem{Torok:2009dg}
A.~Torok {\em et.~al.}, {\it {Meson-Baryon Scattering Lengths from Mixed-Action
  Lattice QCD}},  \href{http://xxx.lanl.gov/abs/0907.1913}{{\tt
  arXiv:0907.1913}}.

\bibitem{Peardon:2009gh}
{\bf Hadron Spectrum} Collaboration, M.~Peardon {\em et.~al.}, {\it {A novel
  quark-field creation operator construction for hadronic physics in lattice
  QCD}},  {\em Phys. Rev.} {\bf D80} (2009) 054506,
  [\href{http://xxx.lanl.gov/abs/0905.2160}{{\tt arXiv:0905.2160}}].

\bibitem{Kaplan:1986yq}
D.~B. Kaplan and A.~E. Nelson, {\it {Strange Goings on in Dense Nucleonic
  Matter}},  {\em Phys. Lett.} {\bf B175} (1986) 57--63.

\bibitem{Nelson:1987dg}
A.~E. Nelson and D.~B. Kaplan, {\it {Strange Condensate Realignment in
  Relativistic Heavy Ion Collisions}},  {\em Phys. Lett.} {\bf B192} (1987)
  193.

\bibitem{Brown:1987hj}
G.~E. Brown, K.~Kubodera, and M.~Rho, {\it {Strangeness condensation and
  'clearing' of the vacuum}},  {\em Phys. Lett.} {\bf B192} (1987) 273--278.

\bibitem{Brown:1992ib}
G.~E. Brown, V.~Thorsson, K.~Kubodera, and M.~Rho, {\it {A Novel mechanism for
  kaon condensation in neutron star matter}},  {\em Phys. Lett.} {\bf B291}
  (1992) 355--362.

\bibitem{Brown:2007ara}
G.~E. Brown, C.-H. Lee, and M.~Rho, {\it {Recent Developments on Kaon
  Condensation and Its Astrophysical Implications}},  {\em Phys. Rept.} {\bf
  462} (2008) 1--20, [\href{http://xxx.lanl.gov/abs/0708.3137}{{\tt
  arXiv:0708.3137}}].

\bibitem{Stephanov:1996ki}
M.~A. Stephanov, {\it {Random matrix model of QCD at finite density and the
  nature of the quenched limit}},  {\em Phys. Rev. Lett.} {\bf 76} (1996)
  4472--4475, [\href{http://xxx.lanl.gov/abs/hep-lat/9604003}{{\tt
  hep-lat/9604003}}].

\bibitem{Barbour:1997ej}
I.~M. Barbour, S.~E. Morrison, E.~G. Klepfish, J.~B. Kogut, and M.-P. Lombardo,
  {\it {Results on finite density QCD}},  {\em Nucl. Phys. Proc. Suppl.} {\bf
  60A} (1998) 220--234, [\href{http://xxx.lanl.gov/abs/hep-lat/9705042}{{\tt
  hep-lat/9705042}}].

\bibitem{Alford:1998sd}
M.~G. Alford, A.~Kapustin, and F.~Wilczek, {\it {Imaginary chemical potential
  and finite fermion density on the lattice}},  {\em Phys. Rev.} {\bf D59}
  (1999) 054502, [\href{http://xxx.lanl.gov/abs/hep-lat/9807039}{{\tt
  hep-lat/9807039}}].

\bibitem{deForcrand:1999cy}
P.~de~Forcrand and V.~Laliena, {\it {The role of the Polyakov loop in finite
  density QCD}},  {\em Phys. Rev.} {\bf D61} (2000) 034502,
  [\href{http://xxx.lanl.gov/abs/hep-lat/9907004}{{\tt hep-lat/9907004}}].

\bibitem{Cox:1999nt}
J.~Cox, C.~Gattringer, K.~Holland, B.~Scarlet, and U.~J. Wiese, {\it
  {Meron-cluster solution of fermion and other sign problems}},  {\em Nucl.
  Phys. Proc. Suppl.} {\bf 83} (2000) 777--791,
  [\href{http://xxx.lanl.gov/abs/hep-lat/9909119}{{\tt hep-lat/9909119}}].

\bibitem{Son:2000xc}
D.~T. Son and M.~A. Stephanov, {\it {QCD at finite isospin density}},  {\em
  Phys. Rev. Lett.} {\bf 86} (2001) 592--595,
  [\href{http://xxx.lanl.gov/abs/hep-ph/0005225}{{\tt hep-ph/0005225}}].

\bibitem{Son:2000by}
D.~T. Son and M.~A. Stephanov, {\it {QCD at finite isospin density: From pion
  to quark antiquark condensation}},  {\em Phys. Atom. Nucl.} {\bf 64} (2001)
  834--842, [\href{http://xxx.lanl.gov/abs/hep-ph/0011365}{{\tt
  hep-ph/0011365}}].

\bibitem{Kogut:2002tm}
J.~B. Kogut and D.~K. Sinclair, {\it {Quenched lattice QCD at finite isospin
  density and related theories}},  {\em Phys. Rev.} {\bf D66} (2002) 014508,
  [\href{http://xxx.lanl.gov/abs/hep-lat/0201017}{{\tt hep-lat/0201017}}].

\bibitem{Kogut:2002zg}
J.~B. Kogut and D.~K. Sinclair, {\it {Lattice QCD at finite isospin density at
  zero and finite temperature}},  {\em Phys. Rev.} {\bf D66} (2002) 034505,
  [\href{http://xxx.lanl.gov/abs/hep-lat/0202028}{{\tt hep-lat/0202028}}].

\bibitem{Kogut:2002cm}
J.~B. Kogut, D.~Toublan, and D.~K. Sinclair, {\it {The phase diagram of four
  flavor SU(2) lattice gauge theory at nonzero chemical potential and
  temperature}},  {\em Nucl. Phys.} {\bf B642} (2002) 181--209,
  [\href{http://xxx.lanl.gov/abs/hep-lat/0205019}{{\tt hep-lat/0205019}}].

\bibitem{Kogut:2003ju}
J.~B. Kogut, D.~Toublan, and D.~K. Sinclair, {\it {The pseudo-Goldstone
  spectrum of 2-colour QCD at finite density}},  {\em Phys. Rev.} {\bf D68}
  (2003) 054507, [\href{http://xxx.lanl.gov/abs/hep-lat/0305003}{{\tt
  hep-lat/0305003}}].

\bibitem{Kogut:2004zg}
J.~B. Kogut and D.~K. Sinclair, {\it {The finite temperature transition for
  2-flavor lattice QCD at finite isospin density}},  {\em Phys. Rev.} {\bf D70}
  (2004) 094501, [\href{http://xxx.lanl.gov/abs/hep-lat/0407027}{{\tt
  hep-lat/0407027}}].

\bibitem{deForcrand:2007uz}
P.~de~Forcrand, M.~A. Stephanov, and U.~Wenger, {\it {On the phase diagram of
  QCD at finite isospin density}},  {\em PoS} {\bf LAT2007} (2007) 237,
  [\href{http://xxx.lanl.gov/abs/0711.0023}{{\tt arXiv:0711.0023}}].

\bibitem{Bernard:1992qa}
V.~Bernard, N.~Kaiser, J.~Kambor, and U.~G. Meissner, {\it {Chiral structure of
  the nucleon}},  {\em Nucl. Phys.} {\bf B388} (1992) 315--345.

\bibitem{Frink:2002ht}
M.~Frink, B.~Kubis, and U.-G. Meissner, {\it {Analysis of the pion kaon
  sigma-term and related topics}},  {\em Eur. Phys. J.} {\bf C25} (2002)
  259--276, [\href{http://xxx.lanl.gov/abs/hep-ph/0203193}{{\tt
  hep-ph/0203193}}].

\bibitem{Tiburzi:2008bk}
B.~C. Tiburzi and A.~Walker-Loud, {\it {Hyperons in Two Flavor Chiral
  Perturbation Theory}},  {\em Phys. Lett.} {\bf B669} (2008) 246--253,
  [\href{http://xxx.lanl.gov/abs/0808.0482}{{\tt arXiv:0808.0482}}].

\bibitem{Jiang:2009sf}
F.-J. Jiang and B.~C. Tiburzi, {\it {Hyperon Axial Charges in Two-Flavor Chiral
  Perturbation Theory}},  {\em Phys. Rev.} {\bf D80} (2009) 077501,
  [\href{http://xxx.lanl.gov/abs/0905.0857}{{\tt arXiv:0905.0857}}].

\bibitem{Mai:2009ce}
M.~Mai, P.~C. Bruns, B.~Kubis, and U.-G. Meissner, {\it {Aspects of
  meson-baryon scattering in three- and two- flavor chiral perturbation
  theory}},  {\em Phys. Rev.} {\bf D80} (2009) 094006,
  [\href{http://xxx.lanl.gov/abs/0905.2810}{{\tt arXiv:0905.2810}}].

\bibitem{Tiburzi:2009cf}
B.~C. Tiburzi, {\it {Two-Flavor Chiral Perturbation Theory for Hyperons}},
  \href{http://xxx.lanl.gov/abs/0908.2582}{{\tt arXiv:0908.2582}}.

\bibitem{Rentmeester:2003mf}
M.~C.~M. Rentmeester, R.~G.~E. Timmermans, and J.~J. de~Swart, {\it
  {Determination of the chiral coupling constants c(3) and c(4) in new p p and
  n p partial-wave analyses}},  {\em Phys. Rev.} {\bf C67} (2003) 044001,
  [\href{http://xxx.lanl.gov/abs/nucl-th/0302080}{{\tt nucl-th/0302080}}].

\bibitem{Fettes:1998ud}
N.~Fettes, U.-G. Meissner, and S.~Steininger, {\it {Pion nucleon scattering in
  chiral perturbation theory. I: Isospin-symmetric case}},  {\em Nucl. Phys.}
  {\bf A640} (1998) 199--234,
  [\href{http://xxx.lanl.gov/abs/hep-ph/9803266}{{\tt hep-ph/9803266}}].

\bibitem{Oller:2000fj}
J.~A. Oller and U.~G. Meissner, {\it {Chiral dynamics in the presence of bound
  states: Kaon nucleon interactions revisited}},  {\em Phys. Lett.} {\bf B500}
  (2001) 263--272, [\href{http://xxx.lanl.gov/abs/hep-ph/0011146}{{\tt
  hep-ph/0011146}}].

\bibitem{Meissner:2004jr}
U.~G. Meissner, U.~Raha, and A.~Rusetsky, {\it {Spectrum and decays of kaonic
  hydrogen}},  {\em Eur. Phys. J.} {\bf C35} (2004) 349--357,
  [\href{http://xxx.lanl.gov/abs/hep-ph/0402261}{{\tt hep-ph/0402261}}].

\bibitem{Borasoy:2005ie}
B.~Borasoy, R.~Nissler, and W.~Weise, {\it {Chiral dynamics of kaon nucleon
  interactions, revisited}},  {\em Eur. Phys. J.} {\bf A25} (2005) 79--96,
  [\href{http://xxx.lanl.gov/abs/hep-ph/0505239}{{\tt hep-ph/0505239}}].

\bibitem{Oller:2005ig}
J.~A. Oller, J.~Prades, and M.~Verbeni, {\it {Surprises in threshold antikaon
  nucleon physics}},  {\em Phys. Rev. Lett.} {\bf 95} (2005) 172502,
  [\href{http://xxx.lanl.gov/abs/hep-ph/0508081}{{\tt hep-ph/0508081}}].

\bibitem{Gasser:1990ce}
J.~Gasser, H.~Leutwyler, and M.~E. Sainio, {\it {Sigma term update}},  {\em
  Phys. Lett.} {\bf B253} (1991) 252--259.

\bibitem{Buettiker:1999ap}
P.~Buettiker and U.-G. Meissner, {\it {Pion nucleon scattering inside the
  Mandelstam triangle}},  {\em Nucl. Phys.} {\bf A668} (2000) 97--112,
  [\href{http://xxx.lanl.gov/abs/hep-ph/9908247}{{\tt hep-ph/9908247}}].

\bibitem{Ohki:2009mt}
H.~Ohki {\em et.~al.}, {\it {Nucleon sigma term and strange quark content in
  2+1-flavor QCD with dynamical overlap fermions}},
  \href{http://xxx.lanl.gov/abs/0910.3271}{{\tt arXiv:0910.3271}}.

\bibitem{WalkerLoud:2008bp}
A.~Walker-Loud {\em et.~al.}, {\it {Light hadron spectroscopy using domain wall
  valence quarks on an Asqtad sea}},  {\em Phys. Rev.} {\bf D79} (2009) 054502,
  [\href{http://xxx.lanl.gov/abs/0806.4549}{{\tt arXiv:0806.4549}}].

\bibitem{WalkerLoud:2008pj}
A.~Walker-Loud, {\it {New lessons from the nucleon mass, lattice QCD and heavy
  baryon chiral perturbation theory}},  {\em PoS} {\bf LATTICE2008} (2008) 005,
  [\href{http://xxx.lanl.gov/abs/0810.0663}{{\tt arXiv:0810.0663}}].

\bibitem{Nielsen:1980rz}
H.~B. Nielsen and M.~Ninomiya, {\it {Absence of Neutrinos on a Lattice. 1.
  Proof by Homotopy Theory}},  {\em Nucl. Phys.} {\bf B185} (1981) 20.

\bibitem{Nielsen:1981xu}
H.~B. Nielsen and M.~Ninomiya, {\it {Absence of Neutrinos on a Lattice. 2.
  Intuitive Topological Proof}},  {\em Nucl. Phys.} {\bf B193} (1981) 173.

\bibitem{Nielsen:1981hk}
H.~B. Nielsen and M.~Ninomiya, {\it {No Go Theorem for Regularizing Chiral
  Fermions}},  {\em Phys. Lett.} {\bf B105} (1981) 219.

\bibitem{Karsten:1981gd}
L.~H. Karsten, {\it {Lattice fermions in Euclidean Space}},  {\em Phys. Lett.}
  {\bf B104} (1981) 315.

\bibitem{Wilczek:1987kw}
F.~Wilczek, {\it {On lattice fermions}},  {\em Phys. Rev. Lett.} {\bf 59}
  (1987) 2397.

\bibitem{Borici:2007kz}
A.~Borici, {\it {Creutz fermions on an orthogonal lattice}},  {\em Phys. Rev.}
  {\bf D78} (2008) 074504, [\href{http://xxx.lanl.gov/abs/0712.4401}{{\tt
  arXiv:0712.4401}}].

\bibitem{Celmaster:1982ht}
W.~Celmaster, {\it {Gauge theories on the body-centered hypercubic lattice}},
  {\em Phys. Rev.} {\bf D26} (1982) 2955.

\bibitem{Celmaster:1983jq}
W.~Celmaster and F.~Krausz, {\it {Fermion mutilation on a body-centered
  tesseract}},  {\em Phys. Rev.} {\bf D28} (1983) 1527.

\bibitem{Pernici:1994yj}
M.~Pernici, {\it {Chiral invariance and lattice fermions with minimal
  doubling}},  {\em Phys. Lett.} {\bf B346} (1995) 99--105,
  [\href{http://xxx.lanl.gov/abs/hep-lat/9411012}{{\tt hep-lat/9411012}}].

\bibitem{Aoki:2009ix}
{\bf CS} Collaboration, S.~Aoki {\em et.~al.}, {\it {Physical Point Simulation
  in 2+1 Flavor Lattice QCD}},  \href{http://xxx.lanl.gov/abs/0911.2561}{{\tt
  arXiv:0911.2561}}.

\bibitem{Lellouch:2009fg}
L.~Lellouch, {\it {Kaon physics: a lattice perspective}},  {\em PoS} {\bf
  LATTICE2008} (2009) 015, [\href{http://xxx.lanl.gov/abs/0902.4545}{{\tt
  arXiv:0902.4545}}].

\bibitem{Golterman:xw}
M.~Golterman, {\it Lattice chiral gauge theories},
  \href{http://xxx.lanl.gov/abs/hep-lat/0011027v1}{{\tt hep-lat/0011027v1}}.

\bibitem{Karsch:1982ve}
F.~Karsch, {\it {SU(N) Gauge Theory Couplings on Asymmetric Lattices}},  {\em
  Nucl. Phys.} {\bf B205} (1982) 285--300.

\bibitem{Burgers:1987mb}
G.~Burgers, F.~Karsch, A.~Nakamura, and I.~O. Stamatescu, {\it {QCD on
  anisotropic}},  {\em Nucl. Phys.} {\bf B304} (1988) 587.

\bibitem{Engels:1981qx}
J.~Engels, F.~Karsch, H.~Satz, and I.~Montvay, {\it {Gauge Field Thermodynamics
  for the SU(2) Yang-Mills System}},  {\em Nucl. Phys.} {\bf B205} (1982) 545.

\bibitem{Engels:1990vr}
J.~Engels, J.~Fingberg, F.~Karsch, D.~Miller, and M.~Weber, {\it
  {Nonperturbative thermodynamics of SU(N) gauge theories}},  {\em Phys. Lett.}
  {\bf B252} (1990) 625--630.

\bibitem{Namekawa:2001ih}
{\bf CP-PACS} Collaboration, Y.~Namekawa {\em et.~al.}, {\it {Thermodynamics of
  SU(3) gauge theory on anisotropic lattices}},  {\em Phys. Rev.} {\bf D64}
  (2001) 074507, [\href{http://xxx.lanl.gov/abs/hep-lat/0105012}{{\tt
  hep-lat/0105012}}].

\bibitem{Levkova:2006gn}
L.~Levkova, T.~Manke, and R.~Mawhinney, {\it {Two-flavor QCD thermodynamics
  using anisotropic lattices}},  {\em Phys. Rev.} {\bf D73} (2006) 074504,
  [\href{http://xxx.lanl.gov/abs/hep-lat/0603031}{{\tt hep-lat/0603031}}].

\bibitem{Morrin:2006tf}
R.~Morrin, A.~O. Cais, M.~Peardon, S.~M. Ryan, and J.-I. Skullerud, {\it
  {Dynamical QCD simulations on anisotropic lattices}},  {\em Phys. Rev.} {\bf
  D74} (2006) 014505, [\href{http://xxx.lanl.gov/abs/hep-lat/0604021}{{\tt
  hep-lat/0604021}}].

\bibitem{Hashimoto:2003fs}
S.~Hashimoto and M.~Okamoto, {\it {Anisotropic lattice actions for heavy
  quark}},  {\em Phys. Rev.} {\bf D67} (2003) 114503,
  [\href{http://xxx.lanl.gov/abs/hep-lat/0302012}{{\tt hep-lat/0302012}}].

\bibitem{Bar:2002nr}
O.~Bar, G.~Rupak, and N.~Shoresh, {\it {Simulations with different lattice
  Dirac operators for valence and sea quarks}},  {\em Phys. Rev.} {\bf D67}
  (2003) 114505, [\href{http://xxx.lanl.gov/abs/hep-lat/0210050}{{\tt
  hep-lat/0210050}}].

\bibitem{Drouffe:1983kq}
J.~M. Drouffe and K.~J.~M. Moriarty, {\it {GAUGE THEORIES ON A SIMPLICIAL
  LATTICE}},  {\em Nucl. Phys.} {\bf B220} (1983) 253--268.

\bibitem{Gasser:1987zq}
J.~Gasser and H.~Leutwyler, {\it {Spontaneously Broken Symmetries: Effective
  Lagrangians at Finite Volume}},  {\em Nucl. Phys.} {\bf B307} (1988) 763.

\bibitem{Beane:2009gs}
S.~R. Beane {\em et.~al.}, {\it {High Statistics Analysis using Anisotropic
  Clover Lattices: (II) Three-Baryon Systems}},  {\em Phys. Rev.} {\bf D80}
  (2009) 074501, [\href{http://xxx.lanl.gov/abs/0905.0466}{{\tt
  arXiv:0905.0466}}].

\bibitem{Bedaque:2009md}
P.~F. Bedaque, M.~I. Buchoff, A.~Cherman, and R.~P. Springer, {\it {Can
  fermions save large N dimensional reduction?}},  {\em JHEP} {\bf 10} (2009)
  070, [\href{http://xxx.lanl.gov/abs/0904.0277}{{\tt arXiv:0904.0277}}].

\bibitem{Bedaque:2009ri}
P.~F. Bedaque, M.~I. Buchoff, and R.~K. Mishra, {\it {Sommerfeld enhancement
  from Goldstone pseudo-scalar exchange}},  {\em JHEP} {\bf 11} (2009) 046,
  [\href{http://xxx.lanl.gov/abs/0907.0235}{{\tt arXiv:0907.0235}}].

\bibitem{Buchoff:2009za}
M.~I. Buchoff, A.~Cherman, and T.~D. Cohen, {\it {Color Superconductivity at
  Large N: A New Hope}},  \href{http://xxx.lanl.gov/abs/0910.0470}{{\tt
  arXiv:0910.0470}}.

\end{thebibliography}\endgroup
%
%
\appendix
\raggedbottom\sloppy
 
 
\appendix
\renewcommand{\thechapter}{A}

\chapter{Scalar Field on the Lattice}\label{ap:Scalar}
As in Ref.~\cite{Sharpe:1993wt}, the partition function for a theory with only a scalar field is given by (neglecting source terms)
\beq
Z = \int d [\phi] e^{iS(\phi)},
\eeq
where $d[\phi]= \prod_n d\phi_n$ represents all possible paths of $\phi$ and 
\beq
 S(\phi) = \int d^4x \mathcal{L}(\phi, \partial\phi).
 \eeq 
 Performing the Euclidean rotation, $x_0 \rightarrow i x_4$,
 \beq
  e^{iS(\phi)} \rightarrow e^{-S_E(\phi)}.
  \eeq
 The Euclidean correlation function is related to the vacuum expectation value of the time ordered fields,
 \beq
 \langle \phi(x_E) \phi(0) \rangle = Z_E^{-1} \int d [\phi] \phi(x_E)\phi(0)e^{-S_E(\phi)}= \langle 0 | T[\hat{\phi}(x_E)\hat{\phi}(0) | 0 \rangle
 \eeq
 The operator $\hat{\phi}(x_E)$ can be written as,
 \beq
 \hat{\phi}(x_E) = e^{-\hat{H}x_4 - i\hat{\mathbb{p}}\cdot\hat{\mathbb{x}}}\hat{\phi}(0) e^{-\hat{H}x_4 + i\hat{\mathbf{p}}\cdot \mathbf{x}}
 \eeq
 which results in the correlation function,
 \beq
  \langle \phi(x_E) \phi(0) \rangle = \langle 0 | \hat{\phi}(0) e^{-\hat{H}x_4 + i\hat{\mathbf{p}}\cdot \mathbf{x}} \hat{\phi}(0) | 0 \rangle .
  \eeq
  Inserting two complete set of states,
  \bea
    \langle \phi(x_E) \phi(0) \rangle &=& \sum_{n,m} N_{n m}\langle 0 | \hat{\phi}(0)|n\rangle\langle n| e^{-\hat{H}x_4 + i\hat{\mathbf{p}}\cdot \mathbf{x}}|m\rangle\langle m| \hat{\phi}(0) | 0 \rangle \nn\\
    &=& \sum_n N_n\big|\langle 0 | \hat{\phi}(0) |n\rangle\big|^2 e^{-E_n x_4} e^{\mathbf{p}\cdot \mathbf{x}}
    \eea
    where $N_n$ represents the proper finite volume normalization.  The key result is that the calculation of the correlation function is a sum of decaying exponentials whose arguments are the energy at each level multiplied by the length in the Euclidean time direction.  Since the ground state is the smallest energy, if we examine the correlation function at long times, the dominant term will be proportional to $e^{-E_0 x_4}$.  Thus, by taking logarithms of appropriate ratios at large $x_4$, we can extract the ground state energy, which for a two point correlator is the mass.
\appendix
\renewcommand{\thechapter}{B}

\chapter{Three or more flavor \CPT at NLO}\label{ap:Three_flav}
The NLO continuum \CPT Lagrangian contains operators at $\mc O(p^4)$, $\mc O(p^2m_q)$, and $O(m_q^2)$.  The continuum Lagrangian is given by
\bea
\mc L_{cont} &=& L_1 \Big(\tr(\partial_\mu \Sigma^\dag \partial^\mu \Sigma)\Big)^2 + L_2 \tr(\partial_\mu \Sigma^\dag \partial_\nu \Sigma)\tr(\partial^\mu \Sigma^\dag \partial^\nu \Sigma) \nn\\
&&+L_3 \tr(\partial_\mu \Sigma^\dag \partial^\mu \Sigma\partial_\nu \Sigma^\dag \partial^\nu \Sigma) \nn\\
&&+ L_4 \tr(\partial_\mu \Sigma^\dag \partial^\mu \Sigma)\tr(m_q \Sigma^\dag +\Sigma m_q) \nn \\
&&+L_5 \tr\big(((\partial_\mu \Sigma^\dag \partial^\mu \Sigma)(m_q \Sigma^\dag +\Sigma m_q)\Big) \nn\\
&&+L_6 \Big(\tr(m_q \Sigma^\dag +\Sigma m_q)\Big)^2+ L_7 \Big(\tr(m_q \Sigma^\dag -\Sigma m_q)\Big)^2\nn\\
&&+L_8 \tr(m_q \Sigma m_q \Sigma + \Sigma^\dag m_q \Sigma^\dag m_q).
\eea



\end{document}